\documentclass[11pt, a4paper]{article}
\usepackage[english]{babel}
\usepackage[utf8]{inputenc}

\usepackage{jheppub}
\usepackage{amsmath}
\usepackage{amssymb}
\usepackage{xcolor}
\usepackage{float}
\usepackage{enumerate}
\usepackage{slashed}

\usepackage{float}
\usepackage[normalem]{ulem}
\usepackage{mathrsfs} 
\usepackage{multicol}
\usepackage{cancel}
\usepackage{mathtools}
\usepackage{hyperref}

\usepackage{subfigure}

\usepackage[compat=1.1.0]{tikz-feynman}
\usepackage{xcolor}
\usepackage[export]{adjustbox}

\usepackage{latexsym}
\usepackage{mathrsfs}
\usepackage{amsthm}
\usepackage{amstext}

\usepackage{graphicx}
\usepackage{slashed}

\makeatletter
\@addtoreset{subfigure}{row}
\makeatother

\usepackage[rightcaption]{sidecap}
\usepackage{caption}

\theoremstyle{plain}
\newtheorem{thm}{Theorem}[section]

\newtheorem{srule}[thm]{Selection rule}

\DeclareMathOperator{\diag}{diag}

\newcommand {\beq} {\begin{equation}}
\newcommand {\eeq} {\end{equation}}
\newcommand{\bea}{\begin{eqnarray}}
\newcommand{\eea}{\end{eqnarray}}
\newcommand{\bit}{\begin{itemize}}
\newcommand{\eit}{\end{itemize}}
\def\nl{\nonumber \\}

\newcommand{\bal}{\begin{align}}
\newcommand{\eal}{\end{align}}
\newcommand{\bca}{\begin{cases}}
\newcommand{\eca}{\end{cases}}
\newcommand{\bmu}{\begin{multline}}
\newcommand{\emu}{\end{multline}}
\newcommand{\bga}{\begin{gather}}
\newcommand{\ega}{\end{gather}}
\def\mc{\mathcal}

\def\mk{\mathfrak}
\def\tx{\text}
\def\nl{\nonumber \\}

\def\bD{\bar{D}}

\def\a{\alpha}
\def\da{{\dot{\alpha}}}
\def\b{\beta}
\def\db{{\dot{\beta}}}
\def\G{\Gamma}
\def\g{\gamma}

\def\d{\delta}

\def\l{\lambda}
\def\bl{\bar{\lambda}}
\def\t{\theta}
\def\bt{\bar{\theta}}
\def\F{\Phi}
\def\bF{\bar{\Phi}}

\def\f{\varphi}
\def\fb{\bar{\varphi}}
\def\o{\omega}
\def\O{\Omega}
\def\e{\varepsilon}
\def\elc{\epsilon}
\def\s{\sigma}
\def\bs{\bar{\sigma}}

\def\k{\kappa}

\def\sq{\square}

\def\p{\partial}

\def\vk{\vec{k}}

\def\vp{\vec{p}}
\def\vx{\vec{x}}

\def\sqt{\sqrt{2}}

\def\p{\partial}

\def\a{\alpha}
\def\da{{\dot{\alpha}}}
\def\b{\beta}

\def\db{{\dot{\beta}}}
\def\c{\gamma}

\def\l{\lambda}
\def\bl{\bar{\lambda}}

\def\s{\sigma}
\def\bs{\bar{\sigma}}

\def\p{\partial}

\def\le{\left(}
\def\ri{\right)}

\def\beq{\begin{equation}}
\def\eeq{\end{equation}}
\def\t{{\theta}}
\def\bt{{\bar{\theta}}}

\title{Supersymmetric Galilean Electrodynamics} 

\author[a]{Stefano Baiguera,} 
\author[b]{Lorenzo Cederle}
\author[b,c]{and Silvia Penati}

\affiliation[a]{Department of Physics, Ben-Gurion University of the Negev,
Beer Sheva 84105, Israel}
\affiliation[b]{Universit\`a degli studi di Milano Bicocca, Piazza della Scienza 3, 20161, Milano, Italy}
\affiliation[c]{INFN, Sezione di Milano - Bicocca, Piazza della Scienza 3, 20161, Milano, Italy}

\emailAdd{baiguera@post.bgu.ac.il}
\emailAdd{l.cederle@campus.unimib.it}
\emailAdd{silvia.penati@mib.infn.it}

\abstract{In 2 + 1 dimensions, we propose a renormalizable non-linear sigma model action which describes the ${\cal N} = 2$ supersymmetric generalization of Galilean Electrodynamics. We first start with the simplest model obtained by null reduction of the relativistic Abelian ${\cal N} = 1$ supersymmetric QED in 3 + 1 dimensions and study its renormalization properties directly in non-relativistic superspace. Despite the existence of a non-renormalization theorem induced by the causal structure of the non-relativistic dynamics, we find that the theory is non-renormalizable. Infinite dimensionless, supersymmetric and gauge-invariant terms, which combine into an analytic function, are generated at quantum level. Renormalizability is then restored by generalizing the theory to a non-linear sigma model where the usual minimal coupling between gauge and matter is complemented by infinitely many marginal couplings driven by a dimensionless gauge scalar and its fermionic superpartner. Superconformal invariance is preserved in correspondence of a non-trivial conformal manifold of fixed points where the theory is gauge-invariant and interacting. }

\begin{document}

\maketitle

\section{Introduction}

Symmetries play a crucial role in modern physics, in that they govern the behaviour of observables and greatly restrict the theoretical models that describe physical phenomena.

In this context, it has been known from a long time that Lorentz invariance is a distinctive trait in the description of electromagnetism, whose theoretical formulation is encoded by Maxwell's theory.
While this model describes several physical phenomena with striking precision, it is still relevant to study its non-relativistic limit \cite{Bellac1973GalileanE,Santos_2004,Festuccia:2016caf}.
One of the reasons is that the investigation of this corner of electromagnetism could teach us lessons on the relativistic case itself.
Indeed, it is sometimes not clear whether certain phenomena, not only the ones involving the electromagnetic interaction, are consequences of the Lorentz invariance of the theory, or if they would manifest even when Galilean symmetry is present.
For example, time dilatation effects and a strong gravitational description of the Schwarzschild black hole can be obtained in a purely non-relativistic setting, using a torsionful connection and a vanishing Newtonian potential, without resorting to full general relativity \cite{VandenBleeken:2017rij,Hansen:2020pqs}.

From a theoretical perspective, the Galilean version of electromagnetism provides a non-trivial example of non-relativistic QFT with massless degrees of freedom, which can be coupled in a covariant way to a curved background described by Newton-Cartan geometry \cite{CartanSurLV,Friedrichs1928}.
The renormalization properties of this theory, which was called Galilean Electrodynamics (GED), were studied in \cite{Chapman:2020vtn}.
GED also arises as the linearized action on D-branes in non-relativistic open string theory \cite{Gomis:2020fui}.
Other investigations of Galilean-invariant gauge theories were considered in \cite{PhysRevD.31.848,Bagchi:2022twx,Banerjee:2022uqj}.

From a condensed matter perspective, there are several reasons to consider non-relativistic limits.
Emergent symmetries arising in the infrared are often different from the invariances of the microscopic description, and in particular the Lorentz group may not be present.
Non-relativistic symmetries govern the realm of cold atoms \cite{Nishida:2010tm}, fermions at unitarity \cite{Son:2008ye}, quantum Hall effect \cite{Geracie:2014nka}, strange metallic phases \cite{Hartnoll:2009ns} and quantum mechanical problems like the Efimov effect \cite{Bedaque:1998km}.

Supersymmetry (SUSY) has been studied for several years as a candidate to uncover physics beyond the Standard Model.
In concrete condensed matter applications, SUSY may arise as an emergent symmetry, \emph{e.g.}, in the tricritical Ising model \cite{FRIEDAN198537}, in topological superconductors \cite{Grover:2013rc}, optical lattices \cite{Yu:2010zv} or other settings.
From a theoretical point of view, supersymmetry strongly constrains the analytic structure of the effective action, and controls the running of physical couplings along the RG flow, leading to exact results and non-renormalization theorems \cite{Grisaru:1979wc,Seiberg:1993vc}.
Furthermore, even if supersymmetry plays an indirect role in holography, most of the examples where the AdS/CFT correspondence is explicitly tested are supersymmetric.

Therefore, it does not appear surprising that several investigations of theories which are both supersymmetric and non-relativistic had a great revival in recent years.
Starting from the first investigations involving the SUSY generalization of the Galilean algebra and limits of the relativistic models \cite{CLARK198491,doi:10.1063/1.529465,Bergman:1995zr}, there have been studies of superconformal anyons \cite{Leblanc:1992wu}, spontaneous SUSY breaking \cite{Meyer:2017zfg}, the analysis of the renormalization properties of supersymmetric Galilean or Lifshitz-invariant models \cite{Auzzi:2019kdd,Arav:2019tqm}, supergravity \cite{Bergshoeff:2014ahw,Bergshoeff:2015uaa} and the study of non-relativistic corners of $\mathcal{N}=4$ super Yang-Mills (SYM) theory \cite{Harmark:2014mpa,Harmark:2019zkn,Baiguera:2020jgy,Baiguera:2020mgk,Baiguera:2021hky}.

In this work, we study the renormalization properties of 2+1 dimensional, ${\cal N}=2$ Supersymmetric Galilean Electrodynamics (SGED), \emph{i.e.}, the supersymmetric generalization of GED obtained from the null reduction of Abelian $\mathcal{N}=1$ SUSY QED in 3+1 dimensions.
Classically, the theory is both supersymmetric and Schr\"{o}dinger-invariant, and the two symmetries combine into the non-relativistic version of superconformal invariance \cite{Sakaguchi:2008ku,Nakayama:2008qm}.
Given the peculiar behaviour of both the Galilean invariance and supersymmetry at quantum level, it is interesting to study the interplay between them in this framework.
Indeed, the global $\mathrm{U}(1)$ particle number symmetry typical of non-relativistic theories, together with the retarded nature of their propagator, are usually responsible for cancellations at quantum level, first observed in \cite{Bergman:1991hf,Klose:2006dd}.
This tendency is also present and enhanced in supersymmetric settings, leading to the one-loop exactness of the Galilean version of the 2+1 dimensional $\mathcal{N}=2$ Wess-Zumino model \cite{Auzzi:2019kdd}, and to similar non-renormalization theorems in the Lifshitz case \cite{Arav:2019tqm}.
On the contrary, GED presents an intricate renormalization structure: the theory is non-renormalizable and an infinite number of marginal deformations with non-trivial $\beta$ functions need to be added to the action.
After deforming the theory with a set of marginal deformations closed at quantum level, one can find the existence of conformal manifolds, where Schr\"{o}dinger symmetry is preserved.
Still, the model admits a non-renormalization theorem, in that the electric charge is protected and does not run at quantum level \cite{Chapman:2020vtn}.

Quite surprisingly, we find that supersymmetry does not significantly improve the renormalization properties observed in the GED case.
First of all, we derive a non-renormalization theorem which protects the coupling constant $g$ of the theory at quantum level, providing a SUSY generalization of the analogous result for the electric charge in GED.
We then find that the model still admits an infinite number of marginal deformations, that are generated by one-loop radiative corrections to the effective action. 
This forces us to deform the original SGED interactions into a non-linear sigma model governed by a single analytic function ${\cal F}$. In a full superspace set-up, the SGED action which is quantistically meaninful is 
\beq
S_{\rm SGED} =  \int d^3 x  d^2 \theta \, W^2 + \int d^3 x  d^4 \theta \; \bF e^{g V} \Phi \, {\cal F}( \bar{D}_2 D_2 V)
\label{eq:non_linear_sigma_model}
\eeq
where the first term resembles the ordinary, relativistic pure gauge action in superspace, whereas the second term describes the coupling between the gauge sector and matter degrees of freedom encoded into (anti)chiral $\Phi, \bar{\Phi}$ superfields. It is interesting to observe that in the Galilean framework the gauge-matter coupling is driven not only by the usual minimal coupling $\bF e^{g V} \Phi$, but also by an infinite number of couplings of the form $\bF \Phi (\bar{D}_2 D_2 V)^n$ coming from the expansion of ${\cal F}$. This is due to the anysotropic assignment of dimensions in Galilean superspace that allows to construct the dimensionless and gauge-invariant superfield $\bar{D}_2 D_2 V$, in terms of the gauge prepotential $V$ and dimensionless SUSY covariant derivatives. 

Equation \eqref{eq:non_linear_sigma_model} is our proposal for a consistent theory of non-relativistic, supersymmetric electrodynamics. We prove that this theory is one-loop renormalizable and compute the one-loop beta functions for the infinite sigma model couplings. We prove that the theory admits an interacting superconformal manifold of fixed points, where the minimal coupling is turned off and the gauge-matter interaction is driven only by the ${\cal F}$ couplings. 

The paper is organized as follows.
In section \ref{sec:nonrelQFT_review} we briefly review the preliminaries on non-relativistic QFTs, focusing on the null reduction method and its application to theories with either supersymmetry or gauge invariance.
Then, in section \ref{sec:SGED_null_reduction} we combine the two symmetries together in a Galilean-invariant setting to define the classical SGED action as obtained by null reduction of the relativistic four dimensional parent.
In section \ref{sec:one_loop_radiative_corrections} we study the corresponding quantum theory and compute the one-loop corrections to the effective action using the superfield formalism.
This procedure generates an infinite series of new UV divergent terms, whose renormalization requires to generalize the original null reduced model to a non-linear sigma model. 
The expert reader may want to jump directly to section \ref{sec:renormalizable_SGED}, where we use a covariant approach to compute one-loop quantum corrections to the new theory, including the original contributions.
We collect some conclusions and a discussion on future developments in section \ref{sec:conclusions}.
Technical details are reported in a few appendices. 
Appendix \ref{app-conv} contains the conventions on spinors, superspace and covariant derivatives.
The anisotropic dimensional counting of Galilean-invariant theories is summarized in appendix \ref{app-dim}.
We collect in appendix \ref{app:math-tools} the relevant mathematical tools used to compute the integrals entering one-loop corrections.
Appendix \ref{app:additional_loop_corrections} is devoted to prove that a certain class of vertices is not generated at quantum level. Finally, in appendix \ref{app:details_covariant} we introduce the essential data for the covariant formalism, where we also list the properties of covariant derivatives needed to compute the diagrams presented in the main text.

\section{Non-relativistic QFT: a short review}
\label{sec:nonrelQFT_review}

In this section we introduce the basic ingredients necessary to investigate Galilean-invariant supersymmetric gauge theories.
We first review the null reduction method, which allows to construct non-relativistic theories starting from a relativistic parent in one higher dimension. We introduce the resulting non-relativistic superconformal algebra and the prescription to define the non-relativistic superspace.
We then use this method to build non-relativistic actions.
In particular, we review two relevant applications of this technique, the Galilean Wess-Zumino model and the Galilean electrodynamics.

\subsection{Null reduction and supersymmetry algebra}
\label{sec:null_reduction}

We start from a $d+2$--dimensional Lorentzian manifold described by spacetime coordinates (here $i \in \lbrace 1,2, \dots, d \rbrace$)
\beq
x^N = (x^-, x^+, x^{i}) \equiv (x^-, x^\mu) \; , \qquad {\rm where} \qquad x^{\pm} = \frac{x^d \pm x^0}{\sqrt{2}}
\eeq
with metric $\eta_{MN} dx^M dx^N = 2 dx^+ dx^- + dx^i dx^i .$
Null reduction consists in the dimensional reduction of a relativistic theory obtained by compactifying the $x^-$ direction along a small circle  \cite{Duval:1984cj,Duval:1993hs,Julia:1994bs,Duval:1990hj}.
This procedure gives a natural embedding of the Schr\"{o}dinger group $\mathrm{Sch}(d)$ in $d$ spatial dimensions inside the conformal group $\mathrm{O}(d,2)$ in $d+2$ spacetime dimensions \cite{Son:2008ye}. 
In the following, we will refer specifically to the $d=2$ case. 

The algebra of Schrodinger generators is given by
\beq
\begin{aligned}
& [P_j, K_k] = i \delta_{jk} M \, , \qquad
[H, K_j] = i P_j \, , & \\ 
& [P_j, J] = -i \epsilon_{jk} P_k \, , \qquad
[K_j, J] = -i \epsilon_{jk} K_k  \, ,   \qquad (j,k=1,2) &
\end{aligned}
\label{Bargmann_algebra} 
\eeq
\beq
\begin{aligned}
& [D, P_j] =  - i P_j \, , \qquad
[D,K_j] = i K_j \, , \qquad
[D,H] = -2 i H \, , & \\
& [D,C] = 2 i C \, , \qquad
[H,C] = i D &
\end{aligned}
\label{eq:Schroedinger_algebra}
\eeq
where $P_j$ are the spatial components of the momentum, $H$ is the Hamiltonian, $ K_j $ are the generators of Galilean boosts, $ J $ is the planar angular momentum, $D$ the dilatation operator and $C$ the generator of special conformal transformations.
The algebra has a $\mathrm{U}(1)$ central extension with central charge $M$ corresponding to the mass or particle number conservation.

The supersymmetric extension of the Schr\"{o}dinger algebra can be obtained by performing null reduction of the four dimensional $\mathcal{N}=1$ supersymmetry algebra. In particular, the fermionic part of the algebra is mapped into \cite{Bergman:1995zr}
\beq
\begin{aligned}
& [J,Q_1] =  \frac12 Q_1 \, , \quad
[J,Q_2] = - \frac12 Q_2 \, , \quad
[Q_1, K_1 - i K_2] = -i Q_2 \, , & \\
&  \lbrace Q_1, Q_1^{\dagger}  \rbrace = \sqrt{2} H \, , \quad
\lbrace Q_2, Q_2^{\dagger}  \rbrace = \sqrt{2} M \, , & \\
& \lbrace Q_1, Q_2^{\dagger} \rbrace = -  (P_1 -i P_2) \, , \quad
\lbrace Q_2, Q_1^{\dagger} \rbrace = -  (P_1 + i P_2) 
\end{aligned}
\label{commu_superGalileo2} 
\eeq
where $Q_{\a}$ with $\alpha \in \lbrace 1,2 \rbrace$ are two complex supercharges. 
Since null reduction does not affect the number of fermionic generators, it follows that starting from the ${\cal N}=1$ superalgebra in four dimensions we obtain the non-relativistic ${\cal N}=2$ superalgebra in three dimensions. 

More generally, one can perform null reduction starting from the superconformal algebra $\mathrm{SU}(2,2|1)$.
In this case there is an additional complex fermionic generator $S,$ which is the superpartner of the bosonic generator $C$ of special conformal transformations.
The commutation relations \eqref{Bargmann_algebra}, \eqref{eq:Schroedinger_algebra} and \eqref{commu_superGalileo2} are supplemented by the following rules, which also include the bosonic generator $R$ of $\mathrm{U}(1)$ R-symmetry \cite{Leblanc:1992wu,Sakaguchi:2008ku,Nakayama:2008qm}
\beq
\begin{aligned}
& \lbrace S, S^{\dagger} \rbrace = C \, , \qquad
\lbrace S, Q_2^{\dagger} \rbrace = - (K_1 + i K_2) \, , \qquad
&  \\
& 
[H, S^{\dagger}] = i Q_1^{\dagger} \, , \qquad
[J,S] = - \frac{1}{2} S \, , \qquad
[D, S] = - i S \, , \qquad
[C, Q_2] = - i S & \\
& \lbrace S, Q_1^{\dagger}  \rbrace = \frac{i}{2} \le i D - J  +\frac{3}{2} R \ri \, , \qquad
[R, Q_{\a}] = - Q_{\a}  \, , \qquad
[R,S] = - S &
\end{aligned}
\label{eq:superconformal_algebra}
\eeq

Supersymmetric theories are naturally formulated in superspace, where representations of the SUSY algebra (supermultiplets) are realized in terms of superfields. 
In the non-relativistic case, the ${\cal N}=2$ Galilean three-dimensional superspace \cite{Auzzi:2019kdd} 
is obtained as the null reduction of the four dimensional ${\cal N}=1$ relativistic superspace described by bosonic coordinates $x^N$, supplemented by spinorial coordinates $(\theta^\a, \bar{\theta}^{\dot\a})$, $\a \in \{1,2\}, \dot{\a} \in \{\dot{1},\dot{2}\}$. 
Null reduction fixes the coordinate dependence of any local superfield $\Psi$ to be
\beq\label{eq:nullred}
\Psi (x^N, \theta^\a, \bar{\theta}^{\da}) = e^{i M x^-} \tilde\Psi (x^{+}\equiv t, x^i, \theta^\a, (\theta^\a)^\dagger\equiv \bar{\theta}^\a) 
\eeq
where $M$ is the dimensionless eigenvalue of the $\mathrm{U}(1)$ mass operator.\footnote{We note that choosing $m$ to be dimensionless requires having rescaled $x^-$ by an energy dimension-one parameter to make it dimensionless, as well. This then implies that $x^+$ acquires scale dimension 2.}
Therefore, in  $2+1$ dimensions we identify 
\beq \label{eq:identifications}
\p_- \rightarrow iM \, , 
\qquad 
\p_+ \rightarrow \p_t 
\eeq
We note that prescription \eqref{eq:nullred} is SUSY preserving, as it assigns the same $\mathrm{U}(1)$ charge (or mass) to all the components. 
The details of the representations on chiral and vector superfields are collected in appendix \ref{app-conv}.

\subsection{Review of the Galilean Wess-Zumino model}
\label{sec:Galilean_WZmodel}

Null reduction not only allows to derive the Galilean algebra and its representations on fields, but it can also be used to build actions invariant under the non-relativistic symmetries, starting from a relativistic parent theory.
This method was used in \cite{Auzzi:2019kdd} to find a Galilean-invariant Wess-Zumino (WZ) model, whose action reads
\beq
\label{non-rel WZ action in superfield formalism} 
S = \int   d^3 x d^4 \theta \le \bar{\Phi}_1 \Phi_1 +  \bar{\Phi}_2 \Phi_2 \ri + g \int   d^3 x  d^2 \theta \,  \Phi_1^2 \Phi_2 + \mathrm{h.c.}
\eeq
with Berezin integration defined in \eqref{Berezin integration null reduction}.

The main feature of this action, compared to its relativistic counterpart, is that the integrand must be uncharged with respect to the global $\mathrm{U}(1)$ symmetry associated to the central extension of the Bargmann algebra.
For this reason, a non-vanishing superpotential exists only when at least two species of (anti)chiral superfields are chosen.
Eq.~\eqref{non-rel WZ action in superfield formalism} implies that the masses of the two matter superfields are $M_1=m$ and $M_2 = -2m.$

On general grounds, supersymmetry poses strong constraints on the dynamics of a theory, and often allows to obtain exact results.
In the relativistic WZ model, the existence of a non-renormalization theorem which states that the superpotential is quantum exact, forces all its loop corrections to vanish \cite{Grisaru:1979wc,Seiberg:1993vc} and perturbative corrections are allowed only for the K\"{a}hler potential.
The non-renormalization theorem for the superpotential remains true also in the non-relativistic version of the model \cite{Auzzi:2019kdd}. However, in this case there are additional constraints which are due to the retarded nature of the (anti)chiral propagator, contrarily to the causal Feynman propagator of the relativistic model. In fact, the particular structure of its poles forces the K\"{a}hler potential to be one-loop exact. We can then conclude that the Galilean WZ model is one-loop exact and the $\beta$-function of the theory is fully determined by \cite{Auzzi:2019kdd}
\beq
\beta_g=\frac{d g } {d \log \mu} =\frac{g^3}{ 4 \pi m}
\label{betafunction_WZ}
\eeq
Its non-vanishing value signals the breaking of the original scale invariance, due to quantum effects.

\subsection{Review of Galilean electrodynamics}
\label{sec:review_GED}

Null reduction can be also applied to gauge-invariant theories, \emph{e.g.} to obtain the Galilean Electrodynamics (GED) \cite{Festuccia:2016caf}.
The reduction of the ordinary relativistic gauge field $A_M$ is performed by requiring it to be $x^-$--independent\footnote{According to the general decomposition \eqref{eq:nullred} applied to fields rather than superfields, this corresponds to assigning it a vanishing mass.} and decomposing it as $A_M = (\varphi, A_{\mu}),$ where $\varphi$ is a real spacetime scalar and $A_{\mu} = (A_t, A_i)$ are the 3D components. The gauge field defines an electric and a magnetic field, according to
\beq\label{eq:emfields}
E_i = \p_t A_i - \p_i A_t \; , \qquad f_{ij} = \p_i A_j - \p_j A_i
\eeq
Once reduced, the relativistic Maxwell action reads
\beq
S_{\rm U(1)} = \int d^3 x \, \left[\frac{1}{2} \le \p_t \varphi \ri^2 + E^i \p_i \varphi - \frac{1}{4} f_{ij} f^{ij}  \right]
\label{eq:GED_action}
\eeq

The non-relativistic electrodynamics has no propagating degrees of freedom, since the speed of light is sent to infinity and the mediation becomes istantaneous.
It is possible to introduce propagating modes by coupling the system to matter fields, \emph{e.g.}, to a Schr\"{o}dinger scalar $\phi.$
The minimal coupling term reads
\beq
S_{\rm min} = \int d^3 x \, \left[ \frac{i}{2} \le \bar{\phi} \nabla_t \phi -  \phi \nabla_t \bar{\phi} \ri - \frac{1}{2  \mathcal{M}} \nabla_i \bar{\phi} \nabla^i \phi  \right]
\label{eq:scalar_Schrodinger_term}
\eeq
where we have introduced covariant derivatives acting on $\phi$ ($\bar{\phi}$) as
\beq
\nabla_t \equiv \p_t \mp i e A_t \, , \qquad
\nabla_i \equiv \p_i \mp i e A_i 
\label{eq:standard_cov_div}
\eeq
and we have defined a covariantized mass $ \mathcal{M} \equiv m - e \varphi.$
It is possible to build such a combination due to the fact that the field $\varphi$ arising from the null component of the parent relativistic connection is a scalar under gauge transformations, and in the anisotropic counting of dimensions it is dimensionless (see appendix \ref{app-dim} for the precise counting).

The sum of terms \eqref{eq:GED_action} and \eqref{eq:scalar_Schrodinger_term} gives the minimal realization of the scalar Galilean Electrodynamics, $S_{\rm GED} = S_{U(1)} + S_{\rm min}$. 

Based on the previous discussion, one could expect the renormalization properties of this theory to be simpler than its relativistic parent.
Surprisingly, this is not the case \cite{Chapman:2020vtn}.
The existence of the covariantized mass $\mathcal{M}$ is responsible for making the theory a $\sigma$--model, since it is necessary to series-expand the kinetic term in order to build standard Feynman rules and compute loop corrections.
It turns out that such a theory is non-renormalizable, instead an infinite number of marginal deformations dependent on the dimensionless scalar $\varphi$ need to be added to make the theory renormalizable. Therefore, the general structure of the renormalizable scalar GED action is $S_{\rm GED} + \Delta S_{\tx{GED}}$, where \cite{Chapman:2020vtn}
\beq
\label{eq:action_extra_shira}
\Delta S_{\tx{GED}} = \int dt \, d^2 x \biggl( \mc{J} [\mathcal{M}] \, \p^i \mathcal{M} \p_i \mathcal{M} \, \bar{\phi} \phi - \frac{\l}{4} \mc{V} [\mathcal{M}] \, (\bar{\phi} \phi)^2 - \mc{E}[\mathcal{M}] \, (\p^i \p_i \mathcal{M} - e^2 \bar{\phi} \phi) \, \bar{\phi} \phi \biggr)
\eeq
While there are still powerful non-renormalization theorems which protect the electric charge $e$ from running, the functionals $\mathcal{J}[\mathcal{M}], \mathcal{V}[\mathcal{M}]$ and $\mathcal{E}[\mathcal{M}]$ renormalize non-trivially \cite{Chapman:2020vtn}.
An interesting feature of scalar GED is that the freedom governed by these functionals allows to find conformal manifolds of fixed points where the Schr\"{o}dinger symmetry is preserved.

In section \ref{sec:SGED_null_reduction} we are going to consider a supersymmetric generalization of GED and investigate if supersymmetry improves the renormalization properties of the model.


\section{SGED from null reduction}
\label{sec:SGED_null_reduction}

The ${\cal N}=2$ supersymmetric generalization of GED can be obtained by performing the null reduction of a 4D  ${\cal N}=1$ supersymmetric gauge theory, directly in superspace. For simplicity, in section \ref{sec:SGED_action_superfields} we will consider the action for an Abelian gauge superfield coupled to matter described by one chiral superfield charged under the $\mathrm{U}(1)$  gauge group and with charge $m$ with respect to the $\mathrm{U}(1)$  central extension. 
We will then project the action in components in section \ref{sec:SGED_action_components}.

\subsection{Action in superfield formulation}
\label{sec:SGED_action_superfields}

Exploiting the definitions in appendix \ref{app-conv}, the simplest action for electrodynamics in Galilean superspace is the one obtained by null reduction of its relativistic counterpart. It reads
\beq\label{action}
S_{\rm nSGED} = \frac{1}{g^2} \int d^3x d^2\theta \;  W^2 + \int d^3x d^4\theta \; \bar{\Phi} e^V \! \Phi
\eeq
where $W^2 = \tfrac12 W^\a W_\a$ with $W_\a = i \bar{D}^2 D_\alpha V$, being $V$ a real scalar prepotential. The Berezin integration is defined in \eqref{Berezin integration null reduction}. 
Formally, this action has the same expression of supersymmetric electrodynamics in relativistic superspace. Differences are hidden in the superspace integrations that are here defined in terms of non-relativistic covariant derivatives \eqref{eq:nonrelD}. 
We will refer to eq.~\eqref{action} as the {\em null SGED} action, as it is what one obtains applying null reduction.

This action is manifestly invariant under supersymmetry. Moreover, it is invariant under the Schr\"{o}dinger group, as it arises from null reduction. 
According to the anisotropic dimensional scaling of Galilean theories, which we discuss in appendix \ref{app-dim}, the coupling constant $g$ is dimensionless, as it happens in the relativistic counterpart of the model.
Indeed, all the previous symmetries combine into the invariance under the full Galilean superconformal group with generators satisfying the algebra presented in eqs.~\eqref{Bargmann_algebra}--\eqref{eq:superconformal_algebra}.

The invariance of the integrand under the action of the $\mathrm{U}(1)$ mass generator requires to assign a vanishing mass to $V$. 
This is equivalent to the statement that under null reduction the prepotential satisfies $V (x^M, \theta^{\alpha}, \bar{\theta}^{\dot{\alpha}}) = V (x^{\mu}, \theta^{\alpha}, \bar{\theta}^{\beta})$, \emph{i.e.}, it originates from a 4D vector superfield that does not depend on the null direction $x^{-}$. 
Taking into account that its $\theta \bar\theta$ component is the gauge field $A_{\a\b}$ (see expansion \eqref{eq:real}), this is the supersymmetric generalization of the requirement that the gauge field of the parent theory is $\p_-$-invariant and splits into $A_M = (\varphi, A_{\mu})$, as reviewed in section \ref{sec:review_GED}. From a physical point of view, this is equivalent to say that it acts as an instantaneous mediator of interactions between the matter (super)fields.

In addition, action \eqref{action} enjoys local invariance under the supergauge transformations 
\beq
\begin{aligned}
\label{eq:gaugetransf}
& V \to V' = V + i (\bar\Lambda - \Lambda) & \\
& \Phi \to \Phi' =  e^{i\Lambda} \Phi & \\
& \bar\Phi \to \bar\Phi' =  \bar\Phi e^{-i\bar\Lambda} &
\end{aligned}
\eeq
driven by chiral and antichiral parameters, $\bar{D}_\a \Lambda = D_\a \bar\Lambda =0$. 

Dimensional analysis in eq.~\eqref{eq:dimension_counting_derivatives} shows that the vector superfield $V$ and the covariant derivatives $D_2, \bar{D}_2$ are dimensionless.
Therefore, it is possible to build infinitely many marginal deformations of the action in \eqref{action}, in analogy with what happens in the non-supersymmetric case (see eq.~\eqref{eq:action_extra_shira}). 
Requiring these deformations to be invariant under gauge transformations \eqref{eq:gaugetransf}, it turns out that an infinite set of deformations of the SGED action consistent with both supersymmetry and gauge invariance can be written in the form
\beq
\Delta S_{\rm nSGED} = \int d^4 \theta \; \bar{\Phi} e^{V} \!\Phi \, \mathcal{F} [\bar{D}_2 D_2 V]
\label{eq:guess_additional_terms}
\eeq
where $\mathcal{F}$ is an analytic function (\emph{i.e.}, it can be expanded in Taylor series) of the gauge-invariant combination $\bar{D}_2 D_2 V.$
We will see in section \ref{sec:one_loop_radiative_corrections} that a specific set of these corrections is indeed generated at quantum level.

\subsection{Action in components}
\label{sec:SGED_action_components}

Applying the prescription in \eqref{Berezin integration null reduction} for the Berezin integration, we obtain the action \eqref{action} in components.\footnote{The relevant definitions required to perform the explicit expansion are collected in appendix \ref{app-conv}, specifically the covariant derivatives in eq.~\eqref{eq:nonrelD} and the superfield expansions \eqref{chiral} and \eqref{eq:real} in the Wess-Zumino gauge \eqref{eq:WZ_gauge}.}
The result reads
\beq
\label{eq:action_splitting}
 S_{\rm nSGED} = S_{\tx{gauge}} + S_{\tx{matter}} + S_{\tx{int}}
 \eeq
with
\beq
\begin{aligned}
S_{\tx{gauge}}  = \frac{1}{g^2} \int d^3x \, & \left[  
\tilde{D}^2 + \sqt i \bar{\lambda}_2 \, \p_t \lambda_2 - i \bar{\lambda}_1 (\p_1 - i \p_2) \lambda_2 - i \bar{\lambda}_2 (\p_1 + i \p_2) \lambda_1 \right. \\
& \left. + \frac12 (\p_t \f)^2 + E^i \p_i \f - \frac14 f^{ij} f_{ij}  \right]
\end{aligned}
\label{eq:gauge_action_components}
\eeq
\begin{multline}
S_{\tx{matter}} = \int d^3 x \,
\left[
\bar{F} F + \bar{\phi} (2im \p_t + \p_i^2) \phi + \sqrt{2} m \bar{\psi_1} \psi_1 + 2 i m \bar{\psi_2} \p_t \psi_2 \right.
\\ 
\left. - 2^{1/4} i \sqrt{m} \bar{\psi_1} (\p_1 - i \p_2) \psi_2 -  2^{1/4} i \sqrt{m} \bar{\psi_2} (\p_1 + i \p_2) \psi_1 
\right]
\label{eq:matter_action_components}
\end{multline}
\begin{multline}
\label{eq:interacting_action_components}
S_{\tx{int}} = \int d^3 x \,
\left[
\bar{\phi} \tilde{D} \phi - 2 i \varphi \, \bar{\phi}   \p_t \phi + 2i A_i \, \bar{\phi} \p_i\phi + 2\bar{\phi} ( m - \f) A_t \phi + \bar{\phi} A_i A_i \phi \right. \\
\left. - \sqrt{2} \varphi \, \bar{\psi}_1  \psi_1 + 2  m \, \bar{\psi}_2 A_t  \psi_2 - 2^{1/4} \sqrt{m} \, \bar\psi_1 (A_1 - i A_2) \psi_2- 2^{1/4} \sqrt{m} \, \bar{\psi}_2 (A_1 + i A_2) \psi_1 \right.
\\ 
\left.  + \bar\phi \l_1 \psi_1 + \bar\psi_1 \bar{\l}_1 \phi + 2^{1/4} \sqrt{m} \, \bar\phi \l_2 \psi_2 + 2^{1/4} \sqrt{m} \bar{\psi_2} \bar{\lambda}_2 \phi 
\right] \\
\end{multline}
For convenience, we summarize here the field content of the theory:
\begin{itemize}
    \item The gauge action in \eqref{eq:gauge_action_components} contains a dynamical real scalar field $\varphi,$ an auxiliary real scalar field $\tilde{D},$ a dynamical complex Grassmann field $\lambda_2$ and an auxiliary one $\lambda_1,$ and a real gauge field $A_{\mu}$ hidden in the $E_i$ and $f_{ij}$ fields defined in \eqref{eq:emfields}.  This is the supersymmetric generalization of the $U(1)$ action \eqref{eq:GED_action}. 
    
    \item The matter action \eqref{eq:matter_action_components} contains a complex scalar field $\phi,$ an auxiliary scalar $F,$ a dynamical complex Grassmann field $\psi_2$ and an auxiliary one $\psi_1.$
    \item The term \eqref{eq:interacting_action_components} contains the interactions between the gauge and matter fields, and defines the Feynman rules for the vertices of the theory. 
\end{itemize}
The action in \eqref{eq:action_splitting} is invariant under SUSY transformations generated by fermionic parameters $\varepsilon_{\a}, \bar{\varepsilon}_{\beta}$, which  on the component fields read
\beq
\begin{aligned}
& \delta \lambda_{\a} = \varepsilon_{\alpha} \tilde{D} + \frac{i}{2} \varepsilon^{\beta} \partial_{\beta \gamma} A_{\alpha}^{\,\,\, \gamma}  \, , \qquad
\delta A_{\a\b} = - \varepsilon_{\alpha} \bar{\lambda}_{\beta} 
+ \bar{\varepsilon}_{\beta} \lambda_{\alpha} \, ,   & \\
& \delta \tilde{D} = \frac{i}{2} \p_{\alpha\beta} \le \varepsilon^{\alpha} \bar{\lambda}^{\beta} + \bar{\varepsilon}^{\beta} \lambda^{\alpha} \ri  \, , \qquad  
\delta \varphi = - \delta A_{22} \, , & \\
& \delta \phi = -\varepsilon^{\alpha} \psi_{\alpha} \, , \qquad
\delta \psi_\a = i \bar{\varepsilon}^{\b} \p_{\alpha \b} \phi + \varepsilon_{\alpha} F   \, , \qquad
\delta F = - i \bar{\varepsilon}^{\b} \p_{\alpha \b} \psi^{\alpha} & 
\end{aligned}
\eeq

Auxiliary fields can be integrated out using their algebraic equations of motion, thus reducing the action only to terms involving dynamical degrees of freedom.
As an example, we restrict to the purely gauge sector and neglect the interactions from $S_{\rm int}$. 
In this case, the integration of the auxiliary fields $\tilde{D},\lambda_1,$ in \eqref{eq:gauge_action_components} leads to the $\mathrm{U}(1)$ action \eqref{eq:GED_action}, plus an additional quadratic term for the dynamical fermion $\lambda_2$
\beq
\begin{aligned}
\label{eq:nonrel_vector_act}
S_{\tx{gauge}} = \frac{1}{g^2} \int d^3 x \,  \left[  \frac12 (\p_t \f)^2 + E^i \p_i \f - \frac14 f^{ij} f_{ij}
+ \sqt i \bar{\lambda}_2 \, \p_t \lambda_2 
\right] 
\end{aligned}
\eeq

Now we restore the interactions and the auxiliary fields, whose constraints are non-trivial.
We can combine the matter and interaction actions as follows
\beq
\begin{aligned}\label{eq:action_components}
S_{\tx{matter}} + S_{\tx{int}} 
= &  \int d^3 x \, \left[ 2 i (m - \varphi) \bar{\phi} \nabla_t \phi 
- (\overline{\nabla_i \phi}) \, (\nabla_i \phi)
 + \bar{F} F 
+ \sqrt{2} (m-\f)  \bar{\psi_1} \psi_1  \right.
\\ 
& \left. + 2 i m \bar{\psi_2} \nabla_t \psi_2  
-  2^{1/4} i \sqrt{m} \, \bar{\psi_1} (\nabla_1 - i \nabla_2) \psi_2 - \, 2^{1/4} i \sqrt{m} \, \bar{\psi_2} (\nabla_1 + i \nabla_2) \psi_1  \right. \\
 & \left. + \bar{\phi} \tilde{D} \phi + \bar\phi \l_1 \psi_1 + \bar\psi_1 \bar{\l}_1 \phi + 2^{1/4} \sqrt{m} \, \bar\phi \l_2 \psi_2 + 2^{1/4} \sqrt{m} \, \bar{\psi_2} \bar{\lambda}_2 \phi \right]
\end{aligned}
\eeq
where we have introduced covariant derivatives $\nabla_\mu = \partial_\mu - i A_\mu$.

In the gauge sector, the complete equations of motion that we have to use for eliminating the auxiliary fields are 
\beq
\begin{cases}
\tilde{D} = - g^2 \bar{\phi} \phi 
\\
\bigl( \p_1 + i \p_2 \bigr) \lambda_1 =  2^{1/4} i \sqrt{m} g^2 \,   \bar{\psi}_2 \phi
\\
\bigl( \p_1 - i \p_2 \bigr) \bar{\lambda}_1 = - 2^{1/4} i \sqrt{m} g^2 \, \bar{\phi} \psi_2
\end{cases} 
\label{eq:integrate_out_gauge}
\eeq 
Similarly, in the matter sector we can get rid of $\psi_1, \bar\psi_1$ matter fermions and the complex scalar $F$ using 
\beq
\begin{cases}
(m - \f) \psi_1 = i \frac{\sqrt{m}}{2^{1/4} } (\nabla_1 - i \nabla_2) \psi_2 + \frac{1}{\sqrt{2}} \bar{\l}_1 \phi
\\
 (m -\f) \bar{\psi}_1 = - i\frac{\sqrt{m}}{2^{1/4} } (\nabla_1 + i \nabla_2) \bar{\psi}_2  + \frac{1}{\sqrt{2}} \bar{\phi} \l_1 \\ 
 F = \bar{F} = 0 
\end{cases} 
\label{eq:integrate_out_matter}
\eeq 
The final expression for the resulting action appears quite cumbersome and does not add any further useful information. Therefore, we avoid reporting it here.

\vskip 10pt
\section{One-loop radiative corrections}
\label{sec:one_loop_radiative_corrections}

We study the renormalization properties of the null SGED action \eqref{action}, working directly in superspace formalism.
In this section we approach the problem by using ordinary  supergraphs and ordinary D-algebra, whereas the more efficient approach of covariant supergraphs \cite{Grisaru:1984ja} is used in section \ref{sec:renormalizable_SGED} to establish renormalizability of the actual SGED sigma-model. The expert reader can go directly to section \ref{sec:renormalizable_SGED}. 

Here we first review the application of the Faddeev-Popov procedure adapting it to the non-relativistic superspace. This leads to the gauge-fixed action \eqref{eq:SGED_action} whose Feynman rules are listed in section \ref{sec:Feynman_rules}.
The particular structure of the non-relativistic propagator gives rise to selection rules which forbid certain configurations of diagrams.
As a consequence, they imply a non-renormalization theorem for the wavefunction of the prepotential $V$ and the coupling constant $g$ of the theory, that we discuss in section \ref{sec:selection_rules}.
We then compute one-loop corrections to the self-energy and to the vertices in sections \ref{sec:loop_self_energy} and \ref{sec:loop_vertices}, respectively.
We find that the wavefunction renormalization of the chiral superfield is not sufficient to get rid of all the UV divergences of the theory, instead infinite new terms are generated at quantum level.
In section \ref{sec:non_renormalizability_theory} we discuss the consequences on the renormalizability of the model, and in section \ref{sec:non_linear_sigma_model} we introduce the renormalizable SGED non-linear sigma model.

\subsection{Faddeev-Popov procedure and gauge fixing}
\label{sec:Faddeev_Popov_procedure}

The kinetic term of gauge-invariant theories is not invertible due to the redundancy of gauge symmetry. 
This statement also holds for the SGED action in eq.~\eqref{action}, whose quadratic part in the prepotential can be written, after integration by parts, as
\beq
S_{\tx{vec}} = - \frac{1}{2 g^2} \int d^3 x \int d^4 \t \, V \le \square_0 \, \Pi_{\frac{1}{2}} V \ri 
\eeq
where $\square_0 \equiv 2 i M \p_t + \p_i^2 $ is the flat Schr\"{o}dinger operator, and we have defined the combination $\Pi_{1/2} \equiv - \square^{-1}_0 D^{\a} \bar{D}^2 D_{\a} .$
The non-invertibility of $\square_0 \, \Pi_{\frac{1}{2}}$ follows from the fact that it annihilates the non-vanishing combination $\square^{-1}_0 \lbrace D^2 , \bar{D}^2 \rbrace V$.

According to the Faddeev-Popov procedure, invertibility is recovered by performing the functional integral only over the set of gauge inequivalent fields. In supersymmetric theories, this method leads to the following functional integral \cite{Gates:1983nr}
\beq
\mathcal{Z} [V] = \int [\mathcal{D} V \mathcal{D} c \mathcal{D} c' \mathcal{D} \bar{c} \mathcal{D} \bar{c}'] \, 
e^{i \le S_{\rm vec} + S_{\rm g.f.} + S_{\rm FP} \ri}
\eeq
where the total action includes a gauge-fixing term (with real parameter $\zeta$)
\beq
S_{\rm g.f.} = - \frac{1}{\zeta g^2}
\int d^3 x d^4 \theta \, (D^2 V)(\bar{D}^2 V) 
\label{eq:gauge_fixing_term}
\eeq
and a ghost action, which in the Abelian case is simply the free action for a pair of anticommuting chiral ghost superfields
\beq
S_{\rm FP} = \int d^3 x \, d^4 \theta \, 
\le \bar{c}' c - c' \bar{c} \ri
\label{eq:ghost_action}
\eeq
Since the ghosts can be integrated separately and decouple from the rest of the action, we will simply ignore them.

The gauge-fixing term makes the vector kinetic operator 
\beq
S_{\rm vec} + S_{\rm g.f.} 
= - \frac{1}{2} \int d^3 x \, d^4 \theta \,
V \,  \square_0 \left[ 1 + \le \frac{1}{\zeta} -1 \ri \Pi_{0} \right] V
\eeq
invertible. In conclusion, reinserting the matter part of the action and the interactions from eq. \eqref{action}, rescaling for convenience the vector superfield as $V \rightarrow gV$, and choosing the Feynman supergauge $\zeta=1$, the total gauge-fixed null SGED action arising from null reduction reads
\beq
\label{eq:SGED_action}
S_{\text{nSGED}} = \int d^3 x \, d^2 \t \, W^2 -\int d^3 x \, d^4 \t \, (D^2 V) (\bD^2 V)+ \int d^3 x \, d^4 \t \, \bF e^{gV} \F 
\eeq
This is the action from which we read Feynman rules to perform loop calculations.

\subsection{Feynman rules}
\label{sec:Feynman_rules}

At quantum level, we consider the generating functional
\beq
\mathcal{Z} [J, \bar{J}, J_V] = \int [\mathcal{D}\Phi \mathcal{D}\bar{\Phi} \mathcal{D} V ] \, 
\exp \left[ i S_{\rm nSGED} + i \int d^3 x \, \le \int d^2 \theta J \Phi + \int d^2 \bar{\theta} \bar{J} \bar{\Phi} + \int d^4 \theta J_V V  \ri   \right]
\label{eq:generating_functional_superfields}
\eeq
where $S_{\rm nSGED}$ is given in \eqref{eq:SGED_action}, $J,\bar{J}$ are (anti)chiral superfields and $J_V$ is a vector superfield, all of them acting as sources.
Correlation functions are obtained from this expression by repeated application of the functional derivatives defined as\footnote{We collectively denote 
$z \equiv (x^{\mu}, \theta^{\a}, \bar{\theta}^{\beta})$ and $\delta^{(7)} (z-z') = \delta^{(3)} (x-x')\delta^{(2)} (\theta-\theta')\delta^{(2)} (\bar{\theta}-\bar{\theta}')$.}
\beq
\frac{\d J(z)}{\d J(z')}  = \bar{D}^2 \d^{(7)} (z-z') \, , \quad
 \frac{\d \bar{J}(z)}{\d \bar{J}(z')} = D^2 \d^{(7)} (z-z') \, , \quad
 \frac{\d J_V (z)}{\d J_V(z')}  = \d^{(7)} (z-z') \, 
 \label{eq:derivative_sources}
\eeq
The additional SUSY covariant derivatives acting on delta functions arise due to the constrained nature of $\Phi, \bar{\Phi}$.

Starting from the gauge-fixed action \eqref{eq:SGED_action}, we read the following Feynman rules in $\mathcal{N}=2$ momentum superspace and in terms of renormalized quantities. 
\begin{itemize}
    \item Chiral propagator
    \beq
\includegraphics[valign=c]{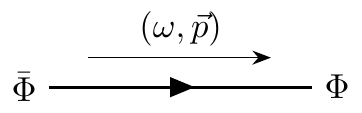}
=
\langle \F (\omega, \vec{p},\theta,\bar{\theta}) \bF (-\omega, -\vec{p}, \theta', \bar{\theta}') \rangle = \frac{i \d^{(4)}(\t'-\t)}{2 m \omega -\vec{p}^2 + i \e} 
\label{eq:chiral_superpropagator}
\eeq
\item Vector propagator 
\beq
\includegraphics[valign=c]{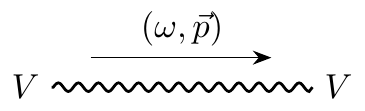}
=
\langle V (\omega,\vp,\theta,\bar{\theta}) V (-\omega,-\vp,\theta',\bar{\theta}') \rangle = - \, \frac{i \d^{(4)}(\t'-\t)}{- \vp^2 + i \e} 
\label{eq:vector_superpropagator}
\eeq
In Feynman gauge ($\zeta=1$) the vector superpropagator does not depend on the energy $\omega$.
This is the supersymmetric generalization of the statement that the mediation of the gauge field is instantaneous, as a consequence of the $c \rightarrow \infty$ limit which features the non-relativistic theory.

\item Vertices: From the series expansion of the exponential in the last term of action \eqref{eq:SGED_action}, we read
\beq
\sum_{n=1}^{\infty} \frac{g^n}{n!} \int d^3 x \int d^4 \theta \, \bar{\Phi}  V^n \Phi \, \equiv \, S_{\rm int} 
\label{eq:interacting_action}
\eeq
These are an infinite number of $(n+2)$--point vertices with one chiral, one antichiral and $n$ vector superfields. In fig.~\ref{fig:original_vertices} we draw the three and four-point vertices. The higher order ones are built in a similar way.
We note that particle number conservation is satisfied, as the numbers of entering and exiting arrows trivially match.
\begin{figure}[H]
    \centering
     \subfigure[]
   {\label{subfig:original_3pt_vertex} \includegraphics[scale=0.92]{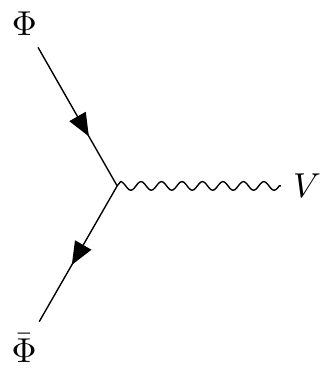}} \qquad
    \subfigure[]{ \label{subfig:original_4p_vertex}
    \includegraphics[scale=0.8]{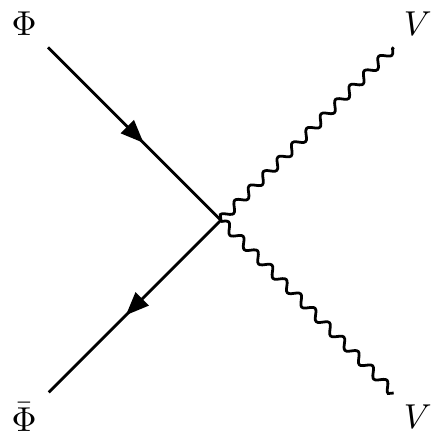}}
    \caption{Vertices with three and four external lines arising from the interacting term in eq.~\eqref{eq:SGED_action}. The vertex (a) has coupling $g,$ while the vertex (b) has coupling $g^2/2.$ }
    \label{fig:original_vertices}
\end{figure}
\item Due to identities \eqref{eq:derivative_sources}, we assign one factor of $\bar{D}^2 (D^2)$ to any chiral (anti-chiral) internal line exiting from a vertex.
\end{itemize}

These Feynman rules can be used to generate supergraphs. Precisely, we build all the topologically inequivalent diagrams involving supervertices, superpropagators and SUSY covariant derivatives acting on them, and impose energy and momentum conservation at each vertex.
We include combinatorial factors arising from the expansion of the interaction lagrangian and we integrate over loop momenta.

After manipulating covariant derivatives in such a way to perform explicitly the integrations along the Grassmannian coordinates ($D$-algebra \cite{Gates:1983nr}), using in particular identities like ~\eqref{eq:useful_identities} and \eqref{rules covariant derivatives on delta functions}, we are left with standard Feynman diagrams in ordinary spacetime, whose corresponding integrals can be evaluated by exploiting standard techniques.

We will use dimensional regularization in $d=2-2\varepsilon$ to regularize the spatial momentum integrals. Therefore, we define renormalized parameters as
\beq\label{eq:renormalized}
g_B = \mu^{\varepsilon} \, Z_g \, g \, , 
\qquad m_B =  \, Z_m \, m \, , 
\qquad V_B = Z_V  \, V
\eeq
\beq
\label{ren_functions}
\Phi_B = Z_{\Phi}^{1/2} \,  \Phi \equiv \le 1 + \frac12 \delta_{\Phi} \ri \Phi  \quad , \quad 
\bar{\Phi}_B = Z_{\bar\Phi}^{1/2} \,  \bar\Phi_B \equiv \le 1+ \frac12 \delta_{\bar\Phi} \ri \bar\Phi_B
\eeq
where $\mu$ is the mass scale of dimensional regularization. 

Although {\em a priori} we require a renormalization also for the mass parameter $m$, the mass does not even enter the SGED superfield action explicitly.
Therefore, a mass counterterm is not necessary to renormalize the effective action.
As it is customary for non-relativistic theories, we then set $Z_m=1$ and identify $m_B=m$ \cite{Bergman:1991hf}.

\vskip 10pt
\noindent
\textbf{Notational remark:} In the rest of this section we denote the dimensionless components of the covariant derivatives as
\beq
D \equiv D_2 \, , \qquad
\bar{D} \equiv \bar{D}_2
\eeq
The notation $D^2, \bar{D}^2$ will be used only to denote the square of the covariant derivatives according to definitions \eqref{eq:combinations_susy_der}.

\subsection{Selection rules and non-renormalization theorems}
\label{sec:selection_rules}

The main distinctive features of non-relativistic theories, compared to relativistic ones, consist in the constraint on particle number conservation related to the U(1) central extension symmetry, and the retarded nature of the propagators in eqs.~\eqref{eq:chiral_superpropagator} and \eqref{eq:vector_superpropagator}. 

The first feature strongly constrains the superpotential \cite{Auzzi:2019kdd}, but in the present case it does not play a crucial role since the interactions between the (anti)chiral superfields and the prepotential are invariant under a global $\mathrm{U}(1)$ symmetry already in the relativistic parent theory.

The second characteristic is responsible for restrictions on the orientation of arrows inside a loop, and gives rise to non-renormalization theorems or powerful resummations \cite{Bergman:1991hf,Klose:2006dd}.
We explain the main idea as follows.
In configuration space, the $ i \varepsilon $ prescription translates into a retarded prescription for the propagators. In fact, the Fourier transforms of superpropagators \eqref{eq:chiral_superpropagator} and \eqref{eq:vector_superpropagator} collectively read 
\beq
G(\vec{x} , t)= \int \frac{d^2 p \,  d \omega}{(2 \pi)^3} 
\frac{i \delta^{(4)} (\theta_1 - \theta_2)}{2 M \, \omega - \vec{p}^2 + i \varepsilon} \, e^{-i (\omega t - \vec{p} \cdot \vec{x})} =
 - \frac{i \, \Theta (t)}{4 \pi \,  t} e^{i\frac{M \vec{x}^2}{2t}} \,  \delta^{(4)} (\theta_1 - \theta_2)
 \label{eq:propagator_position_space}
\eeq
where $\Theta$ is the Heaviside function and $M$ the mass generator ($M=m$ for the chiral propagator, $M=0$ for the vector one).
This feature leads the following result.
\begin{srule}
\label{sel_rule1}
Any 1P-irreducible Feynman diagram with negative superficial degree of divergence in the $\omega$ variable, vanishes identically.  
\end{srule}

\noindent
The superficial degree of divergence $\Delta_{\omega}$, defined as the number of $\omega$  powers at the numerator minus $\omega$  powers at the denominator,
features the UV behaviour of the integrand along the $\omega$ variable.  Its counting has to be performed after D-algebra has been carried out completely, leaving a regular QFT diagram with integrations over momentum variables only.

The prototypical example of Feynman diagram satisfying the hypothesis of Selection rule \ref{sel_rule1} is 
\beq
\int \frac{d\omega \, d^2k}{(2 \pi)^3}  \, \frac{ 1}{[2 m \omega- \vec{k}^2 + i \varepsilon ] [2 m (\omega-\Omega)- (\vec{k} -\vec{p})^2 + i \varepsilon ]}   
\eeq
with $\Delta_{\omega} = -1$. Such a contribution arises for instance in the evaluation of the one-loop correction to the self-energy of the prepotential (see eq.~\eqref{eq:chiral_loop} below). 
To prove that this integral vanishes, we perform the $\omega$ integration first. Since the poles in $\omega$ sit on the same complex half-plane, a simple application of Jordan's lemma allows to close the contour of integration in the half-plane with no poles, and the use of the residue theorem leads to the expected result. In configuration space, the same result follows by observing that expression \eqref{eq:propagator_position_space} for the propagators leads to the product of two Heaviside functions with opposite arguments. Since this has support in a single point, it vanishes by normal ordering. 

The generalization of this proof to a generic loop integral relies on the observation that the request for a meromorphic function to satisfy Jordan's lemma corresponds precisely to the condition $\Delta_{\omega} <0$. Since the retarded nature of the propagators implies that the poles are always in the same half $\omega$-plane, it is straightforward to conclude that closing the contour in the complementary half-plane the integral evaluates to zero.  

Selection rule \ref{sel_rule1} was originally derived for scalar fields in \cite{Bergman:1991hf,Klose:2006dd}. It was later used in the supersymmetric case to show the one-loop exactness of the Galilean Wess-Zumino model \cite{Auzzi:2019kdd}, and applied in the context of supersymmetric Lifshitz theories \cite{Arav:2019tqm} to show similar non-renormalization theorems.
In the case of GED, arguments based on selection rule \ref{sel_rule1} were used in \cite{Chapman:2020vtn} to show that at quantum level the electric charge does not run.
Here we show that this statement has a natural supersymmetric generalization, which leads to

\begin{srule}
\label{sel_rule2}
All loop corrections to the effective action with purely vector external lines vanish
\beq
\Gamma^{(n)} (V) = 0 
\label{eq:all_loop_nonren_theorem}
\eeq
\end{srule}

\noindent
We prove it by starting from the one-loop self-energy correction to the vector superfield and then inferring the general result for diagrams with an arbitrary number of external fields.

\begin{figure}[ht]
    \centering
    \subfigure[]
{\label{subfig:oneloop_vectora}
\includegraphics[scale=1.2]{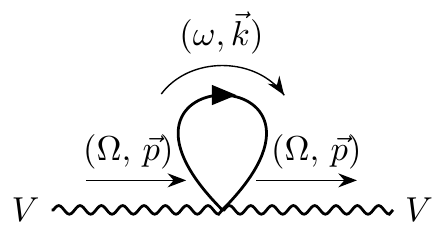}} 
\subfigure[]
{\label{sufig:oneloop_vectorb} \includegraphics[scale=1]{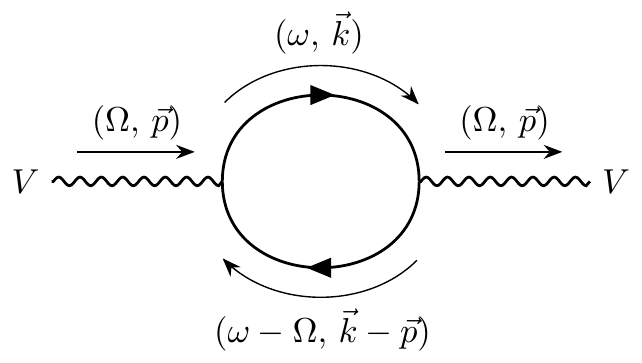}} 
    \caption{One-loop contributions to the self-energy of the vector superfield.}
    \label{fig:oneloop_vector}
\end{figure}

The relevant one-loop Feynman supergraphs contributing to the vector self-energy are depicted in fig.~\ref{fig:oneloop_vector}.
Contribution \ref{subfig:oneloop_vectora} is a tadpole that vanishes in spatial dimensional regularization by using the integrals $I$ in eq.~\eqref{eq:integral_I} and $J_0$ in eq.~\eqref{eq:integral_J_0}. Precisely, 
\beq
i \G^{(2)}_{\rm \ref{subfig:oneloop_vectora}} (V) = - \frac{g^2}{2} \int d^4 \t \, V(\O,\vp,\t) \, V(\O,\vp,\t) \, \int \frac{d \o \, d^2 k}{(2 \pi)^3} \,
\frac{1}{2 m \o - \vk^2 + i \e} = 0 
\label{eq:diagram3a_vector}
\eeq
After evaluating the corresponding D-algebra, diagram \ref{sufig:oneloop_vectorb} reduces to the following spacetime integral
\beq
\label{eq:chiral_loop}
i \Gamma^{(2)}_{\rm \ref{sufig:oneloop_vectorb}} (V) =
\frac{g^2}{2} \int d^4 \t \int \frac{d \o \, d^2 k}{(2 \pi)^3} \,
\frac{V(\O,\vp,\t) \, N(\o-\O,\vk-\vp) \, V(\O,\vp,\t)}{
\bigl[
2 m \o - \vk^2 + i \e
\bigr]
\bigl[
2m(\o - \O) - (\vk - \vp)^2 + i \e
\bigr]
} 
\eeq
where\footnote{The conventions for the generalized Pauli matrices in light-cone coordinates are given in eq.~\eqref{eq:Pauli_lightcone}.}
\beq
\label{eq:numerator}
N(\o,\vk) = \le 2 M \o - \vk^2 \ri + k^{\a \b} \bD_{\a} D_{\b} + \bD^2 D^2
\, , \quad \quad k^{\a \b} \equiv - \sqt M \d_1^{\a} \d_1^{\b} + (\bs^{\mu})^{\a \b} k_{\mu}
\eeq
Potentially non-trivial contributions may only come from terms with $\Delta_{\omega} \geq0$. However, the first contribution in \eqref{eq:numerator} is a tadpole similar to ~\eqref{eq:diagram3a_vector}, which then vanishes. The only other contribution with non-negative $\Delta_\omega$ comes from the second term in
\eqref{eq:numerator} when we take $k^{22} = - \sqrt{2} \omega$ (see the definition of derivatives in eq.~\eqref{eq:3d_derivatives}).   
However, one can easily show that
\beq
 \int \frac{d \o \, d^2 k}{(2 \pi)^3} \,
\frac{2m\o}{
\bigl[
2 m \o - \vk^2 + i \e
\bigr]
\bigl[
2m(\o - \O) - (\vk - \vp)^2 + i \e
\bigr]
} = \int \frac{d \o \, d^2 k}{(2 \pi)^3} \,
\frac{1}{2 m \o - \vk^2 + i \e}= 0
\label{eq:manipulations_srule2}
\eeq
where we have discarded a term with $\Delta_{\omega} = -1,$ and used again the result \eqref{eq:diagram3a_vector}.

In conclusion, there are no one-loop corrections to the vector self-energy
\beq
\label{eq:vector_prop_final}
\G^{(2)} (V) = \G^{(2)}_{\ref{subfig:oneloop_vectora}} (V) + \G^{(2)}_{\ref{sufig:oneloop_vectorb}} (V) = 0
\eeq
\begin{figure}[ht]
    \centering
    \includegraphics[scale=0.7]{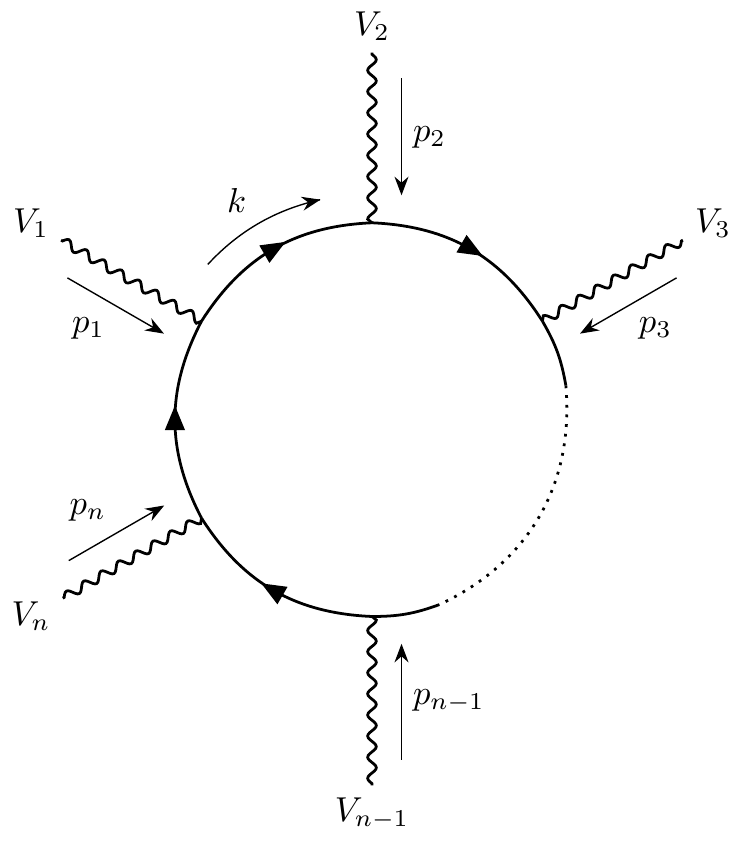}
    \caption{Chiral loop with multiple vector insertions. Here we use the compact notation $k \equiv (\o,\vk)$ and $p_i \equiv (\O_i,\vp_i)$, $i \in \lbrace 1,\dots,n \rbrace $.}
    \label{fig:nonren_theorem}
\end{figure}
Now we generalize the proof to one-loop diagrams with an arbitrary number of external $V$ legs. Neglecting vanishing tadpole contributions, they have the form depicted in figure~\ref{fig:nonren_theorem}.
Defining $\square_k \equiv 2 m \o - \vk^2$ and $\delta_{ij} \equiv \delta^{(4)} (\theta_i - \theta_j)$, in momentum space the contributions of such diagrams read
\begin{multline}
\label{eq:many_vectors}
i \G^{(n)} (V) = \frac{(ig)^n}{n} \, \int d^4 \t_1 \dots d^4 \t_n \int \frac{d \o \, d^2 k}{(2 \pi)^3} \, V_1 \dots V_n
\\
\times 
\frac{i D^2 \bD^2 \d_{12}}{\square_k + i \e} \,
\frac{i D^2 \bD^2 \d_{23}}{\square_{k+p_2} + i \e} \,
\dots \,
\frac{i D^2 \bD^2 \d_{n1}}{\square_{k+p_2+\dots+p_n} + i \e} 
\end{multline}
Since the denominator of the integrand behaves as $\o^{n}$ at large energies, thanks to selection rule \ref{sel_rule1} the integral vanishes, provided that the numerator goes to infinity at most as $\o^{n-2}$. 
To check whether this is the case, we determine the highest powers of $\o$ that can be produced by applying D-algebra to the numerator of~\eqref{eq:many_vectors}. Integrating by parts at vertex 2 (where a vector superfield enters with momentum $p_2$) we obtain
\begin{multline}
\label{eq:basic_D_algebra}
\bigl( D^2 \bD^2 \d_{12} \bigr)  \bigl( D^2 \bD^2 \d_{23} \bigr) V_2 = 
 \d_{23} \, \bigl[ (\square_k \bD^2 D^2 \d_{21}) \, V_2 + (\square_k \bD_{\a} D^2 \d_{21}) \, \bD^{\a} V_2 \\ 
  + (\square_k D^2 \d_{21}) \, \bD^2 V
 + ( k^{\a \b} \bD^2 D^2 \d_{21}) \, \bD_{\a} D_{\b} V_2 + (\bD^2 D^2 \d_{21}) \, \bD^2 D^2 V_2 \bigr] 
\end{multline}
This operation produces terms at most linear in $\omega,$ coming from the kinetic operators $\square_k$ and from $k^{22} = - \sqrt{2} \o$.
We can iterate this procedure for each vertex, until only one fermionic $\d$-function out of the original $n$ delta's in the loop is killed by $\bD^2 D^2$. Performing the spinorial integrations by means of the surviving delta's we obtain a local expression of the form
\beq
\label{eq:formal_sum}
(2 m \o)^{n-1} \, \Biggl( V_1 \dots V_n + V_1 \, \sum_{i=1}^{n-1} \frac{1}{(\sqt m)^i} \, \mk{D}^{(i)} \, \le V_2 \dots V_n \ri \Biggr) \, + \, \dots
\eeq
where dots stand for terms containing lower powers of $\o$ that integrate to zero, according to selection rule \ref{sel_rule1}. 
The $\mk{D}^{(i)}$ operator acting on the product $V_2 \dots V_n$ represents formally the sum over all possible ways to distribute $i$ covariant derivatives $\bD$'s and $i$ $D$'s on the string $V_2 \dots  V_n$, each term being multiplied by the corresponding combinatorial factor, including relative signs coming from the fermionic nature of the derivatives. The precise value of these coefficients is not important for the present derivation.
For clarification, the explicit expression for $i=3$ and $n=5$ is\footnote{We remind the reader that $D, \bar{D}$ refer to the two-components of the covariant SUSY derivatives. }
\begin{align}
 \mk{D}^{(3)} \, \le V_2 \dots V_5  \ri & =  c_1 \, (D \bD V_2) \, V_3 V_4 V_5 +  c_2 \, (D V_2) \, (\bD V_3) \, V_4 V_5 + \dots 
\nl
& \; \; \; \;   + c_3 \, V_2 (\bD V_3) \, V_4 \, (D V_5) + \dots +  c_4 \, V_2 V_3 V_4 \, (D \bD V_5) 
\end{align}
Plugging the resulting expression in \eqref{eq:many_vectors} and performing manipulations similar to the ones in eq.~\eqref{eq:manipulations_srule2}, the final contribution to the effective action takes the form
\begin{align}
i \G^{(n)} (V) & = \frac{(-g)^n}{n} \, \int d^4 \t \, \Biggl( V_1 \dots V_n + V_1 \, \sum_{i=1}^{n-1} \frac{1}{(\sqt m)^i} \, \mk{D}^{(i)} \, V_2 \dots V_n \Biggr)
\nl
& \qquad \qquad \qquad \qquad \qquad \qquad \qquad \qquad
\times \int \frac{d \o \, d^2 k}{(2 \pi)^3} \, 
\frac{1}{2 m \o - \vk^2 + i \e}  =0
\end{align}
where use has been made of the integrals $I,J_0$ in eq.~\eqref{eq:integral_I} and \eqref{eq:integral_J_0}.

This concludes the proof of selection rule \ref{sel_rule2} at one loop.
The generalization to higher loops is easily obtained if we manage to argue that the insertion of extra propagators and extra vertices in a diagram cannot increase its superficial degree of divergence, both in the energy and spatial momentum integrations. This can be understood by taking into account the following observations:
\begin{itemize}
    \item The insertion of additional chiral loop propagators improves the superficial degree of divergence in $\omega$ or at most leaves it unchanged. In fact, their denominator is linear in $\omega$ and at numerator they contribute with a factor $\bar{D}^2 D^2$, which in the worst scenario provides terms of the form $\square_k$ and $k^{22}$ that are at most linear in $\o$. Therefore, a chiral propagator contributes to $\Delta_{\omega}$ at most with $\Delta_{\omega} = 0$. 
    
    Additional vector propagators do not carry any covariant derivative nor factors of momenta, see eq.~\eqref{eq:vector_superpropagator}. Therefore, they have superficial degree of divergence $\Delta_{\omega}=0$.

    \item Concerning the spatial momentum integrations, there is a similar argument.
    Any additional chiral loop propagator brings a quadratic contribution in $\vec{k}$ at the denominator, and in the worst case the covariant derivatives $\bar{D}^2 D^2$ at numerator bring a factor of $\vec{k}^2,$ too. This contributes to the superficial degree of divergence at most with $\Delta_{\vec{k}}=0.$
    
    The inclusion of additional vector propagators is beneficial for the UV convergence of the diagram, since it brings a factor of $\vec{k}^2$ at the denominator, while no other factors appear at the numerator.
    Therefore, it always contributes with a superficial degree of divergence $\Delta_{\vec{k}}=-2.$
\end{itemize}
In conclusion, we have shown that
\beq
\G^{(n)} (V) = 0 \; , \qquad \forall \, n \geq 1
\eeq
This result entails strong consequences on the renormalization properties of the theory.
First of all, it implies that there is no wavefunction renormalization for the prepotential.
Since the renormalization of the gauge coupling $g$ is related to the wavefunction renormalization of $V$ by standard arguments valid for supergauge-invariant theories, we conclude that also $g$ does not renormalize at any loop order.
This generalizes to the supersymmetric scenario the result found in \cite{Chapman:2020vtn} regarding the electric charge in GED.

\subsection{One-loop corrections to the chiral self-energy}
\label{sec:loop_self_energy}

The existence of powerful selection rules does not prevent UV divergences from appearing in diagrams with external (anti)chiral superfields. 
Here we study these contributions for the null SGED theory in \eqref{eq:SGED_action}, by computing one-loop diagrams with one chiral and one antichiral external superfields.

As already mentioned, we compute the loop integrals by performing the integration over the $\omega$ variable first, and then using dimensional regularization in $d=2-2\varepsilon$ to regularize the spatial momentum integrals.
We subtract UV divergences by defining renormalized quantities \eqref{ren_functions}.
According to the assignments in \eqref{ren_functions}, the total Lagrangian must be supplemented by counterterms of the form
\beq\label{eq:conterterm}
\mathcal{L}_{\rm nSGED} + \int d^4 \theta \, \delta_{\Phi} \, \bar{\Phi} e^{gV} \Phi 
\eeq
where we have set $Z_{\Phi} = Z_{\bar\Phi}$.

We adopt the minimal subtraction scheme, which gets rid of the UV divergences without including any finite part.
For this reason, we focus only on the evaluation of the UV divergent part of a given graph, which we denote with the symbol $\simeq.$

\begin{figure}[ht]
    \centering
 \includegraphics[scale=1]{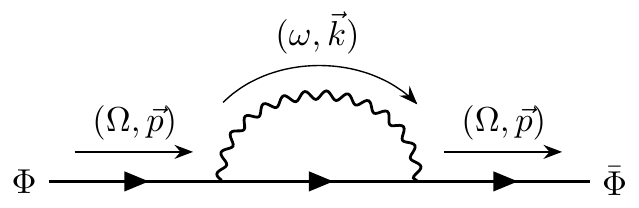}
    \caption{Supergraph contributing to the one-loop corrections to the self-energy of the chiral superfield.}
    \label{fig:1loop_chiral}
    \end{figure}

We start with one-loop corrections to the self-energy of the chiral superfield.
Neglecting vanishing tadpoles, the only relevant diagram is the one depicted in fig.~\ref{fig:1loop_chiral}.
The corresponding D-algebra is trivial, since the covariant derivatives $\bar{D}^2 D^2$ are entirely used to get rid of one of the two delta functions over the grassmannian coordinates, while the remaining delta is used to perform one spinorial integration, so obtaining an expression which is local in the $\t$-variables.
The rest reduces to an ordinary momentum integral which reads
\beq
\begin{aligned}
\label{eq:sunset}
i \Gamma^{(2)}_{\ref{fig:1loop_chiral}}  & =
- g^2 \int d^4 \t \int \frac{d\o \, d^2 k}{(2\pi)^3} \,
\frac{\bF (\O,\vp,\t) \, \F (\O,\vp,\t)}{
\bigl[
- \vk^2 + i \e
\bigr]
\bigl[
2m(\O - \o) - (\vp - \vk)^2 + i \e
\bigr]
} = \\
& = - g^2 \, I \, J_1 \, \int d^4 \t \, \bF (\O,\vp,\t) \, \F (\O,\vp,\t)
\end{aligned}
\eeq
where we have performed the change of variable $\o \longrightarrow \O - \o - (\vp-\vk)^2/2m + \vk^2/2m$ to recognize the appearance of the $I, J_1$ integrals in eqs.~\eqref{eq:integral_I} and \eqref{eq:integral_J_1}.
Inserting their explicit values, we finally obtain the following UV divergent contribution 
\beq
 \Gamma^{(2)} (\F,\bF) 
 \simeq - \frac{g^2}{16 \pi m \e} \int d^4 \t \, \bF  \F 
 \label{eq:total_oneloop_correction_chiral_prop}
\eeq
In minimal subtraction scheme, this requires choosing the following one-loop counterterm in \eqref{eq:conterterm}
\beq
\delta_{\Phi}^{\rm (1)} = \frac{g^2}{16 \pi m} \frac{1}{\varepsilon}
\label{eq:wavefunction_ren_chiral}
\eeq

\vskip 10pt

\subsection{One-loop corrections to the vertices}
\label{sec:loop_vertices}

We proceed with the computation of one-loop corrections to the
vertices in \eqref{eq:interacting_action}.

\subsubsection*{Three-point vertex}

\begin{figure}[ht]
    \centering
    \subfigure[]{\label{subfig:one_loop_three_vertex_b} \includegraphics[scale=0.75]{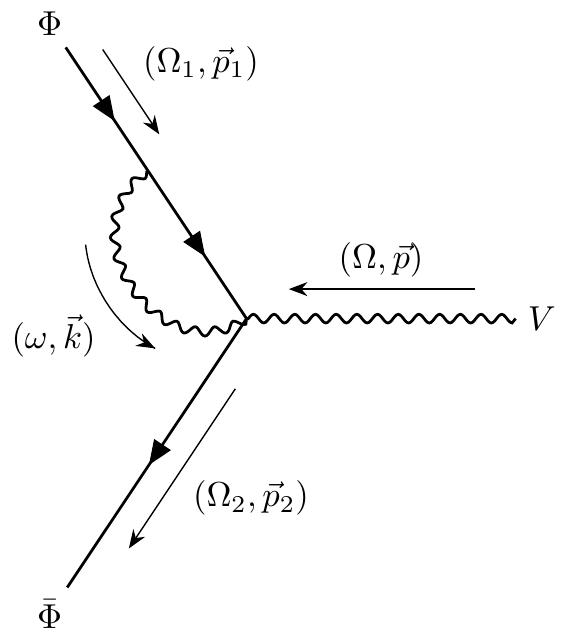}} \quad
    \subfigure[]{\label{subfig:one_loop_three_vertex_c} \includegraphics[scale=0.75]{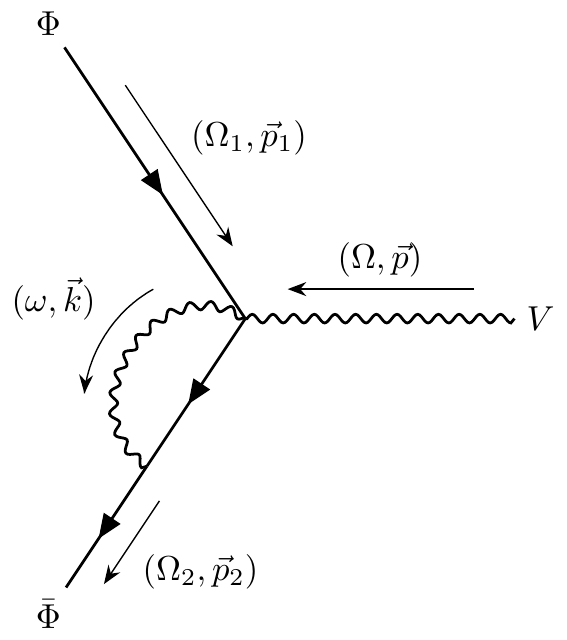}} \quad
    \subfigure[]{\label{subfig:one_loop_three_vertex_d} \includegraphics[scale=0.72]{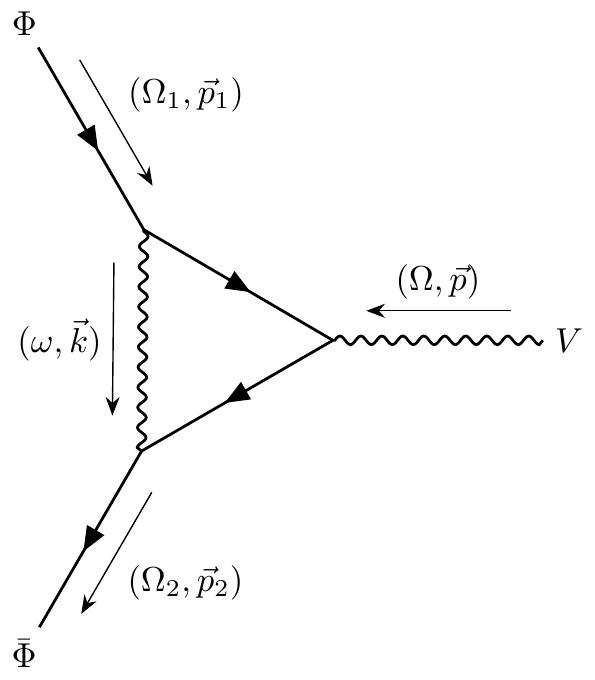}}
    \caption{One-loop diagrams contributing to the quantum corrections of the three-point vertex $\langle \bar{\Phi} V \Phi \rangle.$ The graphs (a) and (b) are mapped into each other by reversing the arrows along the chiral propagators.}
    \label{fig:one_loop_3ptvertex}
\end{figure}

We consider one-loop corrections to the three-point vertex $\langle \bar{\Phi} V \Phi \rangle$ depicted in fig.~\ref{subfig:original_3pt_vertex}. Omitting vanishing tadpole diagrams, the relevant Feynman supergraphs are collected in fig.~\ref{fig:one_loop_3ptvertex}. We analyze them case by case.

\noindent
The first two graphs are mapped one into the other by reversing the orientation of the arrows in the chiral path.
It is then easy to see that they give the same contribution. For these diagrams the D-algebra is trivial, since the covariant derivatives coming from the chiral propagator are directly used to kill one grassmannian delta function, so obtaining a local expression in $\t$. The momentum integral gives
\beq
\begin{aligned}
\label{eq:cancel}
 i \Gamma^{(3)}_{\ref{subfig:one_loop_three_vertex_b}} & = i \Gamma^{(3)}_{\ref{subfig:one_loop_three_vertex_c}}  =
-  g^3 \int d^4 \t \int \frac{d \o \, d^2 k}{(2 \pi)^3} \,
  \frac{\bF(\O_2,\vp_2,\t) \, V(\O,\vp,\t) \, \F(\O_1,\vp_1,\t) }{
\bigl[
- \vk^2 + i \e
\bigr]
\bigl[
2m(\O_1 - \o) - (\vp_1 - \vk)^2 + i \e
\bigr]} \\
& \simeq
- \frac{i g^3}{16 \pi m \e} \int d^4 \t \,
\bF  V \F  
\end{aligned}
\eeq
where we have performed the change of variable $\o \longrightarrow \O_1 - \o - (\vp_1-\vk)^2/2m + \vk^2/2m$ to recognize the appearance of the $I, J_1$ integrals in eqs.~\eqref{eq:integral_I} and \eqref{eq:integral_J_1}.
 
The last contribution comes from diagram \ref{subfig:one_loop_three_vertex_d}, whose D-algebra is less trivial. A series of integration by parts on spinorial derivatives leads to
\beq
\begin{aligned}
\label{eq:canceling}
i \Gamma^{(3)}_{\ref{subfig:one_loop_three_vertex_d}} & =
g^3 \int d^4 \t \int \frac{d \o \, d^2 k}{(2 \pi)^3} \, \\
& \times \frac{ \bF(\O_2,\vp_2,\t) \, \F(\O_1,\vp_1,\t) \, N(\O_1-\o,\vp_1-\vk) \, V(\O,\vp,\t)}{
\bigl[
- \vk^2 + i \e
\bigr]
\bigl[
2m(\O_1 - \o) - (\vp_1 - \vk)^2 + i \e
\Bigr]
\bigl[
2m(\O_2 - \o) - (\vp_2 - \vk)^2 + i \e
\bigr]} 
\end{aligned}
\eeq
where $N(\omega,\vec{p})$ is given in eq.~\eqref{eq:numerator}. Using its explicit expression and discarding terms with $\Delta_{\omega} < 0$, in agreement with selection rule \ref{sel_rule1}, we eventually obtain
\beq
i \Gamma^{(3)}_{\ref{subfig:one_loop_three_vertex_d}} 
\simeq \frac{i g^3}{16 \pi m \e} \int d^4 \t \, \bF  \F \, \biggl( 1 - \frac{1}{\sqt m} \bD D \biggr) V
\eeq
Therefore, summing all the results, we are left with the following divergent contributions to the effective action
\beq
\label{eq:total_three_point}
\Gamma^{(3)} (\F,\bF,V) \simeq
- \frac{g^3}{16 \pi m \e}
\int d^4 \t \, \bF  \F  \,
\biggl( 
1 + \frac{1}{\sqt m} \bD D
\biggr) V 
\eeq
While the first term gets cancelled by wavefunction renormalization \eqref{eq:wavefunction_ren_chiral} of the chiral superfield, the second term is a new contribution which is not present in the original action \eqref{eq:SGED_action}.

\subsubsection*{Four-point vertex}

\begin{figure}[ht]
    \centering
\subfigure[]{\label{subfig:1loop_4ptvertex_b}  \includegraphics[scale=0.65]{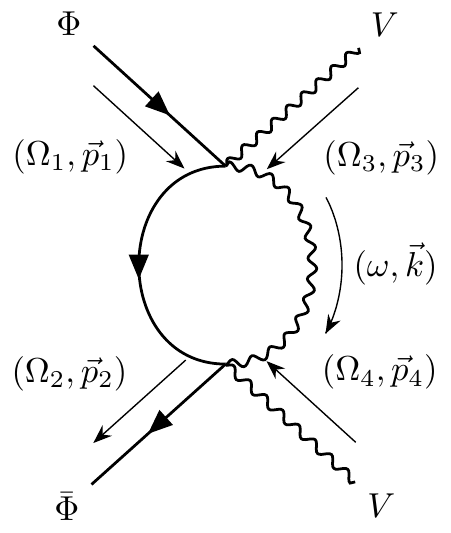}} \quad
\subfigure[]{ \label{subfig:1loop_4ptvertex_c}  \includegraphics[scale=0.64]{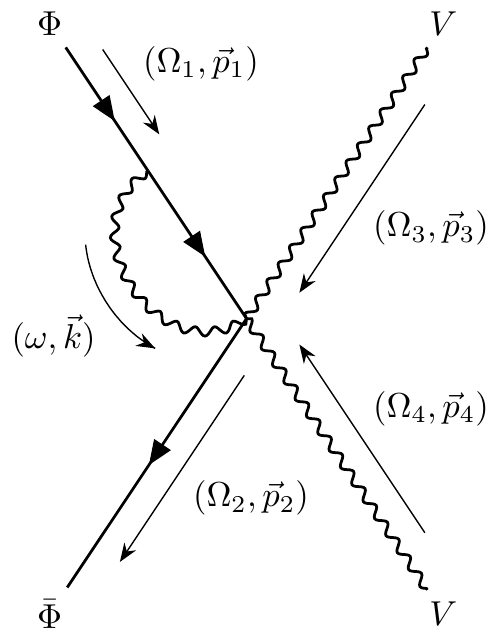}} 
\subfigure[]{\label{subfig:1loop_4ptvertex_d}  \includegraphics[scale=0.64]{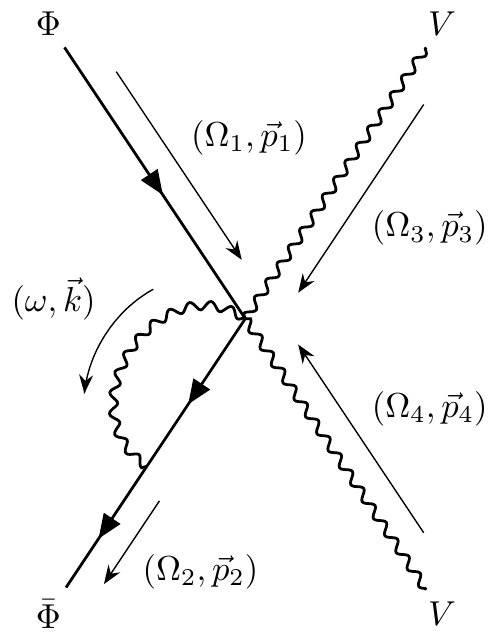}} 
\subfigure[]{ \label{subfig:1loop_4ptvertex_e}   \includegraphics[scale=0.57]{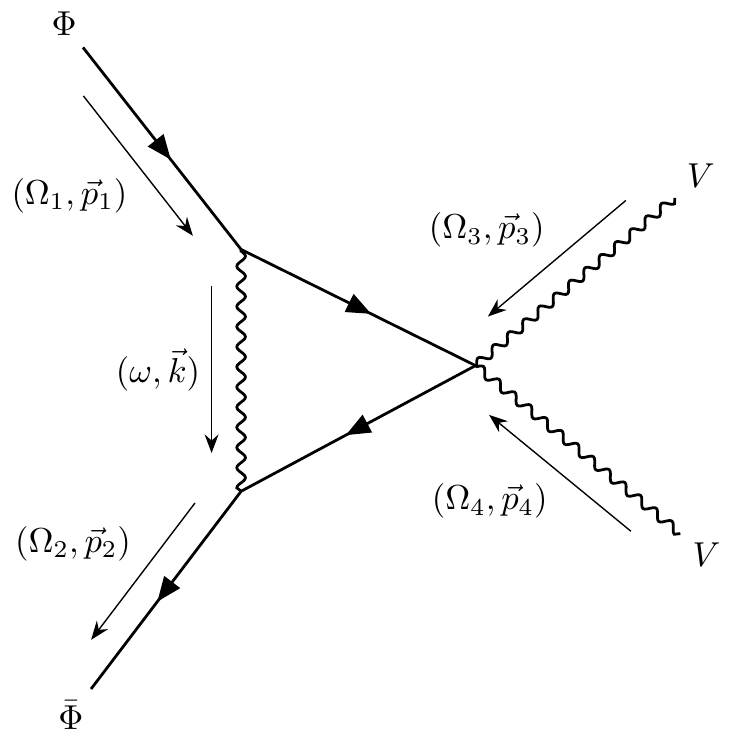} } \\
\subfigure[]{\label{subfig:1loop_4ptvertex_f}  \includegraphics[scale=0.85]{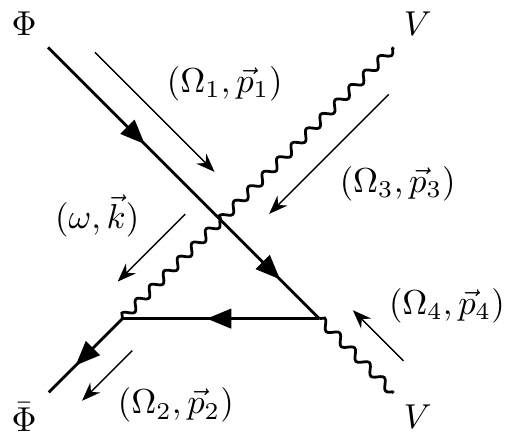}} 
\subfigure[]{\label{subfig:1loop_4ptvertex_g}  \includegraphics[scale=0.85]{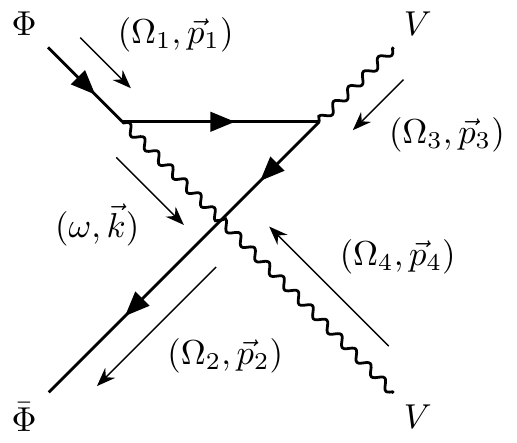}} 
\subfigure[]{\label{subfig:1loop_4ptvertex_h}  \includegraphics[scale=0.75]{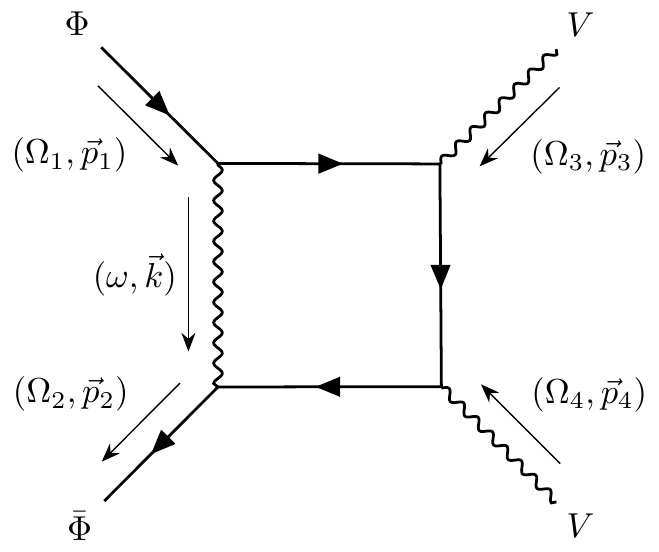}}
    \caption{One-loop diagrams contributing to the quantum corrections of the four-point vertex $\langle \bar{\Phi} V^2 \Phi \rangle.$ The graphs (b),(c) and the graphs (e),(f) are mapped into each other by reversing the arrows along the chiral propagators.}
    \label{fig:1loop_4pt_vertices}
\end{figure}

We proceed the study of one-loop corrections to the vertices in the null SGED model by considering the four-point function $\langle \bar{\Phi} V^2 \Phi \rangle.$
The relevant supergraphs are collected in fig.~\ref{fig:1loop_4pt_vertices}. We briefly go through the computation of these diagrams.

Contributions \ref{subfig:1loop_4ptvertex_b}, \ref{subfig:1loop_4ptvertex_c} and \ref{subfig:1loop_4ptvertex_d} have trivial D-algebra, and can be directly evaluated using the elementary integrals \eqref{eq:integral_I} and \eqref{eq:integral_J_1}.
The explicit results are
\beq
i \Gamma^{(4)}_{\ref{subfig:1loop_4ptvertex_b}} \simeq 
- \frac{i g^4}{16 \pi m \e} \int d^4 \t \, \bF V^2 \F  
\eeq
\beq
i \Gamma^{(4)}_{\ref{subfig:1loop_4ptvertex_c}}  = 
i \Gamma^{(4)}_{\ref{subfig:1loop_4ptvertex_d}}  \simeq
- \frac{i g^4}{32 \pi m \e} \int d^4 \t \, \bF V^2  \F 
\eeq

The next category of Feynman diagrams are triangle diagrams \ref{subfig:1loop_4ptvertex_e}, \ref{subfig:1loop_4ptvertex_f} and \ref{subfig:1loop_4ptvertex_g}, which are obtained from the three-point vertex \ref{subfig:one_loop_three_vertex_d} by adding one external vector superfield in all possible ways.
Consequently, their D-algebra gives the term \eqref{eq:numerator} at numerator. 
Using selection rule \ref{sel_rule1} to get rid of several terms and performing steps similar to the ones used in the previous computations, we find
\beq
i \Gamma^{(4)}_{\ref{subfig:1loop_4ptvertex_e}}
\simeq
\frac{i g^4}{32 \pi m \e} \int d^4 \t \, \bF  \F \, 
\biggl( 
1 - \frac{1}{\sqt m} \bD D
\biggr) V^2
\eeq
\beq
i \Gamma^{(4)}_{\ref{subfig:1loop_4ptvertex_f}}
= i \Gamma^{(4)}_{\ref{subfig:1loop_4ptvertex_g}}  \simeq  
\frac{i g^4}{16 \pi m \e} \int d^4 \t \, \bF  \F \, V
\biggl(
1 - \frac{1}{\sqt m} \bD D
\biggr)
V
\eeq
The square diagram in fig.~\ref{subfig:1loop_4ptvertex_h} corresponds to a new topology that appears for the first time in the four-point function.
The D-algebra in this case is more involved, since the number of covariant derivatives along the internal propagators increases.
After manipulating them by integrating by parts at the vertices and using their algebra, we obtain the following factor at numerator
\beq
\begin{aligned}
\mathcal{N} \bigl( V_3,V_4 \bigr) & = 
\square_l^2  \, V_3 V_4 
+ \square_l \, l^{\a \b} \, V_3 \, \bD_{\a} D_{\b} V_4 - \square_l \, l^{\a \b} \, D_{\b} V_3 \, \bD_{\a} V_4 \\
& - \square_l \, l^{\a \b} \, D_{\b} \bD_{\a} V_3 \, V_4
- l^{\a \b} l^{\c \d} \, D_{\b} \bD_{\a} V_3 \, \bD_{\c} D_{\d} V_4 
 + \dots  
\label{eq:square_numerator}
\end{aligned}
\eeq
where $l^{\mu}=p_1^{\mu}+p_3^{\mu}-k^{\mu}$. 
The subscripts for the prepotentials refer to the assignment of external momenta given in fig.~\ref{subfig:1loop_4ptvertex_h}, $V_j \equiv V(\Omega_j, \vec{p}_j)$.
The ellipsis denote several other terms that have $\Delta_{\omega}<0$, and therefore give vanishing contributions.
The full calculation of the surviving terms gives
\beq
i \Gamma^{(4)}_{\ref{subfig:1loop_4ptvertex_h}}  \simeq
- \frac{i g^4}{16 \pi m \e} 
\int d^4 \t \, \bF  
\biggl[ 
V^2 - \frac{1}{\sqt m} 
\Bigl(
2 V ( \bD D V) - (D V) (\bD V)
\Bigr)
- \frac{1}{2 m^2}
(D \bD V)^2 
\biggr] \F
\eeq
Summing all the results of the diagrams in fig.~\ref{fig:1loop_4pt_vertices}, we obtain the following contribution to the effective action 
\beq
\Gamma^{(4)} (\F,\bF,V) \simeq
- \frac{g^4}{16 \pi m \e} \int d^3 x \int d^4 \t \, \bF  
\biggl[
V^2
+ 
\frac{1}{\sqrt{2} m} V ( \bD D V )
+
\frac{1}{2 m^2}  ( \bD D V)^2
\biggr] \F
\label{eq:total_1loop_correction_4pt_function}
\eeq
The first term gets cancelled by the wavefunction renormalization in eq.~\eqref{eq:wavefunction_ren_chiral}.
However, we find two new terms which were not included in the original SGED action \eqref{eq:SGED_action}, confirming the pattern that we observed for the three-point vertex.

\subsection{Non-renormalizability of the theory}
\label{sec:non_renormalizability_theory}

The results obtained in eqs.~\eqref{eq:total_three_point} and \eqref{eq:total_1loop_correction_4pt_function} pose serious doubts about the renormalizability of the theory.
Indeed, we now show that the action \eqref{eq:SGED_action} is non-renormalizable, as an infinite number of new counteterms need to be added in order to cancel UV divergences at one loop.

First of all, as we argue in appendix \ref{app:additional_loop_corrections}, new vertices containing more than two (anti)chiral superfields are not generated at quantum level (no UV divergences come from diagrams with more than two (anti)chiral external legs).
Therefore, the problematic terms for the renormalizability of the theory only come from diagrams with exactly one chiral and one antichiral superfield, and any number of external vector superfields.

\begin{figure}[ht]
    \centering
    \includegraphics[scale=0.7]{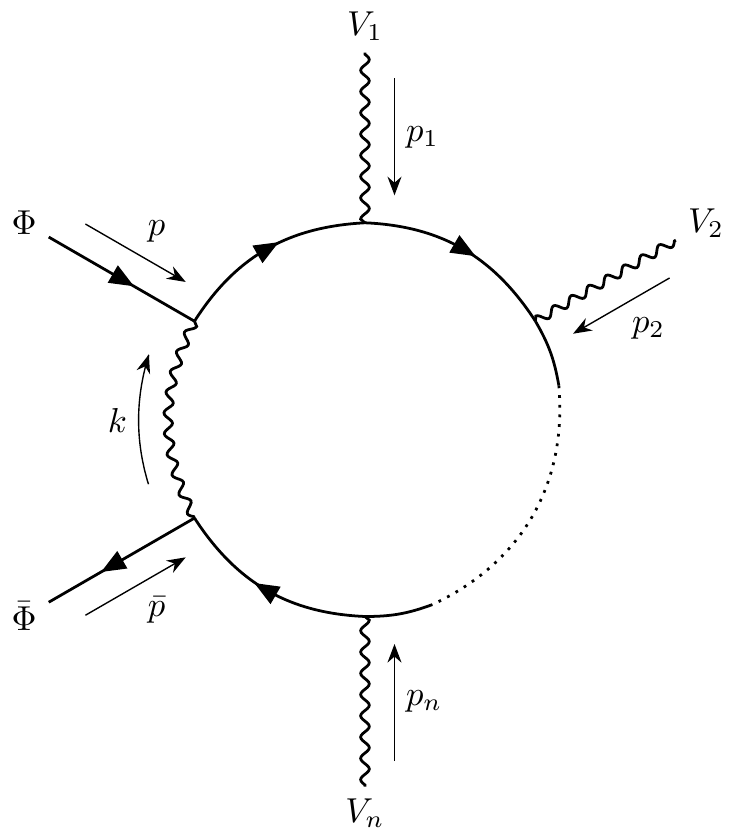}
    \caption{One-loop Feynman supergraph contributing to the vertex with one external chiral, one antichiral and $n$ vector superfields. Here $p_j \equiv (\Omega_j, \vec{p}_j)$ and $k \equiv (\o, \vec{k})$.}
    \label{fig:lack_renormalization}
\end{figure}

We then consider a generic supergraph with one chiral, one antichiral and $n$ vector external legs, \emph{i.e.}, we consider one-loop contributions to the correlator $\langle \bar{\Phi} V^n \Phi  \rangle$, with $n>2$. 
The corresponding diagram is depicted in fig.~\ref{fig:lack_renormalization}.

\noindent
Its contribution to the effective action reads
\begin{multline}
\label{eq:many_vectors_chiral}
i \G^{(n+2)} (\F,\bF,V) = (ig)^{n+2} \, \int d^4 \t_{\F} \, d^4 \t_{\bF} \, d^4 \t_1 \dots d^4 \t_n \int \frac{d \o \, d^2 k}{(2 \pi)^3} \; \bF V_1 \dots V_n  \F  
\\
\frac{- i \, \d_{\bF \F}}{- \vk^2 + i \e} \,
\frac{i D^2 \bD^2 \d_{\F 1}}{\square_{k+p} + i \e} \,
\frac{i D^2 \bD^2 \d_{1 2}}{\square_{k+p+p_1} + i \e} \,
\dots \,
\frac{i D^2 \bD^2 \d_{n \bF}}{\square_{k+p+p_1+\dots+p_n} + i \e} 
\end{multline}
where $\square_k \equiv 2 m \o - \vk^2$ and $\delta_{ij} \equiv \delta^{(4)} (\theta_i - \theta_j)$.
The denominator of the integrand goes to infinity as $\o^{n+1}$, while the numerator is dominated by terms whose UV divergences scale as $\o^n$. 
Therefore, its superficial degree of divergence is $\Delta_{\omega}=0,$ and selection rule \ref{sel_rule1} cannot be applied.
Instead, working in a way similar to the one used in section \ref{sec:selection_rules}, we find that D-algebra manipulations give 
\beq
\begin{aligned}
i \G^{(n+2)} (\F,\bF,V) & = - (-g)^{n+2} \, \int d^4 \t \; \bF  \Biggl( \sum_{i=0}^{n} \frac{1}{(\sqt m)^i} \mk{D}^{(i)} \, V_1 \dots V_n \Biggr) \F \\
&  \times
\int \frac{d \o \, d^2 k}{(2 \pi)^3} \,
\frac{1}{
\bigl[- \vk^2 + i \e \bigr]
\bigl[ 2 m \o- \vk^2 + i \e \bigr]}
\end{aligned}
\eeq
where the differential operator $\mathfrak{D}^{(i)}$ was formally defined in eq.~\eqref{eq:formal_sum}. 
This time the spacetime integral we are left with is divergent, and it evaluates to
\beq\label{eq:newterms}
\G^{(n+2)} (\F,\bF,V) \simeq - \frac{(-g)^{n+2}}{16 \pi m \e} \, \int d^4 \t \; \bF  \Biggl( \sum_{i=0}^{n} \frac{1}{(\sqt m)^i} \mk{D}^{(i)} \, V_1 \dots V_n \Biggr) \F
\eeq
This means that quantum corrections generate novel non-renormalizable terms where covariant derivatives $D$ and $\bD$ act on external vector legs.
For a Feynman supergraph with $n$ external prepotentials, one chiral and one antichiral superfields, the maximum number of covariant derivatives acting on the external vector superfields is $n$ $D$'s plus $n$ $\bar{D}$'s. 

Working out the general combinatorial factors and signs in $\mathfrak{D}^{(i)}$ would require a systematic development of the D-algebra with an increasing number of covariant derivatives acting on the internal lines, therefore with an increasing level of difficulty. Nonetheless, we can infer the general pattern of \eqref{eq:newterms} by means of the following reasoning.

First of all, gauge invariance greatly restricts the spectrum of possible $D$-structures in $\mk{D}^{(i)} \, V_1 \dots V_n$. In fact, all the combinations of the form
\beq
\int d^4 \theta \; \bar{\Phi}  V^a (D V)^b (\bar{D} V)^c (\bar{D} D V)^d \F \qquad
\forall \,\, a,d\geq 0 \,\, \wedge \,\,  b,c > 0
\eeq
with one single covariant derivative $D$ or $\bar{D} $ acting on the prepotential are ruled out since they are not invariant under gauge transformations \eqref{eq:gaugetransf}.
On the contrary, the block $\bar{D} D V$ is dimensionless (see dimensional counting in appendix \ref{app-dim}) and gauge-invariant, therefore it can enter any new non-renormalizable combination at an arbitrary power.
Furthermore, powers of $V$ without derivatives must reorganize in a gauge-invariant structure, precisely they need to resum to the exponential factor $e^{gV}$.
The three- and four-point correlators that we have computed in  eqs.~\eqref{eq:total_oneloop_correction_chiral_prop}, \eqref{eq:total_three_point} and \eqref{eq:total_1loop_correction_4pt_function} indeed satisfy these requirements.

Restricting the evaluation of the combinatorial factors only to the expected structures makes the development of D-algebra much easier. We eventually find that the Feynman supergraph in fig.~\ref{fig:lack_renormalization} for generic $n$ produces the following contribution
\beq
\Gamma^{(n+2)} (\bar{\Phi}, \Phi, V^n) \simeq  - \frac{g^{n+2}}{16 \pi m \varepsilon} \int d^4 \theta \; \bar{\Phi} \le  \frac{\bar{D} D V}{\sqrt{2}m}  \ri^n \F  
\eeq
and no other one-loop diagram modifies this factor.

Combining all these results, one can infer that the total one-loop divergent correction to the effective action sums into a geometric series as
\beq\label{eq:1loop_corrections_geometric_series}
\Gamma = - \frac{g^2}{16 \pi m \varepsilon} \int d^4 \theta \; \bF e^{g V} \Phi \, \frac{1}{1-\frac{g}{\sqrt{2}m}\bar{D} D V}  
\eeq
Since this does not reproduce the structure of the original action \eqref{eq:SGED_action}, the null SGED model is not renormalizable.

\subsection{Non-linear sigma model}
\label{sec:non_linear_sigma_model}

The appearance of the UV divergent term \eqref{eq:1loop_corrections_geometric_series} at one loop suggests that the interacting part of the original action \eqref{eq:SGED_action} has to be promoted to a non-linear sigma model of the form
\beq\label{eq:newaction}
 \int d^3 x  d^4 \theta \; \bF e^{g V} \Phi \, {\cal F}\left( \bar{D} D V \right)
\eeq
where ${\cal F}$ is a generic smooth function of its argument. Upon Taylor expansion of ${\cal F}$ \footnote{By suitably rescaling the (anti)chiral superfields, we choose the normalization ${\cal F}(0) = 1$ to reproduce the ordinary kinetic term for the scalars.}, this action exhibits an infinite number of new couplings weighted by the ${\cal F}$ derivatives
\beq \label{eq:Fderivatives}
{\cal F}^{(n)} \equiv \frac{d^n {\cal F}(x)}{d x^n}\Big|_{x=0}
\eeq
As anticipated in eq.~\eqref{eq:guess_additional_terms}, this marginal deformation is allowed by symmetries and dimensional analysis. It represents the supersymmetric version of the arbitrary functions $\mathcal{J}, \mathcal{V}, \mathcal{E}$ that enter the GED action at quantum level (see  eq.~\eqref{eq:action_extra_shira}). 
In fact, our finding is the supersymmetric version of what already emerges in the GED case \cite{Chapman:2020vtn}. The simplest non-relativistic version of (super)electrodynamics fails to be renormalizable, leading to the necessity to replace it with a more general non-linear sigma model. 
However, while in the GED case three arbitrary functions are required, in the present case supersymmetry forces only one function to appear.

To make a better comparison with the non-SUSY case, it is worth reducing \eqref{eq:newaction} to components. By using the prescription for Berezin integration \eqref{Berezin integration null reduction} and the component projections of the vector superfield \eqref{eq:components_new_terms}, and taking into account that $\mathcal{F}$ is killed by $D, \bar{D}$, we find
\beq
\begin{aligned}\label{eq:sigma_components}
\int d^3x & \Big[ \mathcal{F}(-\varphi) \,  \bar{D}^2 D^2 \le \bF e^{gV} \F \ri\Big|   - \lambda_2 \mathcal{F}'(-\varphi) \,  D_1 \bar{D} D (\bF e^{gV} \F)\Big|
+ \bar{\lambda}_2 \mathcal{F}'(-\varphi)
 \bar{D}_1 (\bF e^{gV} \F)\Big| 
\\
&  
+  \mathcal{F}'(-\varphi) \, 
\le  \tilde{D} + \frac{i}{2} \le \sqrt{2} \partial_t + \partial_{11} \ri \varphi  \ri
\, (\bF e^{gV} \F)\Big|
- \lambda_2 \bar{\lambda}_2  \mathcal{F}''(-\varphi) (\bF e^{gV} \F)\Big| \,   \Big]
\end{aligned}
\eeq
where $\mathcal{F}', \mathcal{F}''$ denote the first and second derivatives respect to $(\bar{D} D V)$, respectively, and the symbol $|$ means the function evaluated at $\theta=\bar{\theta}=0.$
In order to derive the previous expressions, use has been made of the identities \eqref{eq:components_new_terms}, together with 
the non-vanishing projections of $\mathcal{F}$ that explicitly  read
\beq\label{eq:sigma_model_components}
\begin{aligned}
& \mathcal{F}| = \mathcal{F} \le -  \varphi \ri \, , 
\qquad \qquad
 D_1 \mathcal{F}| = -  \bar{\lambda}_2 \,
\mathcal{F}' \le - \varphi \ri
\, , &  \\
&  \bar{D}_1 \mathcal{F}| = \lambda_2 \,
\mathcal{F}' \le - \varphi \ri
\, , \quad
 [\bar{D}_1, D_1] \mathcal{F}| =    \le 2 \tilde{D} + \sqrt{2}i \partial_t \varphi \ri \mathcal{F}' \le -  \varphi \ri &  \\
\end{aligned}
\eeq
These identities show that the non-linear sigma model \eqref{eq:newaction} corresponds to the original component action \eqref{eq:action_splitting} multiplied by the overall function ${\cal F}(-\varphi)$ (first term in the above expansion) plus additional terms proportional to higher components of the gauge multiplet. Consequently, the equations of motion \eqref{eq:integrate_out_gauge}  and \eqref{eq:integrate_out_matter} for the auxiliary fields get modified, now containing contributions proportional to ${\cal F}$ and its derivatives. 
The function ${\cal F}(-\varphi)$ generalizes the covariantized mass $\mathcal{M} = (m - \varphi)$ appearing in the GED action \cite{Chapman:2020vtn} (see also \eqref{eq:action_components}).

\section{A renormalizable SGED}
\label{sec:renormalizable_SGED}

The previous analysis reveals that at quantum level new infinite typologies of UV divergences arise, which combine into the geometric series \eqref{eq:1loop_corrections_geometric_series}.
This forces us to modify the original SGED action and consider the new model 
\beq\label{eq:action_new}
S_{\rm SGED} =  \int d^3 x  d^2 \theta \, W^2 + \int d^3 x  d^4 \theta \; \bF e^{g V} \Phi \, {\cal F}( \bar{D} D V)
\eeq
To make sense to this new action, we need to investigate its renormalizability properties. 

In principle, we could approach the problem using the same supergraph techniques adopted in section \ref{sec:one_loop_radiative_corrections}.
The disadvantage of such a method is that in order to compute contributions to the effective action we need to Taylor expand the exponential factor $e^{gV}$, so temporarily breaking gauge invariance of the sigma-model action. Restoring gauge invariance would require computing infinitely many correlators with an increasing number of external $V$ legs, or rely on the structure of the first few correlation functions, typically three- and four-point correlation functions, and infer the general one-loop result advocating gauge invariance. In the present situation, this procedure would be further complicated by the need to expand the sigma-model function ${\cal F}$, too.
\\
A more convenient, and eventually more compact way to organize the computation is to use a formalism where both supersymmetry and gauge invariance are manifest at each step.
The natural framework to achieve this task is the background field formalism \cite{Gates:1983nr,Grisaru:1984ja}.
We review this formalism in section \ref{sec:background_field_method} by adapting it to the Galilean superspace, and introduce the deformed SGED action with the inclusion of the new vertices, see eq.~\eqref{eq:new_SGED_Action}. 
We collect the corresponding Feynman rules in section \ref{sec:covariant_Feynman_rules}.
This formulation allows first to quickly re-derive the results obtained in section \ref{sec:one_loop_radiative_corrections}.
Then, in section \ref{sec:cov_radiative_corrections} it is used to evaluate one-loop corrections to the effective action of the new sigma-model \eqref{eq:action_new}, whereas its renormalization is discussed in section \ref{sec:renormalization_action}.

\subsection{Covariant approach and background field method}
\label{sec:background_field_method}

We begin by briefly introducing the background field method in non-relativistic ${\cal N}=2$ superspace. This is just an adaptation of the well-known method of ${\cal N}=1$ relativistic superspace in four dimensions \cite{Gates:1983nr}. 

Supergauge covariant derivatives are defined, in gauge chiral representation as\footnote{See appendix \ref{app:details_covariant} for more details on the definition and properties of gauge covariant derivatives in superspace.} \cite{Gates:1983nr}
\beq
\nabla_A = (\nabla_{\alpha}, \bar{\nabla}_{\beta}, \nabla_{\alpha \beta}) =
(e^{-V} D_{\alpha} \, e^V, \bar{D}_{\beta}, i \lbrace \nabla_{\alpha}, \bar{\nabla}_{\beta} \rbrace) 
\label{eq:chiral_cov_div}
\eeq 
and in vector representation as
\beq
\nabla_A = (\nabla_{\alpha}, \bar{\nabla}_{\beta}, \nabla_{\alpha \beta}) =
(e^{-V/2} D_{\alpha} \, e^{V/2}, e^{V/2} \bar{D}_{\beta} \, e^{-V/2}, i \lbrace \nabla_{\alpha}, \bar{\nabla}_{\beta} \rbrace) 
\label{eq:chiral_cov_div2}
\eeq 

Background-quantum splitting in covariant formalism is realised by replacing
\beq
e^V \rightarrow e^{\mathbf{\Omega}} e^V e^{\bar{\mathbf{\Omega}}} 
\label{eq:general_splitting_expV}
\eeq
where for the Abelian theory under consideration, we choose $\mathbf{\Omega}=\bar{\mathbf{\Omega}}= V_0/2$. This corresponds to a linear splitting 
\beq
V \rightarrow V_0 + V
\eeq
between the background prepotential $V_0$ and its quantum fluctuation $V$.

By inserting the splitting \eqref{eq:general_splitting_expV} inside the definitions \eqref{eq:chiral_cov_div}, and applying a similarity transformation, that is we multiply all the expressions by $e^{\bar{\mathbf{\Omega}}}$ on the left and $e^{- \bar{\mathbf{\Omega}}}$ on the right,
we obtain the background-quantum splitting of the covariant derivatives. They read
\beq
\nabla_{\alpha} = e^{-V} \boldsymbol{\nabla}_{\alpha} e^{ V} \, , \qquad
\bar{\nabla}_{\beta} = \boldsymbol{\bar{\nabla}}_{\beta} \, , \qquad
\nabla_{\alpha \beta} = i \lbrace \nabla_{\alpha}, \bar{\nabla}_{\beta}  \rbrace
\label{eq:quantum_splitting_covdiv}
\eeq
where we have defined background covariant derivatives in background-vector representation as
\beq
\boldsymbol{\nabla}_{\alpha} = e^{- V_0/2} D_{\alpha} \, e^{V_0/2} \, , \qquad
\bar{\boldsymbol{\nabla}}_{\beta} = e^{V_0/2} \bar{D}_{\beta} \,  e^{-V_0/2} \, 
\label{eq:back_cov_div}
\eeq 
The $\nabla$ derivatives transform covariantly under:
\begin{itemize}
    \item Quantum transformations parametrized by covariantly (anti)chiral superfields $\Lambda, \bar{\Lambda}$ ($\bar{\boldsymbol{\nabla}}_\alpha \Lambda = \boldsymbol\nabla_\alpha \bar{\Lambda} = 0$)
\beq
e^{V} \rightarrow e^{i \bar{\Lambda}} e^V e^{-i \Lambda} \, , \qquad
\boldsymbol{\nabla}_A \rightarrow \boldsymbol{\nabla}_A 
\eeq
\item Background transformations parametrized by real $K = \bar{K}$
\beq\label{eq:bkg_transfs}
e^{V} \rightarrow e^{i K} e^V e^{-i K} \, , \qquad
\boldsymbol{\nabla}_A \rightarrow e^{i K} \boldsymbol{\nabla}_A e^{-i K} 
\eeq
\end{itemize}

In terms of the background covariant derivatives \eqref{eq:back_cov_div}, we define covariantly (anti)chiral superfields 
\beq
\tilde{\Phi} = e^{V_0/2} \Phi \, , \qquad
\bar{\tilde{\Phi}} =  \bar{\Phi} e^{V_0/2} 
\label{eq:covariantly_chiral_superfields}
\eeq
that satisfy
\beq
\boldsymbol{\bar{\nabla}}_{\beta} \tilde{\Phi} = 0 \, , \qquad
\bar{\tilde{\Phi}} \overset{\leftarrow}{\boldsymbol{\nabla}}_{\alpha} = 0 
\eeq
In addition, we perform the linear background-quantum splitting 
\beq
\tilde{\Phi} \rightarrow \tilde{\Phi}_0 + \tilde{\Phi} \, , \qquad
\bar{\tilde{\Phi}} \rightarrow \bar{\tilde{\Phi}}_0 + \bar{\tilde{\Phi}} 
\label{eq:quantum_splitting_chirals}
\eeq
Equations~\eqref{eq:general_splitting_expV}, \eqref{eq:quantum_splitting_covdiv} and \eqref{eq:quantum_splitting_chirals} collect the necessary ingredients to perform the background field expansion of action \eqref{eq:action_new}, in the covariant formalism. To fit the previous calculations we shift $V_0 \to gV_0$ and $V \to gV$ in all these identities.

\subsection{Covariant Feynman rules}
\label{sec:covariant_Feynman_rules}

We start writing the full SGED action \eqref{eq:newaction} in terms of background and quantum superfields. This amounts to first re-writing it in terms of supergauge covariant derivatives and covariantly (anti)chiral superfields, and performing the background-quantum splitting as described above. 

Starting with the gauge sector, the splitting of the gauge superfield strength reads
\beq
W_\a \equiv ig \bar{D}^2 D_\a V = i \bar{D}^2 (e^{-gV} D_\a e^{gV}) \, \to \, i \boldsymbol{\bar{\nabla}}^2 (e^{-gV} \boldsymbol{\nabla}_{\alpha} e^{gV})
\eeq
Concerning the ${\cal F}$ argument, for the Abelian theory we can first of all write
\beq
\bar{D}_{\beta} D_{\alpha} (gV) = \bar{D}_{\beta} \le e^{-gV} D_{\alpha} e^{gV} \ri = \bar{\nabla}_{\beta} \nabla_{\alpha} {\mathbb I}
\eeq
Performing the quantum splitting \eqref{eq:general_splitting_expV} and the similarity transformation, we obtain
\beq
\bar{\nabla}_{\beta} \nabla_{\alpha} {\mathbb I} \rightarrow
\boldsymbol{\bar{\nabla}}_{\beta} ( e^{-gV} \boldsymbol{\nabla}_{\alpha} e^{gV} ) = \boldsymbol{\bar{\nabla}}_{\beta} 
\boldsymbol{\nabla}_{\alpha} (g V) 
\eeq
Finally, the gauge-fixing procedure can be easily covariantized and leads to the covariantized version of the action \eqref{eq:gauge_fixing_term} \cite{Gates:1983nr}. 

Regarding the matter sector, we perform the splitting in \eqref{eq:general_splitting_expV} and write 
\beq
\bar\Phi \, e^{gV} \Phi \to \bar\Phi \, e^{gV_0/2} e^{gV} e^{gV_0/2} \Phi = \bar{\tilde\Phi} \, e^{gV} \tilde\Phi
\eeq
and then split the superfields according to prescription \eqref{eq:quantum_splitting_chirals}.

Collecting all the terms, the gauge-fixed action we start from, reads  
\beq
\begin{aligned}
 S_{\rm SGED} & = \frac{1}{2} \int d^3 x  d^4 \theta \; \left[ \le e^{-gV} \boldsymbol{\nabla}^{\alpha} e^{gV}  \ri \boldsymbol{\bar{\nabla}}^2 \le e^{-gV} \boldsymbol{\nabla}_{\alpha} e^{gV} \ri  
+ \frac{1}{ \zeta}  V \le \boldsymbol{\nabla}^{2} \boldsymbol{\bar{\nabla}}^2 + \boldsymbol{\bar{\nabla}}^2 \boldsymbol{\nabla}^{2} \ri V \right] \\
&   \quad +  \int d^3 x \, d^4 \theta \;   
\le  \bar{\tilde{\Phi}} +  \bar{\tilde{\Phi}}_0 \ri e^{gV} \le \tilde{\Phi} + \tilde{\Phi}_0 \ri {\cal F}\le\boldsymbol{\bar{\nabla}}_{2} 
\boldsymbol{\nabla}_{2}  V \ri 
\end{aligned}
\label{eq:new_SGED_Action}
\eeq
At quantum level, the generating functional \eqref{eq:generating_functional_superfields} gets replaced by its covariant version
\beq
\mathcal{Z} [\tilde{J}, \bar{\tilde{J}}, J_V] = \int [\mathcal{D}\tilde\Phi \mathcal{D}\bar{\tilde \Phi} \mathcal{D} V] \, 
\exp \left[ i S_{\rm SGED} + i \int d^3 x \, \le \int d^2 \theta \tilde{J} \tilde{\Phi} + \int d^2 \bar{\theta} \bar{\tilde{J}} \bar{\tilde{\Phi}} + \int d^4 \theta J_V V  \ri   \right]
\eeq
where $\tilde{J},\bar{\tilde{J}}$ are covariantly (anti)chiral superfields and $J_V$ is a vector superfield, all of them acting as sources.
The corresponding functional derivatives bring factors of background covariant derivatives
\beq
\frac{\d \tilde{J}(z)}{\d \tilde{J}(z')}  = \boldsymbol{\bar{\nabla}}^2 \d^{(7)} (z-z') \, , \quad
 \frac{\d \bar{\tilde{J}}(z)}{\d \bar{\tilde{J}}(z')} = \boldsymbol{\nabla}^2 \d^{(7)} (z-z') \, , \quad
 \frac{\d J_V (z)}{\d J_V(z')}  = \d^{(7)} (z-z') 
 \label{eq:new_derivative_sources}
\eeq
Starting from the gauge-fixed action \eqref{eq:new_SGED_Action}, we can extract covariant Feynman rules. Here we simply list the results, while 
details on the differential operators and the derivatives introduced in the covariant formalism are collected in appendix \ref{app:details_covariant}.
In particular, the kinetic differential operators for the covariant superfields can be compactly written in terms of operators $\square_{\pm}$ and $\hat{\square}$ defined in eqs.~\eqref{eq:prop_pm} and \eqref{eq:prop_hat}, respectively.

\begin{itemize}
    \item Chiral propagator
    \beq
\langle \tilde{\Phi} (\omega, \vec{p},\theta,\bar{\theta}) \bar{\tilde{\F}} (-\omega, -\vec{p}, \theta', \bar{\theta}') \rangle =
  \frac{1}{\square_+} \,  \d^{(4)}(\t'-\t)  
\label{eq:new_chiral_superpropagator}
\eeq
\item Vector propagator (in Feynman gauge, $\zeta=1$)
\beq
\langle V (\omega,\vp,\theta,\bar{\theta}) V (-\omega,-\vp,\theta,\bar{\theta}) \rangle = - \frac{1}{\hat{\square}} \,  \d^{(4)}(\t'-\t)   
\label{eq:new_vector_superpropagator}
\eeq
\item Vertices: they arise from the series expansion of the exponential and the ${\cal F}$ function in the interacting term in eq.~\eqref{eq:new_SGED_Action}, which reads
\beq
\begin{aligned}
S_{\rm int} = 
 \sum_{\substack{m,n=0 \\ m+n>0}}^{\infty}  \frac{g^m}{m! \, n!} \, {\cal F}^{(n)} \, \int d^3 x \int d^4 \theta \, \le \bar{\tilde{\Phi}} +  \bar{\tilde{\Phi}}_0 \ri  V^m \le \tilde{\Phi} + \tilde{\Phi}_0 \ri
\le \boldsymbol{\bar{\nabla}}_{2} \boldsymbol{\nabla}_{2}  V \ri^n
\end{aligned}
\label{eq:new_interacting_action}
\eeq
where the ${\cal F}^{(n)}$ couplings have been defined in \eqref{eq:Fderivatives}.

There are in principle vertices with one chiral, one antichiral and an arbitrary number of vector superfields. 
However, at one loop only the three-point vertices depicted in fig.~\ref{fig:new_cov_vertices} are needed.
The right vertex is a new contribution proportional the coupling ${\cal F}^{(1)}$, while the left vertex is proportional to $g$ and resembles the one already present in the original SGED action \eqref{eq:SGED_action}. However, since the legs correspond to covariant superfields, it comprises all the vertices of the form $\bar{\Phi} V_0^n \Phi $ in the formulation of section \ref{sec:one_loop_radiative_corrections}.

\begin{figure}[ht]
    \centering
    \includegraphics[scale=1]{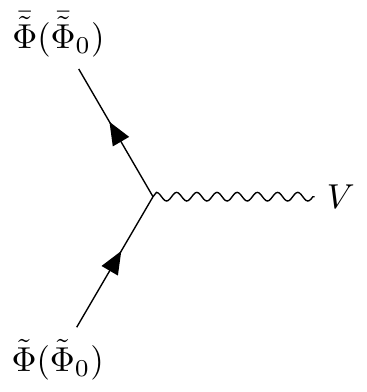} \qquad
     \includegraphics[scale=1]{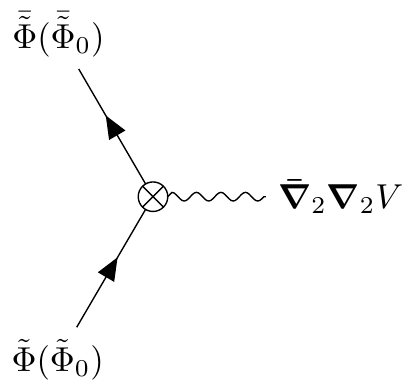}
    \caption{Three-point vertices arising from the interaction terms of the SGED action, eq.~\eqref{eq:new_interacting_action}. While the left vertex is the covariantized version of the original one, the right one is genuinely new.}
    \label{fig:new_cov_vertices}
\end{figure}

\item Due to the identities \eqref{eq:new_derivative_sources}, we assign one factor of $\boldsymbol{\bar{\nabla}}^2 (\boldsymbol{\nabla}^2)$ to any chiral (anti-chiral) internal line exiting from a vertex.
\end{itemize}
In addition, we include the usual combinatorial factors coming from the expansion of the interacting action, and we impose energy and momentum conservation at each vertex.

In order to reduce supergraphs contributions to ordinary Feynman integrals, we perform $\nabla$-algebra manipulations until we are left with a single grassmannian integral of a local function in superspace.
Explicitly, this is realized by first moving all the covariant $\boldsymbol{\nabla}, \bar{\boldsymbol{\nabla}}$ derivatives distributed on the diagram to act on a single vertex,  
then integrating by parts at that vertex. Moving covariant derivatives requires commuting them with $1/\square_{\pm}$ and $1/\hat\square$ operators using identities in \eqref{eq:Dalgebra_covdiv}. The procedure stops when we reach a configuration where exactly two $\boldsymbol{\nabla}$'s and two $\bar{\boldsymbol{\nabla}}$'s survive in each loop.
Whenever we end up with a lower number of covariant derivatives in a loop, the corresponding diagram vanishes due to the fermionic nature of the Berezin integration.

After ${\nabla}$-algebra, the original integral reduces to a linear combination of standard Feynman integrals in configuration space, and ordinary QFT methods can be applied to compute them.
A novelty of this approach, compared to the non-covariant superspace formalism, is that the differential operators $\square_{\pm}, \hat{\square}$ defined in eqs.~\eqref{eq:prop_pm} and \eqref{eq:prop_hat} have a non-trivial dependence on the background fields. This turns out to be crucial when evaluating covariant supergraphs, as we are going to discuss.

We proceed with the one-loop renormalization of the SGED action defined in eq. \eqref{eq:new_SGED_Action}. 
To this end, we consider renormalized quantities as defined in eq.~\eqref{ren_functions}, complemented by the renormalization of the
function $\mathcal{F}$ given by
\beq
\mathcal{F}_B = \mathcal{F} + \delta \mathcal{F} = 
\mathcal{F} + \sum_{l=1}^{\infty} \delta {\cal F}_l
\label{eq:Fbare}
\eeq
where $\delta {\cal F}_l$ is the counterterm at loop order $l$. 
It is important to observe that while the bare sigma-model function depends on $\bar{D} D V$, the functional dependence of the renormalized ${\cal F}$ and its counterterm $\d {\cal F}$ is on $\mu^{\varepsilon} \bar{D} D V$, since this is the correct dimensionless quantity in $d=2-2\varepsilon$.
Requiring that all the coupling constants $\mathcal{F}^{(n)}$ induced by its Taylor expansion remain dimensionless after the regularization implies that they acquire a dimensional deficit as follows
\beq
{\cal F}^{(n)}_B = \mu^{n \varepsilon} \le {\cal F}^{(n)} + \sum_{l=1}^{\infty}\delta {\cal F}^{(n)}_l \ri =
\mu^{n \varepsilon} Z_{\mathcal{F}^{(n)}} \mathcal{F}^{(n)}
\label{eq:Fbare_components}
\eeq
where we have defined $Z_{\mathcal{F}^{(n)}} \equiv 1 + \sum_{l=1}^{\infty}\delta {\cal F}^{(n)}_l/{\cal F}^{(n)}$.

\subsection{Original covariant one-loop radiative corrections}
\label{sec:original_cov_radiative_corrections}

As a warming-up, we begin by discussing how the covariant formalism allows to rederive in a very efficient and compact way the results of section \ref{sec:one_loop_radiative_corrections} regarding the one-loop behavior of the original action \eqref{eq:SGED_action}. The main advantages of the present formalism are twofold. On the one hand, external covariantly (anti)chiral superfields are dressed with factors $e^{g V_0},$ therefore a single Feynman diagram with external (anti)chiral legs encodes infinitively many terms arising from the series expansion of the exponential. On the other hand, all the dependence on the background gauge sector is encoded, in a gauge covariant way, in the covariant propagators \eqref{eq:new_chiral_superpropagator} and \eqref{eq:new_vector_superpropagator}. Therefore, expanding them in powers of the external gauge superfields, produces infinitely many diagrams with an increasing number of external gauge legs (see expansions \eqref{eq:pictorial_expansion_hatprop} and \eqref{eq:pictorial_chiral_expansion}). Moreover, gauge invariance respect to background gauge transformations \eqref{eq:bkg_transfs} is manifest at each step.

The nice consequence is that at one loop we have to consider only two diagrams: a vacuum diagram made by closing a covariant chiral superpropagator, which after expansion would give rise to pure vector contributions, and a self-energy diagram $\langle \bar{\tilde{\Phi}} \tilde{\Phi} \rangle$ which encodes the infinite sum of terms of the form $\langle \bar{\Phi} V_0^n \Phi \rangle$.

\subsubsection*{Vacuum diagram}

\begin{figure}[ht]
    \centering
    \includegraphics[scale=1]{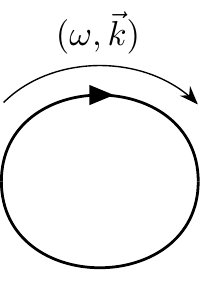}
    \caption{One-loop vacuum diagram.}
    \label{fig:vacuum_1loop}
\end{figure}

We first consider the vacuum diagram given by a single chiral superpropagator, whose edges are identified (see fig.~\ref{fig:vacuum_1loop}).

\noindent
Since there is a single propagator, the $\nabla$--algebra is trivial. We only need to use the identity
\beq
\boldsymbol{\nabla}^2 \frac{1}{\square_+} \boldsymbol{\bar{\nabla}}^2 = \frac{1}{\square_-} \boldsymbol{\nabla}^2 \boldsymbol{\bar{\nabla}}^2 
\label{eq:covDalgebra_single_prop}
\eeq
to make the covariant derivatives act on the $\delta$ function of the Grassmann variables. We are then left with a spacetime integral with propagator $\square_-^{-1}.$ 

Using expansion \eqref{eq:expansion_squarem},
at lowest order this differential operator gives an ordinary scalar propagator $1/\square_0,$ which in spatial dimensional regularization integrates to zero (see appendix \ref{app:math-tools}).
We then consider higher-other terms in the expansion. At each order the number of propagators increases, leading to non-tadpole, potentially non-vanishing contributions. Setting to zero the momenta of the external fields, at order $n$ in the expansion of $1/\square_-$ the worst UV behavior corresponds to a contribution of the form 
\beq
\int \frac{d\omega d^2 k}{(2\pi)^3} \frac{\omega^n}{(2 m \omega - \vec{k}^2 + i \varepsilon)^{n+1}} 
\label{eq:vacuum_diagrams_cov}
\eeq
where powers of $\o$ at numerator come from the term
\beq\label{eq:nasty_insertion}
 i \Gamma^{\alpha \beta} \p_{\alpha \beta} \rightarrow   i \Gamma^{11} \p_{11} =    \sqrt{2} g (\bar{D} D V_0) \,  \omega 
\eeq
appearing in the expansion of the covariant propagator $1/\square$ (see expansion \eqref{eq:expansion_cov_dAlembert}). 
However, using Sokhotski-Plemelj formula and  spatial dimensional regularization, it is easy to see that this contribution vanishes. In fact, upon substituting $\o^n \to (2 m \omega - \vec{k}^2)^n$ at numerator (all the extra terms vanish, having $\Delta_\o < 0$) we can simplify $n$ propagators, thus obtaining a tadpole of the form \eqref{eq:diagram3a_vector}.  

Any other configuration of external fields coming from the expansion of $1/\square_-$ leads to integrals with less powers of $\o$ at numerator, thus having $\Delta_\o < 0$. Therefore, they all vanish, thanks to selection rule \ref{sel_rule1}. 
This is nothing but an alternative proof of selection rule \ref{sel_rule2}.

\subsubsection*{Self-energy corrections}

We now evaluate the one-loop self-energy corresponding to the diagram in figure \ref{fig:1loop_chiral} but this time with (anti)chiral $\bar{\tilde{\Phi}}, \tilde{\Phi}$ superfields on the external legs and internal lines corresponding to covariant superpropagators. 

After using identity \eqref{eq:covDalgebra_single_prop}, $\nabla$--algebra is completely solved and we are left with internal $1/\square_-$ and $1/\hat\square$ propagators. 
As discussed in appendix \ref{app:details_covariant}, expanding these differential operators produces all possible insertions of the background gauge field. 

The lowest order term in these expansions corresponds to replacing the covariant propagators with flat Schr\"{o}dinger operators. The corresponding contribution to the effective action evaluates to
\beq
 - g^2 \int d^4 \theta \int \frac{d\omega d^2 k}{(2 \pi)^3} \frac{\tilde{\Phi}_0(\Omega,\vec{p}, \theta) \bar{\tilde{\Phi}}_0 (\Omega, \vec{p}, \bar{\theta})}{(- \vec{k}^2+ i \varepsilon) [ 2 m (\Omega- \omega) - (\vec{p} - \vec{k})^2 + i \varepsilon ]} \simeq  -  \frac{i g^2}{16 \pi m \varepsilon} \int d^4 \theta \; \bar{\Phi}_0 e^{g V_0} \Phi_0 
 \label{eq:cov_correction_original_vertex}
\eeq

Further potentially divergent contributions come from higher order terms in the propagators expansions, as long as the corresponding integral has $\Delta_{\omega}\geq 0$. Since in general the insertion of external legs increases the number of propagators, then improving the degree of convergence of the integral, the only possibility to obtain non-trivial UV divergent contributions is when extra powers of $\omega$ get produced at numerator, which exactly compensate for the additional propagators.

Considering first higher order terms in the expansion of the $\square_-^{-1}$ propagator, this happens only for insertions of type \eqref{eq:nasty_insertion}. The integral corresponding to the expansion at order $n$ reads
\begin{multline}
\label{eq:many_props}
 -g^2 \, \int d^4 \t \int \frac{d \o \, d^2 k}{(2 \pi)^3} \; 
  \frac{\bar{\tilde{\Phi}}_0 (\Omega, \vec{p}, \bar{\theta})\tilde{\Phi}_0(\Omega,\vec{p}, \theta)}{(- \vec{k}^2+ i \varepsilon) \left[ 2 m (\Omega- \omega) - (\vec{p} - \vec{k})^2 + i \varepsilon \right]}
\\
\times  
\frac{\sqrt{2} g (\bar{D} D V_0) \, \omega  }{2 m (\Omega+\Omega_1-\omega) - (\vec{p}+\vec{p}_1-\vec{k})^2 + i \varepsilon} \,
\frac{\sqrt{2} g (\bar{D} D V_0) \, \omega }{2 m (\Omega+\Omega_1+\Omega_2-\omega) - (\vec{p}+\vec{p}_1+\vec{p}_2-\vec{k})^2 + i \varepsilon} \\
\times  \, \dots \, \times \frac{\sqrt{2} g (\bar{D} D V_0) \, \omega }{2 m (\Omega+\Omega_1+ \cdots + \Omega_n-\omega) - (\vec{p}+\vec{p}_1+ \cdots + \vec{p}_n-\vec{k})^2 + i \varepsilon}
\end{multline}
To extract the UV divergence we set all the external momenta to zero, \emph{i.e.}, $\Omega_1 = \Omega_2 = \dots =0$ and $\vec{p}_1 = \vec{p}_2 = \dots = 0 .$
The integrals can be the evaluated by performing the integration over $\omega$ first, and then using dimensional regularization along the remaining spatial directions.
Summing the contributions at any order in the expansion, the final result reads
\beq
\begin{aligned}\label{eq:1loop_corrections_geometric_series2}
  - \frac{ig^2}{16 \pi m \varepsilon} \sum_{n=0}^{\infty} \int d^4 \theta \;  \bar{\tilde{\Phi}}_0 \tilde{\Phi}_0\le \frac{g}{\sqrt{2}m} \bar{D} D V_0 \ri^n = - \frac{ig^2}{16 \pi m \varepsilon}  \int d^4 \theta \;  \frac{ \bar{\Phi}_0 e^{gV_0} \Phi_0 }{1- \frac{g}{\sqrt{2}m} \bar{D} D V_0} 
\end{aligned}
\eeq
and corresponds precisely to the geometric series obtained in eq.~\eqref{eq:1loop_corrections_geometric_series}. 

In order to complete the calculation, we need to consider contributions from the expansion of the vector superpropagator $1/\hat\square$. However, as shown in appendix \ref{app:convergence_vec_insertions}, none of these terms give rise to divergent integrals. Therefore, we have found confirmation that the one-loop self-energy counterterm is  \eqref{eq:1loop_corrections_geometric_series}. 

\subsection{New covariant self-energy corrections}
\label{sec:cov_radiative_corrections}

\begin{figure}[ht]
    \centering
\subfigure[]{ \label{subfig:prop_cov2}  \includegraphics[scale=0.9]{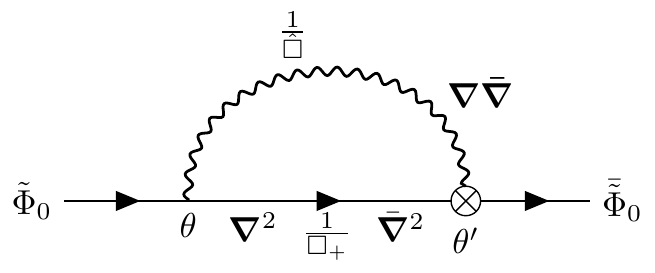}} 
\subfigure[]{\label{subfig:prop_cov3}  \quad \includegraphics[scale=0.9]{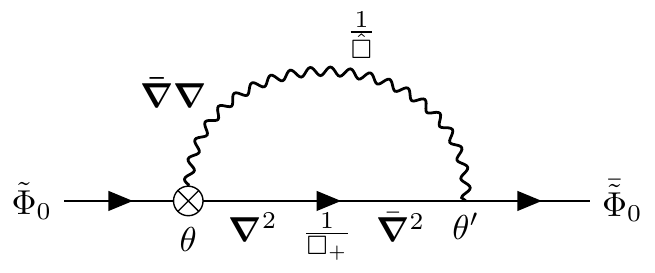}} 
\subfigure[]{ \label{subfig:prop_cov4}   \includegraphics[scale=0.9]{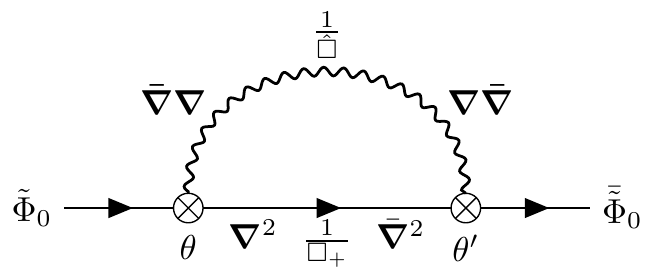} } 
    \caption{Novel one-loop self-energy diagrams containing three-point vertices originating from the expansion of the geometric series \eqref{eq:new_interacting_action}. The assignment of momenta is the same as in figure \ref{fig:1loop_chiral}.}
    \label{fig:1loop_cov_prop}
\end{figure}

For the non-linear sigma model \eqref{eq:new_SGED_Action}, the result in \eqref{eq:1loop_corrections_geometric_series2} needs to be completed with extra corrections corresponding to supergraphs in figure \ref{fig:1loop_cov_prop} that 
involve at least one new vertex carrying coupling ${\cal F}^{(1)}$ (right vertex in figure \ref{fig:new_cov_vertices}).

We begin by computing diagrams \ref{subfig:prop_cov2} and \ref{subfig:prop_cov3} that can be easily seen to give the same contribution. Therefore, we simply  double the result of the first one. 

In diagram \ref{subfig:prop_cov2} $\nabla$--algebra is made more complicated by the presence of an additional pair of background covariant derivatives $\bar{\boldsymbol{\nabla}} \boldsymbol{\nabla}$ inserted at the right vertex\footnote{To avoid confusion, we denote $ \nabla \equiv \nabla_2, \, 
\bar{\nabla} \equiv \bar{\nabla}_2$
while $\nabla^2, \bar{\nabla}^2$ denote the square of the covariant derivatives, according to the definitions in \eqref{eq:combinations_covdiv}.}. Integrating by parts at the right vertex to move derivatives from the vector to the chiral propagator, the resulting string of manipulations goes as follows
\beq
\begin{aligned}
& \left[ \bar{\boldsymbol{\nabla}} \boldsymbol{\nabla} \delta^{(4)} (\theta' - \theta) \right] 
\left[ \bar{\boldsymbol{\nabla}}^{2} \boldsymbol{\nabla}^2 \delta^{(4)} (\theta' - \theta) \right]  \bar{\tilde{\Phi}}_0  = \\
& = - \delta^{(4)} (\theta' - \theta) 
\left[  \le \boldsymbol{\nabla}  \bar{\boldsymbol{\nabla}}^{2} \boldsymbol{\nabla}^2  \delta^{(4)} (\theta' - \theta)  \ri  
\bar{\boldsymbol{\nabla}}  \bar{\tilde{\Phi}}_0  
 +  \le  \bar{\boldsymbol{\nabla}}^{2} \boldsymbol{\nabla}^2  \delta^{(4)} (\theta' - \theta)  \ri  
\lbrace \boldsymbol{\nabla}, \bar{\boldsymbol{\nabla}} \rbrace   \bar{\tilde{\Phi}}_0   \right]
\end{aligned}
\eeq
The first term vanishes due to $\nabla$--algebra.
The second one can be further manipulated using 
\beq
\lbrace \boldsymbol{\nabla} , \bar{\boldsymbol{\nabla}} \rbrace   \bar{\tilde{\Phi}}_0  = i \boldsymbol{\nabla}_{22} \bar{\tilde{\Phi}}_0  = 
 \sqrt{2}m \le 1 - \frac{g}{\sqrt{2}m} \bar{D} D V_0 \ri \bar{\tilde{\Phi}}_0  
\eeq
Except for this additional factor multiplying the external antichiral superfield, the rest of the integral is exactly the same that led to result \eqref{eq:1loop_corrections_geometric_series2}.
Therefore, we obtain
\begin{align} \label{eq:1loop_new}
i \Gamma_{\ref{subfig:prop_cov2} + \ref{subfig:prop_cov3}} & \simeq \frac{g}{16 \pi  m \varepsilon} \, 2\sqrt{2}m \, {\cal F}^{(1)} \,  \le 1 - \frac{g}{\sqrt{2}m} \bar{D} D V_0 \ri
\int d^4 \theta \; \frac{\bar{\tilde{\Phi}}_0 \tilde{\Phi}_0 }{1- \frac{g}{\sqrt{2} m} \bar{D} D V_0} \nonumber \\
& =  \frac{g }{16 \pi  m \varepsilon}  \, 2\sqrt{2}m \, {\cal F}^{(1)} \,  
\int d^4 \theta \; \bar{\Phi}_0 e^{g V_0} \Phi_0 
\end{align}

We then consider diagram \ref{subfig:prop_cov4}.
To perform $\nabla$-algebra we first bring all the covariant derivatives at the left (or equivalently at the right) of the propagators. On the chiral line this is trivial since $[\frac{1}{\square_+} , \bar{\boldsymbol{\nabla}}^2 ] =0$. On the vector propagator, we can exchange derivatives with $1/\hat\square$ up to an extra term proportional to their commutator. However, due to the identity
\beq
\left[\bar{\boldsymbol{\nabla}} \boldsymbol{\nabla}  , \frac{1}{\hat{\square}} \right] = \mathcal{O} \le \frac{1}{\hat{\square}} \ri^2
\eeq
one can show that the contributions arising from the commutator are convergent. At this point we can simply apply the transfer rule to the derivatives on the upper vector propagator, and obtain
\beq
\bar{\boldsymbol{\nabla}} (\theta)  \boldsymbol{\nabla} (\theta)
\bar{\boldsymbol{\nabla}} (\theta')  \boldsymbol{\nabla} (\theta')
\delta (\theta - \theta')  = 
- \bar{\boldsymbol{\nabla}} (\theta')  \boldsymbol{\nabla} (\theta')
\boldsymbol{\nabla} (\theta')  \bar{\boldsymbol{\nabla}} (\theta')
\delta (\theta - \theta') = 0 
\eeq
Therefore, diagram \ref{subfig:prop_cov4} does not contribute due to the fermionic nature of the covariant derivatives.

In conclusion, summing contribution \eqref{eq:1loop_new} coming from the new ${\cal F}^{(1)}$-vertex to the old contribution \eqref{eq:1loop_corrections_geometric_series2}, we obtain the following  one-loop correction to the chiral self-energy (from now on, we remove the subscript from background fields)
\beq
\Gamma^{(2)} (\tilde{\Phi}, \bar{\tilde{\Phi}}) \simeq 
- \frac{g}{16 \pi m \varepsilon}   \int d^4 \theta \, 
 \bar{\Phi} e^{g V} \Phi \, \le \frac{ g}{1- \frac{g}{\sqrt{2}m} \bar{D} D V} - 2 \sqrt{2}m \, {\cal F}^{(1)} \ri
 \label{eq:total_selfenergy_cov}
\eeq

\vskip 10pt
\subsection{Renormalization of the action}
\label{sec:renormalization_action}

The structure of the counterterm action which arises from the assignments \eqref{ren_functions} and \eqref{eq:Fbare} is 
\begin{align}
& \qquad \mathcal{L}_{\rm SGED}  + \int d^4 \theta \, \delta_{\Phi} \, \bar{\Phi} e^{g V} \Phi +
\int d^4 \theta \, \delta_{\Phi} \, \bar{\Phi} e^{g V} \Phi 
 \, \le {\cal F} - 1 \ri  
 + \int d^4 \theta \;  (1 + \d_{\Phi}) \bar{\Phi} e^{g V} \Phi  \,  \delta {\cal F}  \nonumber \\
& \hspace{-0.3cm} \underset{\rm 1 \, loop}{\longrightarrow} \; \mathcal{L}_{\rm SGED}  + \int d^4 \theta \, \delta_{\Phi}^{1 \rm L} \, \bar{\Phi} e^{g V} \Phi + \int d^4 \theta \,  \bar{\Phi} e^{g V} \Phi \, \delta {\cal F}_{1 \rm L}
\end{align}
On the other hand, the one-loop divergences collected in eq.~\eqref{eq:total_selfenergy_cov} can be rewritten as
\begin{align}
\Gamma^{(2)} (\tilde{\Phi}, \bar{\tilde{\Phi}}) \simeq 
& - \frac{g}{16 \pi m \varepsilon} (g - 2\sqrt{2}m \, {\cal F}^{(1)})  \int d^4 \theta \, 
 \bar{\Phi} e^{g V} \Phi \nonumber \\
& + \frac{g^2}{16 \pi m \varepsilon} \int d^4 \theta \, \bar{\Phi} e^{g V} \Phi \, \le  1 - \frac{1}{1 - \frac{g}{\sqrt{2}m} \bar{D} D V} \ri 
\end{align}
where we have added and subtracted the term $-\frac{g^2}{16 \pi m \varepsilon}\int d^4 \theta \bar{\Phi} e^{g V} \Phi$.
Therefore, summing these two expressions, we find that renormalization at one loop requires to fix
\beq
\delta_{\Phi}^{1 \rm L} =  \frac{g}{16 \pi m} \le g - 2\sqrt{2} m \, {\cal F}^{(1)} \ri \frac{1}{\varepsilon} + {\rm finite \; terms}
\label{eq:counterterm_Phi}
\eeq
\beq
\delta {\cal F}_{1 \rm L}= -\frac{g^2}{16 \pi m} \,  \le  1 - \frac{1}{1 - \frac{g}{\sqrt{2}m} \bar{D} D V} \ri \frac{1}{\varepsilon} + {\rm finite \; terms}
\label{eq:counterterm_F1}
\eeq
It can be easily seen that the second identity implies the following one-loop renormalization for the $n$-coupling
\beq
\delta {\cal F}_{1\rm L}^{(n)} =  g^{n+2} \, \frac{n!}{16\pi m (\sqrt{2} m)^{n}} \, \frac{1}{\varepsilon} + {\rm finite \; terms}
\eeq

\begin{figure}[ht]
    \centering
    \includegraphics[scale=1]{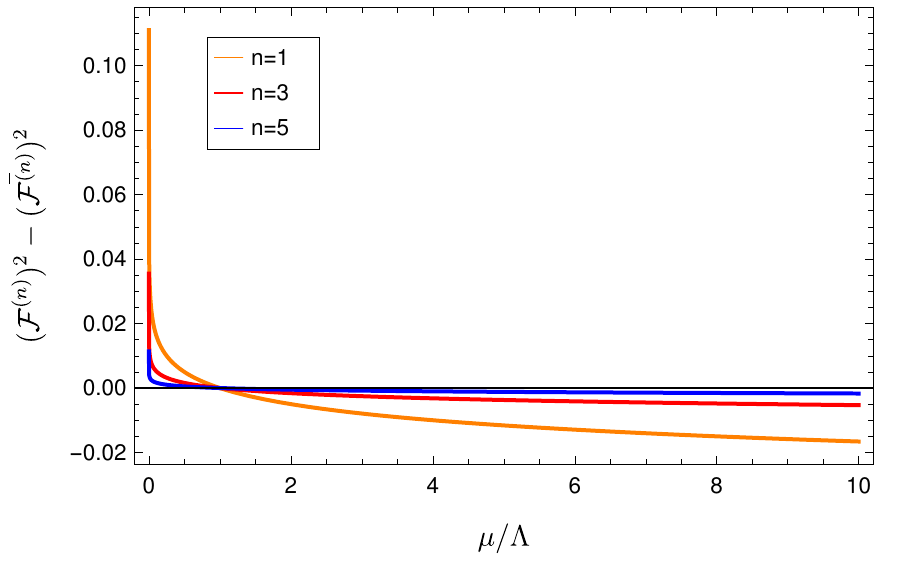}
    \caption{Plot of the left-hand side of eq.~\eqref{eq:F2subtracted} as a function of $\mu,$ for fixed $g=0.8,m=1$ and various choices of $n$.}
    \label{fig:F2subtracted}
\end{figure}

The beta function for the $n$--th coupling can be easily evaluated by deriving eq. \eqref{eq:Fbare_components} respect to $\log \mu$\footnote{From now on we omit the subscript specifying the loop order, so avoiding unnecessary clutter.} 
\beq
0 = \frac{d \mathcal{F}^{(n)}_B}{d \log \mu} = \mu^{n \varepsilon}
Z_{\mathcal{F}^{(n)}} \mathcal{F}^{(n)} \left[ n \varepsilon+ \frac{\mu}{Z_{\mathcal{F}^{(n)}}} \frac{d Z_{\mathcal{F}^{(n)}}}{d \mu} + \frac{\mu}{\mathcal{F}^{(n)}} \frac{d \mathcal{F}^{(n)}}{d \mu}  \right]
\label{eq:derivative_Fbare}
\eeq
At lowest order we simply approximate $Z_{\mathcal{F}^{(n)}}=1,$ then obtaining
\beq
\frac{d \mathcal{F}^{(n)}}{d \log \mu} = - n \varepsilon \mathcal{F}^{(n)}
\label{eq:leading_beta_Fn}
\eeq
At the next order, we plug this identity into eq.~\eqref{eq:derivative_Fbare} and we use result \eqref{eq:counterterm_F1}. After performing the $\varepsilon \rightarrow 0$ limit, we obtain the beta-functions of the SGED theory at one-loop
\beq
\beta_{\mathcal{F}^{(n)}} = \frac{d \mathcal{F}^{(n)}}{d \log \mu} = - g^{n+2} \frac{n! \; n}{16 \pi m (\sqrt{2} m)^{n} \, \mathcal{F}^{(n)}} 
\label{eq:beta_Fn}
\eeq
supplemented by $\beta_g = 0 $. Thanks to the $g$ independence on the mass scale, this equation can be easily integrated, leading to the following behavior for the square of the running ${\cal F}^{(n)}$ coupling
\beq
({\cal F}^{(n)})^2(\mu)  - (\bar{\cal F}^{(n)})^2  =
- g^{n+2} \frac{n! \; n}{8 \pi m (\sqrt{2} m)^{n}} 
\log{\le \frac{\mu}{\Lambda} \ri}
\label{eq:F2subtracted}
\eeq
where $\bar{\cal F}^{(n)}$ is the value of the coupling at the substraction scale $\Lambda$. 
We depict the difference in the left-hand side of eq.~\eqref{eq:F2subtracted} in fig.~\ref{fig:F2subtracted}, as a function of $\mu$ and at fixed $g >0.$

\noindent
Finally, the anomalous dimension of the (anti)chiral superfield is easily obtained from the definition
\beq
\gamma_{\Phi} \equiv \frac{1}{2} \frac{d \log Z_{\Phi}}{d \log \mu} = \frac{1}{2} \frac{\partial \log Z_{\Phi}}{\partial \mathcal{F}^{(1)}} \frac{d \mathcal{F}^{(1)}}{d \log \mu} =
\frac{g}{8 \pi \sqrt{2}} \, \mathcal{F}^{(1)}
\label{eq:wavefunction_ren}
\eeq
where in the last step we used eqs.~\eqref{eq:counterterm_Phi} and \eqref{eq:leading_beta_Fn}, together with the non-renormalization of the mass and the coupling $g.$

A priori, the above results are valid for generic values of $g$, since we derived them using the background field method that automatically accounts for contributions to all orders in this coupling.
However, in order to make corrections \eqref{eq:counterterm_Phi} and \eqref{eq:counterterm_F1} perturbatively meaningful, as well as the beta function in \eqref{eq:beta_Fn}, we need to require
\beq
g \ll 1 \, , \qquad
g\mathcal{F}^{(1)} \ll 1 \, , \qquad
g^{n+2}/{\cal F}^{(n)} \ll 1 
\label{eq:assumptions_perturbative}
\eeq

The beta function of any coupling constant $\mathcal{F}^{(n)}$ entering the non-linear sigma model vanishes when $g=0$, where also the wavefunction renormalization vanishes, see eq.~\eqref{eq:wavefunction_ren}.
At this value of the coupling, the gauge transformations of matter fields become trivial and the minimal coupling $e^{gV}$ disappears, while a $V$ dependence survives in the analytic function $\mathcal{F}$. At $g=0$ the theory exhibits a non-trivial infrared fixed point where it enjoys the full non-relativistic superconformal invariance and corresponds to an interacting field theory with action
\beq
S_{\rm fixed} = \int d^3 x  d^2 \theta \, W^2 + \int d^3 x  d^4 \theta \; \bF \Phi \, {\cal F}( \bar{D} D V)
\eeq
The analysis of the infinite-dimensional matrix $\partial_I \beta^J$ with $I,J \in \lbrace g, \mathcal{F}^{(1)}, \mathcal{F}^{(2)}, \dots  \rbrace$ reveals that its eigenvalues are given by the set
\beq
  \frac{1}{16 \pi m}  \left\lbrace 0, \frac{1}{\sqrt{2} m} \frac{g^{3}}{(\mathcal{F}^{(1)})^2}  , \dots ,
\frac{n! \, n}{(\sqrt{2} m)^{n}} \frac{g^{n+2}}{(\mathcal{F}^{(n)})^2} \right\rbrace
\eeq
When evaluated at the fixed point $g=0$, we clearly find a vanishing matrix, which corresponds to stating that there is an infinite-dimensional superconformal manifold parametrized by arbitrary values of the couplings $\mathcal{F}^{(n)}.$

The RG flow of the $\mathcal{F}^{(n)}$ couplings depicted in fig. \ref{fig:F2subtracted} and the topology of the conformal manifold that we have found are peculiar of the one-loop approximation. At higher orders, we expect different couplings to mix at different orders, possibly leading to a richer spectrum of fixed points. In that case, the existence of a conformal manifold would constrain the functional form of the ${\cal F}$ function non-trivially.

\section{Conclusions}
\label{sec:conclusions}

In this paper we have studied a non-relativistic version of three dimensional quantum electrodynamics with ${\cal N}=2$ supersymmetry. 
We have found that the retarded nature of the non-relativistic propagator entails the existence of a non-renormalization theorem which protects the coupling constant $g$ from acquiring quantum corrections.
This result generalizes to the supersymmetric setting what has been observed for the electric charge in GED \cite{Chapman:2020vtn}.
However, despite this partial quantum protection, an infinite set of marginal deformations, consistent with supersymmetry and gauge invariance, are generated at quantum level.
These quantum corrections deform the simplest action, obtained by null reduction from four dimensions, into a non-linear sigma model of the form \eqref{eq:non_linear_sigma_model}, which contains infinite new couplings.
The one-loop analysis of the deformed action reveals that the beta functions of these new couplings vanish when $g=0$, see eq.~\eqref{eq:beta_Fn}.
This case identifies an interacting non-trivial superconformal manifold where the minimal coupling term in the SGED action is turned off, but infinite marginal deformations parametrized by the Taylor series of the analytic function $\mathcal{F}$ survive.

The SGED provides a first example of a theory where supersymmetry does not significantly improve the renormalization properties of its purely scalar counterpart (GED).
While it restricts the class of marginal deformations to a single analytic function, its functional expression is not fixed at one loop level.

The approach and the results derived in this work open the possibility for several future directions.
First of all, it would be interesting to exploit the background field method described in section \ref{sec:renormalizable_SGED} to study higher-loop corrections.
It is clear that the fixed point $g=0$ will survive, because all the counterterms and the beta functions are proportional to the coupling constant $g.$
However, it is non-trivial to understand if higher loops will allow for other fixed points, characterized by choosing specific values of the couplings $\mathcal{F}^{(n)}$ in eq.~\eqref{eq:Fderivatives}.
Furthermore, we expect that higher loops should impose further constraints on the functional form of the sigma model, thus restricting the choice of the marginal deformations.

Another possible future direction is the study of renormalization properties of non-Abelian Galilean-invariant gauge theories, which were considered at classical level in \cite{Bagchi:2015qcw,Bagchi:2022twx}.
Their supersymmetric generalization would also provide a non-relativistic version of SYM theories, which play a fundamental role in high energy physics.
In particular, $\mathcal{N}=4$ SYM is superconformal invariant at the full quantum level and it allows to perform precise matchings of the AdS/CFT correspondence.
In recent years, it has been shown that decoupling limits of $\mathcal{N}=4$ SYM lead to quantum mechanical models, called Spin Matrix Theories (SMT), which have non-relativistic symmetries and are closed under the action of the one-loop dilatation operator \cite{Harmark:2007px}.
The effective Hamiltonian of these sectors in the near-BPS limit have been derived in \cite{Harmark:2019zkn,Baiguera:2020jgy,Baiguera:2020mgk,Baiguera:2021hky}, leading in some cases to field theory and superfield formulations.
It would be interesting to find a link between these decoupling limits of $\mathcal{N}=4$ SYM and a null reduction procedure, and study the quantum properties of both theories.
In holography, SMTs provide a simpler setting where precise matching of the AdS/CFT duality can be performed even at finite number $N$ of colours.
The dual gravitational models are described by non-relativistic string theories \cite{Gomis:2000bd,Andringa:2012uz,Bergshoeff:2018yvt,Harmark:2017rpg,Harmark:2018cdl,Harmark:2019upf,Oling:2022fft}, and the properties of worldvolume actions in this framework were studied in \cite{Gomis:2020fui,Ebert:2021mfu}.
These investigations aim at deepening our understanding of non-relativistic holography.
Another approach in this direction has been recently carried out in \cite{Bizon:2018frv,Maxfield:2022hkd}, where a non-relativistic limiting procedure of the AdS/CFT correspondence is proposed.

It would be interesting to consider other gauge-invariant actions, in particular Chern-Simons theories.
Since they are topological, one could couple them to non-relativistic matter without changing the form of the gauge action.
Investigations of these models have been performed in \cite{Jackiw:1990mb,Bergman:1993kq,Nishida:2007pj,Son:2013rqa,Geracie:2014nka}. 
Supersymmetric $\mathcal{N}=2$ Chern-Simons models were considered in \cite{Leblanc:1992wu,Bergman:1995zr}.
It would be interesting to derive these models from null reduction, couple them to supersymmetric non-relativistic matter and study their quantum properties.

While in the present work we focused on the renormalization properties of field theories in flat space, it is possible to couple GED to Newton-Cartan geometry \cite{Festuccia:2016caf}.
It would be therefore natural to extend the present investigation to the case of SGED coupled to non-relativistic supergravity \cite{Bergshoeff:2014ahw,Bergshoeff:2015uaa,Bergshoeff:2015ija}.
Finally, it would be interesting to study the general structure of terms that must be added to the flat space theory in order to preserve supersymmetry in a non-relativistic and curved setting, along the lines of what has been done for relativistic theories \cite{Festuccia:2011ws}.

\vskip 25pt

\acknowledgments

We thank Roberto Auzzi and Giuseppe Nardelli for collaboration at the first stage of this work. We thank Shira Chapman for valuable discussions.
This work has been supported in part by Italian Ministero dell'Universit\`a e Ricerca (MUR), and by Istituto Nazionale di Fisica Nucleare (INFN) through the ``Gauge Theories, Strings, Supergravity'' (GSS) research project.
SB is supported by the Israel Science Foundation (grant No. 1417/21), by the German Research Foundation through a German-Israeli Project Cooperation (DIP) grant “Holography and the Swampland”, by the Azrieli Foundation and by the Kreitmann School of Advanced Graduate Studies.


\addtocontents{toc}{\protect\setcounter{tocdepth}{1}}
\appendix

\section{Galilean $ \mathcal{N}=2 $ superspace}
\label{app-conv}

In this appendix we collect the conventions on spinors, superspace and the Berezin integration, following strictly the notations used in \cite{Auzzi:2019kdd}.

In four-dimensions, the ${\cal N}=1$ relativistic superspace is parametrized by coordinates $(x^M, \theta^\alpha, \bar{\theta}^{\dot\alpha})$, where $M\in \lbrace 0, 1,2,3 \rbrace$, $\alpha \in \lbrace 1,2 \rbrace$ and $\dot\alpha \in \lbrace \dot{1},\dot{2} \rbrace$. 
We work with \emph{mostly positive} Lorentzian metric $\eta_{MN} = \diag{(-1,1,\dots,1)}$.
By means of the null reduction prescription described in section \ref{sec:null_reduction}, we obtain a non-relativistic three-dimensional ${\cal N}=2$ superspace, whose spacetime and spinor coordinates are denoted by $x^{\mu} \equiv (x^+, x^i)$ with $i = 1,2$ and $\theta^\alpha, \bar\theta^\alpha \equiv (\theta^\alpha)^\dagger, \alpha = 1,2$, respectively. 

\vskip 10pt

\subsection*{Double spinor notation}  

Before performing null reduction, it is convenient to trade the components of a generic four-dimensional one-form field $A_M$ with its components in double spinor notation, defined as
\beq
\begin{aligned}
\label{eq:dsi}
&& \sqrt{2} {\cal V}_{\alpha \da} \equiv   (\sigma^-)_{\a \da} \, \varphi + (\sigma^+)_{\alpha \da} \, {\cal V}_t + (\sigma^1)_{\alpha \da} \, {\cal V}_1 + (\sigma^2)_{\alpha \da} \, {\cal V}_2  \\
&& \sqrt{2} {\cal V}^{\alpha \da} \equiv   (\bar\sigma^-)^{\da \a} \, \varphi + (\bar\sigma^+)^{\da \a} \, {\cal V}_t + (\bar\sigma^1)^{\da \a} \, {\cal V}_1 + (\bar\sigma^2)^{\da \a} \, {\cal V}_2
\end{aligned}
\eeq
where in light-cone coordinates $ \f \equiv A_-$, ${\cal V}_t \equiv A_+$ and $\mathcal{V}_i \equiv A_i.$
The Pauli matrices $\sigma^M = (\sigma^-, \sigma^+, \sigma^i)$ are given by
\beq
\begin{aligned}
&& \sigma^{\pm} = \frac{1}{\sqrt{2}} (\sigma^{3} \pm {\mathbb I}) \, , \qquad
\bar{\sigma}^{\pm} = \frac{1}{\sqrt{2}} (-{\sigma}^{3} \pm {\mathbb I}) \, , \qquad \bar{\sigma}^i = - \sigma^i 
\nl
&& \sigma^- = - \bar{\sigma}^+ = \sqrt{2}
\begin{pmatrix}
0 & 0 \\
0 & -1 
\end{pmatrix} \, , \qquad
\sigma^+ = - \bar{\sigma}^-= \sqrt{2}
\begin{pmatrix}
1 & 0 \\
0 & 0 
\end{pmatrix} 
\end{aligned}
\label{eq:Pauli_lightcone}
\eeq
Under null reduction, definitions \eqref{eq:dsi} give rise to analogous identities in $2+1$ dimensions. Relabelling the Pauli matrix components as $(\s^M)_{\a\b}$ and $(\bar\s^M)^{\a\b}$, with $M \in \lbrace +, -, 1,2 \rbrace $, in $2+1$ dimensions we define
\beq
\begin{aligned}
\label{eq:dsi3d}
&& \sqrt{2} {\cal V}_{\alpha \b} \equiv   (\sigma^-)_{\a \b} \, \varphi + (\sigma^+)_{\alpha \b} \, {\cal V}_t + (\sigma^1)_{\alpha \b} \, {\cal V}_1 + (\sigma^2)_{\alpha \b} \, {\cal V}_2 \\
&& \sqrt{2} {\cal V}^{\alpha \b} \equiv   (\bar\sigma^-)^{\a\b} \, \varphi + (\bar\sigma^+)^{\a\b} \, {\cal V}_t + (\bar\sigma^1)^{\a\b} \, {\cal V}_1 + (\bar\sigma^2)^{\a\b} \, {\cal V}_2
\end{aligned}
\eeq
Similarly, in double spinor notation spacetime coordinates are given by
\beq
x^{\alpha \beta} = - \frac12 (\bar{\sigma}_M)^{\beta \alpha} x^M \, , \qquad
x^M = (\sigma^M)_{\alpha \beta} x^{\alpha \beta} 
\label{eq:coordinates_double_spinor}
\eeq

\vskip 10pt

\subsection*{Fermions}

Complex non-relativistic fermions in $2+1$ dimensional Galilean geometry are given in terms of two complex Grassmann scalars $\psi_\a = (\psi_1, \psi_2)$. They can be obtained from null reduction of the relativistic four-dimensional Weyl spinors, according to the prescription in \eqref{eq:nullred}. 
Spinorial indices are raised and lowered as
\beq\label{raise&lower}
\psi^{\alpha} = \epsilon^{\alpha \beta} \psi_{\beta} \, , \qquad
\psi_{\alpha} = \epsilon_{\alpha \beta} \psi^{\beta}
\eeq
where the Levi-Civita symbol is
\beq \label{eq:LC}
\elc^{\a \b}  = - \elc_{\a \b} =  
\begin{pmatrix}
0    &   1 
\\
-1    &    0
\end{pmatrix} 
\eeq
The same rules apply to complex conjugate fermions $\bar{\psi}$, where 
the prescription for complex conjugation is 
\beq
(\psi^{\alpha})^{\dagger} =  \bar{\psi}^{\alpha} \, , \qquad
(\psi_{\alpha})^{\dagger} =  \bar{\psi}_{\alpha} \, , \qquad
(\bar{\psi}^{\alpha})^{\dagger} =  \psi^{\alpha} \, , \qquad
(\bar{\psi}_{\alpha})^{\dagger} =  \psi_{\alpha} 
\label{complex conjugation 3+1 spinors}
\eeq
We contract fermionic quantities according to the convention
\beq \label{eq:contractions}
 \chi \cdot \psi \equiv  \chi^{\alpha} \psi_{\alpha} 
 = \psi \cdot \chi \, ,  \qquad
 \bar{\chi} \cdot \bar{\psi} \equiv \bar{\chi}_{\alpha} \bar{\psi}^{\alpha}
  = \bar{\psi} \cdot \bar{\chi}  
  \eeq
  
\vskip 10pt

\subsection*{Superspace derivatives}

In the Galilean setting, spacetime derivatives in double spinor notation are defined as
\beq
\p_{\alpha \beta} = (\sigma^M)_{\alpha \beta} \p_M \, ,  \qquad
\p_M = - \frac12 (\bar{\sigma}_M)^{\beta \alpha} \p_{\alpha \beta}
\label{eq:derivatives_double_spinor}
\eeq
and are subject to rules \eqref{raise&lower} for raising and lowering spinorial indices. Therefore, we can explicitly write
\beq
\label{eq:3d_derivatives}
\p_{\a \b} =
\begin{pmatrix}
\sqrt{2} \p_t     &    \p_1 - i \p_2 
\\
\p_1 + i \p_2    &    - i \sqrt{2} M
\end{pmatrix} \, ,
\qquad
\p^{\a \b} = 
\begin{pmatrix}
- i\sqrt{2} M          &    -(\p_1 - i \p_2) 
\\
-(\p_1 + i \p_2)     &    \sqrt{2} \p_t 
\end{pmatrix} 
\eeq
where $M$ denotes the central charge in the Bargmann algebra  associated with the $\mathrm{U}(1)$ mass eigenvalue. 

We define the action of spinorial derivatives on the Grassmann variables as
\beq
\p_{\alpha} \theta^{\beta} = \delta_{\alpha}^{\,\,\, \beta} \, , \qquad
\p^{\beta} \theta_{\alpha} = - \delta_{\alpha}^{\,\,\, \beta} \, , \qquad
\bar{\p}_{\alpha} \bar{\theta}^{\beta} =  \delta_{\alpha}^{\,\,\, {\beta}} \, , \qquad
\bar{\p}^{\beta} \bar{\theta}_{\alpha} = - \delta^{\alpha}_{\,\,\, {\beta}} 
\eeq
While spacetime derivatives are anti-hermitian ( $ (\p_{M})^{\dagger} = - \p_{M} $), the spinorial ones are hermitian, \emph{i.e.}, $ (\p_{\alpha})^{\dagger} = \bar{\p}_{\alpha}$.  

Supercharges and SUSY covariant derivatives in Galilean superspace can be obtained by null-reducing their relativistic counterparts in four dimensions. Starting from 4D supercharges 
\beq
{\cal Q}_{\alpha} = i \le \p_{\alpha} + \frac{i}{2} \bar{\theta}^{\dot{\alpha}}\p_{\alpha {\dot{\alpha}}} \ri \, , \qquad
\bar{\cal Q}_{\dot{\alpha}} =  - i \le \bar{\p}_{\dot{\alpha}} + \frac{i}{2} \theta^{\alpha} \p_{\alpha {\dot{\alpha}}} \ri  
\eeq
and covariant derivatives 
\beq
\label{eq:covder4d}
{\cal D}_\a = \frac{\p}{\p \theta^\a} - \frac{i}{2} \bar{\theta}^\db \p_{\a \db} \; , \qquad
\bar{\cal D}_\da = \frac{\p}{\p \bar{\theta}^\da} - \frac{i}{2} \theta^\b \p_{\b \da}
\eeq
in three dimensions we obtain 
\beq
\begin{aligned}
& Q_1 = i \frac{\p}{\p \theta^1} - \frac12  \bar{\theta}^2 (\p_1 - i \p_2) - \frac{1}{\sqrt{2}} \bar{\theta}^1  \p_t \, \quad 
\bar{Q}_1 = -i \frac{\p}{\p \bar{\theta}^1} + \frac12 \theta^2 (\p_1 + i \p_2) + \frac{1}{\sqrt{2}} \theta^1 \p_t   \\
& Q_2 = i \frac{\p}{\p \theta^2} - \frac12  \bar{\theta}^1 (\p_1 + i \p_2) - \frac{i}{\sqrt{2}} \bar{\theta}^2 M \, \, \quad
\bar{Q}_2 = -i  \frac{\p}{\p \bar{\theta}^2} + \frac12 \theta^1 (\p_1 - i \p_2) - \frac{i}{\sqrt{2}} \theta^2 M 
\end{aligned}
\eeq
\beq
\begin{aligned}
\label{eq:nonrelD}
& D_1 =  \frac{\p}{\p \t^1} - \frac{i}{2} \bt^2 (\p_1 - i \p_2) - \frac{i}{\sqrt{2}} \bt^1 \p_t 
\, , \quad 
\bD_1 =   \frac{\p}{\p \bt^{1}} - \frac{i}{2} \t^2 (\p_1 + i \p_2) - \frac{i}{\sqrt{2}} \t^1 \p_t  \\
& D_2 =  \frac{\p}{\p \t^2} - \frac{i}{2} \bt^1 (\p_1 + i \p_2) - \frac{1}{\sqrt{2}} \bt^2 \, M \, , \quad 
\bD_2 =   \frac{\p}{\p \bt^2} - \frac{i}{2} \t^1 (\p_1 - i \p_2) - \frac{1}{\sqrt{2}} \t^2 \,  M
\end{aligned}
\eeq
For a generic superfield $\Phi$ we define SUSY transformations as
\beq\label{eq:SUSYtransf}
\delta \Phi =  [i \le \varepsilon^\a Q_\a + \bar\varepsilon_\a \bar{Q}^\a \ri , \Phi ]
\eeq
Covariant derivatives satisfy the following algebra 
\beq
\{ D_\a , \bD_\b \} = - i \p_{\a \b}  \; , \qquad \{ D^\a , \bD^\b \} = - i \p^{\b \a} 
\eeq
whereas $\{ D_\a , D_\b \} = \{ \bD_\a , \bD_\b \}  =0$. 
We list here further identities which turn out to be useful when doing D-algebra computations on supergraphs
\beq 
\label{eq:useful_identities}
\begin{aligned}
& 
[ D^\a , \bD^2 ] = i \p^{\b \a} \bD_\b \, , 
\qquad
[ \bD^\a , D^2 ] = -i \p^{\a \b} D_\b 
\\
& 
\{ D^2 , \bD^2 \} = (2iM\p_t + \p_i^2) + D^\a \bD^2 D_\a = (2iM\p_t + \p_i^2)  + \bD_\a D^2 \bD^\a
\end{aligned}
\eeq
Here we have used
\beq
\square_0 \equiv - \tfrac12 \p^{\a\beta} \p_{\a\beta}  = 2iM\p_t + \p_i^2    \, , \quad 
D^2 \equiv \tfrac12 D^\a D_\a = D_2 D_1 \, , \quad 
\bar D^2 \equiv \tfrac12 \bar D_\a \bar D^\a= \bar{D}_1\bar{D}_2
\label{eq:combinations_susy_der}
\eeq

When acting on a Grassmannian delta function $ \delta_{ij} \equiv \delta^{(2)} (\theta_i - \theta_j) \, \delta^{(2)} (\bar{\theta}_i - \bar{\theta}_j ) ,$ the covariant derivatives give
\begin{align}
& \delta_{ij} \delta_{ij} = 0 \, , \quad
\delta_{ij} D^{\alpha} \delta_{ij} = 0 \, , \quad
\delta_{ij} D^{2} \delta_{ij} = 0 \, , \quad
\delta_{ij} D^{\alpha} \bar{D}^{\dot{\alpha}} \delta_{ij} = 0 \, , \quad \delta_{ij} D^{\alpha} \bar{D}^2 \delta_{ij} = 0 \, ,
 &  \nonumber \\
 &   \delta_{ij} D^{\alpha} \bar{D}^2 D^{\beta} \delta_{ij} = -\epsilon^{\alpha \beta} \delta_{ij} \, , \quad   \delta_{ij} D^{2} \bar{D}^2  \delta_{ij} = \delta_{ij} \bar{D}^2 D^{2} \delta_{ij} =  \delta_{ij} \frac{ D^{\alpha} \bar{D}^2 D_{\alpha}}{2}  \delta_{ij} =  \delta_{ij}   &
 \label{rules covariant derivatives on delta functions}
\end{align}

\vskip 10pt

\subsection*{Berezin integration}

Manifestly supersymmetric actions can be constructed by using the Berezin integral on spinorial coordinates. In relativistic superspace, for a generic superfield $\Psi$ we define
\beq \label{relberezin}
\int d^4x d^4\theta \, \Psi = \int d^4x \, {\cal D}^2 \bar{\cal D}^2 \Psi \Big|_{\theta = \bar{\theta}=0}
\eeq
with covariant derivatives given in eq.~\eqref{eq:covder4d}. 
Performing null reduction and extracting the $x^-$ dependence of the superfield as in eq.~\eqref{eq:nullred}, we obtain the prescription for the Berezin integrals in Galilean superspace
\beq \label{Berezin integration null reduction}
\begin{aligned}
& \int d^4x d^4\theta \, \Psi = \int d^4x \, {\cal D}^2 \bar{\cal D}^2 \Psi \Big|_{\theta = \bar{\theta}=0} \;  \longrightarrow  \\
&  \int d^3x D^2 \bar{D}^2 \tilde{\Psi} \Big|_{\theta = \bar{\theta}=0} \; \times \frac{1}{2\pi} \int_0^{2\pi}  dx^- \,  e^{iMx^-}   \equiv \int d^3x d^4\theta \, \tilde{\Psi}  \; \times \frac{1}{2\pi} \int_0^{2\pi}  dx^- \,  e^{iMx^-}
\end{aligned}
\eeq
where spinorial derivatives are now the ones introduced in eq.~\eqref{eq:nonrelD}. 

It is immediate to observe that if $M \neq 0$ we obtain a trivial reduction due to the vanishing of the $x^-$ integral. Therefore, non-vanishing expressions arise only if the super-integrand $\Psi$ is uncharged with respect to the mass generator, or in other words if it is invariant under the global U(1) symmetry \cite{doi:10.1063/1.529465}.

\vskip 15pt

\subsection*{Superfield expansion}

\noindent
The null reduction prescription \eqref{eq:nullred} of four-dimensional relativistic superfields leads to a natural definition of superfields in Galilean superspace.
A chiral superfield $\Phi$ satisfying the constraint $\bar{D}_\a \Phi = 0$ with spinorial derivatives given in eq.~\eqref{eq:nonrelD} has an expansion of the form 
\beq\label{chiral}
\Phi (x, \theta, \bar{\theta}) = \phi(x_L) + \theta^1 \psi_1(x_L) + \theta^2 \, 2^{1/4} \sqrt{m} \, \psi_2(x_L) - \theta^2F(x_L)  \, , \quad x_L^{\alpha\beta} =  x^{\alpha\beta} - \frac{i}{2} \theta^{\alpha}  \bar{\theta}^{\beta} 
\eeq
where $\psi_\a$ is a non-relativistic fermion, whereas $\phi$ and $F$ are complex scalars, the latter being an auxiliary field.
Similarly, an antichiral superfield satisfiying $D_\a \bar{\Phi} = 0$ reads
\beq\label{antichiral}
\bar\Phi (x, \theta, \bar{\theta}) = \bar{\phi}(x_R) + \bar{\theta}_1 \bar{\psi}^1(x_R) + \bar{\theta}_2 \,  2^{1/4} \sqrt{m} \, \bar{\psi}^2(x_R) - \bar\theta^2F(x_R) \, , \quad x_R^{\alpha\beta} =  x^{\alpha\beta} + \frac{i}{2} \theta^{\alpha} \bar{\theta}^{\beta}  
\eeq
In both cases we have suitably normalized the fermions, in order to obtain a standard kinetic term in the action in components (see eq. \eqref{eq:action_components}). 

The single components are given by (we denote with $|$ the evaluation of a quantity at $\theta^\alpha=\bar{\theta}^\alpha=0$)
\beq
\begin{aligned}
&& \f = \Phi| \; , \qquad \psi_1 = D_1\Phi| \; , \quad \quad \psi_2 = \frac{1}{2^{1/4} \sqrt{m}} D_2\Phi| \; , \quad \; \; F = D^2 \Phi|  \\
&& \fb = \bar\Phi| \; , \qquad \bar\psi_1 = -\bar{D}_1\bar\Phi| \; , \quad \bar\psi_2 = -\frac{1}{2^{1/4} \sqrt{m}} \bar{D}_2\bar\Phi| \; , \quad \bar{F} = \bar{D}^2 \bar\Phi|
\end{aligned}
\eeq
A real vector superfield in non-relativistic superspace has the following expansion
\begin{multline}
\label{eq:real}
V = C + \t^{\a} \chi_{\a} + \bt_{\a} \bar{\chi}^{\a} - \t^2 N - \bt^2 \bar{N} + \t^{\a} \bt^{\b} A_{\a \b}
\\
- \bt^2 \t^{\a} \Bigl(\l_{\a} + \frac{i}{2} \p_{\a \b} \bar{\chi}^{\b} \Bigr) - \t^2 \bt_{\a} \Bigl(\bl^{\a} + \frac{i}{2} \p^{\b \a} \chi_{\b}\Bigr) + \t^2 \bt^2 \Bigl(\tilde{D} + \frac12 \sq_0 C\Bigr) \, ,
\end{multline}
where $N, \bar{N}$ are complex scalars, $C,\tilde{D}$ are real scalars, $\chi_{\a}, \lambda_{\a} $ and their hermitian conjugates are non-relativistic fermions, while $A_{\a\b}$ is a $2+1$ vector potential written in double spinor notation (see definition in eq.~\eqref{eq:dsi3d})
\beq\label{Vexpansion}
\sqrt{2} A_{\alpha \beta}  =
\left( \begin{matrix}
\sqrt{2} A_t & A_1 - i A_2 \\
A_1 + i A_2 & - \sqrt{2} \f 
\end{matrix} \right) \; , \qquad 
\sqrt{2} A^{\alpha \beta}  =
\left( \begin{matrix}
- \sqrt{2} \f & -(A_1 - i A_2) \\
-(A_1 + i A_2) & \sqrt{2} A_t 
\end{matrix} \right)
\eeq
The entries of these matrices can be explicitly obtained by projecting the prepotential as follows
\bea
&& \varphi = - \frac{1}{2} [\bar{D}_2, D_2 ] V| \qquad \quad A_t = \frac{1}{2} [\bar{D}_1, D_1 ] V| \nonumber \\
&& A_1 -iA_2= \frac{1}{\sqrt{2}} [\bar{D}_2, D_1 ] V| \qquad A_1 + iA_2 = \frac{1}{\sqrt{2}} [\bar{D}_1, D_2 ] V|
\eea
These can be compactly written as $A_{\a\b} = \tfrac12 [\bar{D}_\b , D_\a ] V|$ and $A^{\a\b} = \tfrac12 [\bar{D}^\a , D^\b ] V|$. 

The remaining $V$ components read
\beq
\label{eq:projections_vector}
\begin{aligned}
&& C = V| \; , \quad \chi_\a = D_\a V| \; , \quad \bar\chi_\a = - \bar{D}_\a V| \; , \quad N = D^2V| \; , \quad \bar{N} = \bar{D}^2 V| \\
&& \l_\a = \bar{D}^2 D_\a V| \; , \quad \bar{\l}_\a = - D^2 \bar{D}_\a V| \; , \quad  \tilde{D}= \frac12 D^\a \bar{D}^2 D_\a V| 
\end{aligned}
\eeq
We normalize the fermionic components as
\bea
\label{eq:nonrel_vector_projections}
& \chi_{\a}  = (\chi_1 ,2^{1/4} \sqrt{m} \, \chi_2) \, ,
\qquad
\bar{\chi}_{\a} = (\bar{\chi}_1,2^{1/4} \sqrt{m} \, \bar{\chi}_2) &
\\
& \l_{\a}  = (\lambda_1,\lambda_2) \, ,
 \qquad
\bl_{\a} = (\bar{\lambda}_1,\bar{\lambda}_2)  &
\eea
The absence of the  rescaling factor $2^{1/4} \sqrt{m}$ in the second line is due to the fact that the gaugino field $\l_{\a} $ is massless.

The Wess-Zumino gauge used to reduce superspace actions to components is defined by the following conditions 
\beq 
V| \!= \!  D_\a V| \!= \! \bar{D}_\a V| \!= \! D^2V| \!= \! \bar{D}^2 V| \!= \! 0 
\label{eq:WZ_gauge}
\eeq
Non-relativistic supersymmetric Abelian gauge theories are described in terms of a couple of chiral and antichiral superfield strengths, $W_\a = i \bar{D}^2 D_\a V$ and $\bar W_\a = - i D^2 \bar{D}_\a V$, whose expansion in the Wess-Zumino gauge reads 
\beq
W_\a =  i \lambda_\a + \theta^\b (f_{\a\b} - i \e_{\a\b} \tilde{D}) - \theta^2 \partial_{\alpha\beta} \bar{\lambda}^{\beta} \; , \quad \quad
\bar W_\da = i \bar\lambda_\da - \bar\theta^\b (\bar{f}_{\a\b} + i \e_{\a\b} \bar{D}) + \bar\theta^2 \partial_{\beta\alpha} \lambda^{\beta}
\eeq
with the $\mathrm{U}(1)$ field strength given by $F_{\a\b, \g\d} = \p_{\a\b} A_{\g \d} - \p_{\g\d} A_{\a\b} = \e_{\a\g} \bar{f}_{\b \d} + \e_{\b\d} f_{\a\g}$, while $\l_\a$ is the gaugino. 

\vskip 10pt
As discussed in the main text, the $\bar{D}_2D_2 V$ superfield plays a relevant role in the SGED construction. Its non-vanishing components read
\beq
\begin{aligned}
&(\bar{D}_2 D_2 V) | = - \varphi  \; , \quad D_1(\bar{D}_2 D_2 V) | = - \bar{\lambda}_2 \; , \quad \bar{D}_1(\bar{D}_2 D_2 V) |  = \lambda_2 \, ,& \\
& [\bar{D}_1, D_1](\bar{D}_2 D_2 V) |  =  2\tilde{D}
+ \sqrt{2} i \partial_t \varphi &
\end{aligned}
\label{eq:components_new_terms}
\eeq
Projecting SUSY transformation \eqref{eq:SUSYtransf} on the 
$D_2 \bar{D}_2 V$ components, we obtain their supersymmetry trasformations
\beq
\delta \varphi = - \le \varepsilon_2 \bar{\lambda}_2 - \bar\varepsilon_2 \lambda_2 \ri \, , \quad  
\delta \lambda_2 = - \varepsilon_2 \tilde{D} \quad , \quad 
\delta \bar\lambda_2 = - \bar\varepsilon_2 \tilde{D}  \quad , \quad \delta \tilde{D} = 0
\eeq

\section{Dimensional analysis}\label{app-dim}

In this appendix we perform dimensional analysis in  non-relativistic frameworks, where the scalings of time and space coordinates differ by the dynamical exponent $z$ defined by scale transformations
\beq
t \rightarrow e^{z \sigma} t \, , \qquad
x^i \rightarrow e^{\sigma} x^i 
\label{eq:dynamical_exponent}
\eeq
The relativistic case corresponds to $z=1,$ while in the Galilean setting under investigation, $z=2.$
Since the speed of light is formally sent to infinity, in this context we do not associate to mass and length opposite dimensions, rather we perform a  counting of dimensions consistent with the anisotropic scaling \eqref{eq:dynamical_exponent}.
Equivalently, from null reduction \eqref{eq:nullred} we observe that we can choose the mass $M$ to be dimensionless.

In the resulting $\mathcal{N}=2$ superspace we count energy dimensions as follows
\beq
[t] = -2 \; \Rightarrow \; [\partial_t] = 2 \, , \qquad  \qquad [x^i] = -1 \; \Rightarrow \; [\partial_i] = 1  \qquad i=1,2
\eeq
From the ordinary definition of gauge covariant derivatives, $D_\mu = \p_\mu - i A_\mu$, it follows that gauge connections have the same dimensions as derivatives. 
Consistency of these dimensions with the requirement $[V]=0$ implies that superspace spinorial coordinates have different dimensions. Precisely,
\beq
\begin{aligned}
&& [\theta^1] = [\bar{\theta}^1] = -1 \; \; \Rightarrow \; \; [D_1] = [\bar{D}_1] = 1  \\
&& [\theta^2] = [\bar{\theta}^2] = 0 \; \quad \Rightarrow \; \; [D_2] =[\bar{D}_2] = 0 
\end{aligned}
\label{eq:dimension_counting_derivatives}
\eeq
The appearance of dimensionless derivatives has very important consequences in our construction. In fact, it allows to build a series of infinite marginal deformations using the dimensionless superfield $\bar{D}_2 D_2 V$.

Consistently with the previous assignments, we obtain that the $x_{L,R}$ components defined in eq.~\eqref{chiral} have dimensions
\beq
[x_{L,R}^{11}] = -2 , \qquad [x_{L,R}^{12}] = [x_{L,R}^{21}] = -1 , \qquad [x_{L,R}^{22}] = 0
\eeq
Moreover, it follows that the measure of the Berezin integration satisfies
\beq
[d^2\theta] = [D^2] = 1 \qquad \quad   [d^2\bar{\theta}^2] = [\bar{D}^2] = 1 \qquad \quad  [d^4\theta] = [D^2 \bar{D}^2] = 2
\eeq

\noindent
We list explicitly the energy dimensions of the relevant component fields entering the SGED action.
\begin{itemize}
\item \underline{Gauge fields}

From the previous dimensional analysis it follows that the dimensions of the $A_{\a\b}$ matrix summarize as
\beq
[A_t] = 2 \, , \qquad [A_1]=[A_2]=1 \, , \qquad [\varphi] = 0
\label{eq:dimension_counting_gauge_field}
\eeq
whereas for the superfield strengths we obtain
\beq
[W_1] = [\bar{D}^2 D_1 V] = 2 \, , \qquad 
[W_2] = [\bar{D}^2 D_2 V] = 1 
\eeq
It follows that $[W^2] = [\frac12 (W_2W_1 - W_1 W_2)] =3$. Therefore, from the gauge action in eq.~\eqref{action} we obtain $[g] = 0$.

\item
\underline{Matter fields}

Since $[d^3x \, d^4\theta] = -2,$ the matter Lagrangian in eq.~\eqref{action} needs to have dimension 2. 
This implies that $[\Phi]=1$. From its expansion \eqref{chiral} we then read the dimensions of the field components and their complex conjugates
\beq
[\phi] = [\bar{\phi}] = 1 \, , 
\quad [\psi_1] = [\bar{\psi}^1] = 2 \, , 
\quad [\psi_2] = [\bar{\psi}^2] =  1 \, \quad 
[F] = [\bar{F}]= 2
\eeq 
\end{itemize}

\section{Mathematical tools}
\label{app:math-tools}

In this appendix we compute the relevant momentum integrals entering the evaluation of one-loop radiative corrections.

When dealing with three-dimensional Galilean theories, the typical integration over momentum space is of the form
\beq
\int \frac{d \o \, d^2 k}{(2 \pi)^3} \, f(\o,\vk) 
\eeq
The strategy that we are going to adopt is the following: we perform the $\omega$ integration first, and then use dimensional regularization in $d = 2 - 2\varepsilon$ to evaluate the spatial $d^2 k$ integral. 
This method has the advantage that it is not necessary to move to Euclidean space and there are no ambiguities in taking the $\varepsilon \rightarrow 0$ limit.\footnote{For an alternative approach which introduces a dimensional deficit both along temporal and spatial coordinates, the reader can look at the dimensional splitting techinique described in \cite{Arav:2016akx}.}
In the following, we will denote with $\simeq$ the evaluation of expressions discarding finite terms in the UV cutoff.

\subsection*{Standard integral over $\omega$}

The main mathematical tool that we need for the evaluation of the integrals along $\omega$ is the Sokhotski-Plemelj formula \cite{Gelfand:105396}. 
In the theory of distributions, it states that 
\beq
\label{eq:sp_formula}
\lim_{\e \to 0^+} \int_a^b \, d \o \, \frac{f(\o)}{\o \pm i \e} = \mp i \pi f(0) + \mathcal{P} \int_a^b \, d \o \, \frac{f(\o)}{\o} 
\eeq
where $f \in \mathcal{S}(\mathbb{R}, \mathbb{C}) $ is a test function in the Schwartz space, $a<0<b$ are real-valued constants and $\mathcal{P}$ denotes the Cauchy principal value.

All the $\omega$ integrations that we encounter in this paper can be written as
\beq
\label{eq:integral_I}
I \equiv \int_{-\infty}^{\infty} \frac{d\o}{2\pi} \, \frac{1}{2m \o - \vk^2 + i \e} = 
\frac{1}{4 \pi m} \int_{-\infty}^{\infty} d\o \, \frac{1}{\o + i \e} 
\eeq
where we used the change of variables $\o \longrightarrow \o + \vk^2/2m$ to bring the integral into a convenient form to apply the formula \eqref{eq:sp_formula}.
We interpret the integrand in distributional sense with $f(\o)=1$ and $a \to - \infty$, $b \to + \infty.$
Since the principal value of this expression vanishes, we simply obtain
\beq
\label{eq:integral_I_final}
I = - \frac{i}{4 m} 
\eeq
As a further check of the correctness of this result, we observe that the integrand in eq.~\eqref{eq:integral_I} is precisely the Galilean propagator.
Its general expression in configuration space reads
\beq
G(t,\vx) = \int \frac{d \o \, d^2 k}{(2 \pi)^3} \, \frac{i}{2 m \o - \vk^2 + i \e} e^{-i(\o t - \vk \cdot \vx)} = \frac{\t(t)}{4 \pi i t} e^{i \frac{i m \vx^2}{2t}} 
\eeq
If we take a step back in the computation and we consider the expression of the propagator after the computation of the $\o$-integral, but before that of the $\vk$-integral, we can write
\beq
\int \frac{d^2 k}{(2 \pi)^2} \, e^{i \vk \cdot \vx} 
\Biggl(
\int_{-\infty}^{\infty} \frac{d \o}{2 \pi} \, \frac{i e^{-i \o t}}{2 m \o - \vk^2 + i \e} - \frac{\t(t)}{2m} e^{-i \frac{\vk^2}{2m} t}
\Biggr) = 0 
\eeq
which implies
\beq
\int_{-\infty}^{\infty} \frac{d \o}{2 \pi} \, \frac{i e^{-i \o t}}{2 m \o - \vk^2 + i \e} = \frac{\t(t)}{2m} e^{-i \frac{\vk^2}{2m} t} 
\eeq
This expression, evaluated at $t=0,$ precisely gives\footnote{Note that we interpreted $\theta (0) = 1/2,$ which is the standard choice in the theory of distributions. One can interpret this result as the average between the $t \rightarrow 0^{\pm}$ limits.}
\beq
\int_{-\infty}^{\infty} \frac{d \o}{2 \pi} \, \frac{1}{2 m \o - \vk^2 + i \e} = - \frac{i}{4m}
\eeq
which is in agreement with the previous result in eq.~\eqref{eq:integral_I}.

\vskip 10pt

\subsection*{Integrals over the spatial momenta}

We now deal with the integrals along the spatial momenta $\vec{k}$ using dimensional regularization in the two-dimensional Euclidean space that is left after the integration along the temporal direction. 
Although we focus on the UV divergences, in some cases we also need to regularize the integrals in the IR.

\paragraph{Integral $\mathbf{J_0}$.} The simplest integral corresponds to
\beq
\label{eq:integral_J_0}
J_0 \equiv \int \frac{d^2 k}{(2 \pi)^2} = 0
\eeq
which is well known to vanish in dimensional regularization~\cite{Leibbrandt:1975dj}.

\paragraph{Integral $\mathbf{J_1}$.} 
The next integral that we consider is
\beq
\label{eq:integral_J_1}
J_1 \equiv \int \frac{d^2 k}{(2 \pi)^2} \frac{1}{-\vk^2 + i \e} 
\eeq
Using dimensional regularization, we introduce the energy scale $\mu,$ which compensates the dimensional deficit of the integral analytically continued to $d$ dimensions. Since the integral is IR divergent, we also introduce a mass scale $\kappa^2$ at the denominator.
The integral evaluates to
\beq
\label{eq:integral_J_1_reg}
J_1 = - \mu^{2-d} \int \frac{d^d k}{(2\pi)^d} \, \frac{1}{\vk^2 + \kappa^2} = - \frac{1}{2^d \pi^{d/2}} \biggl( \frac{\kappa}{\mu} \biggr)^{d-2} \Gamma \biggl( 1 - \frac{d}{2} \biggr) 
\eeq
where we have used the result for the volume of the unit $d$--dimensional sphere $\O_d = \frac{2 \pi^{d/2}}{\G (d/2)}$, 
and we have applied standard integration techniques to express the result in terms of the Euler $\Gamma$ function.

Setting $d = 2 - 2 \e$, the Laurent series expansion for the $\Gamma$ function around $\varepsilon=0$ leads to
\beq
J_1 = - \frac{1}{4 \pi} \biggl[ \frac{1}{\e} + \log{\biggl( \frac{4 \pi \mu^2}{\kappa^2} \biggr)} - \c + \mathcal{O}(\e) \biggr] \simeq 
- \frac{1}{4 \pi \e} 
\eeq
where in the last step we have kept only the term relevant for the minimal subtraction scheme. 

{\em En passant}, we note that an alternative but equivalent way to regularize the IR divergence of the massless vector superfield in eq.~\eqref{eq:integral_J_1} is to subtract it via a shift
\beq
\int \frac{d^d k}{(2 \pi)^d} \, f(k) \biggl( \frac{1}{\vk^2} - a \, \d^{(2)} (k) \biggr) 
\eeq
where $k \equiv |\vec{k}|$. This integral is finite for any test function $f(k)$ which vanishes at infinity. Choosing a particular functional form for $f(k)$, it is possible to determine the value of $a$ and correctly regularize the IR divergence \cite{Penati:1996nt}.

\paragraph{Integral $\mathbf{J_2}$.} 
We define the integral
\beq
\label{eq:integral_J_2}
J_2 (\vp) \equiv \int \frac{d^2 k}{(2 \pi)^2} \, \frac{1}{\bigl[ - \vk^2+i \e \bigr] \bigl[ - (\vp - \vk)^2 + i \e \bigr]} \longrightarrow
\int \frac{d^2 k}{(2 \pi)^2} \, \frac{1}{\bigl[ \vk^2+\k^2 \bigr] \bigl[ (\vp - \vk)^2 + \k^2 \bigr]} 
\eeq
This integral is UV finite by power counting, but we need to introduce an IR regulator.
Using the Feynman parametrization formula and performing the change of variables $\vk \to \vk + x \vp$, we obtain
\beq
J_2 (\vp) 
= \int_0^1 d x \int \frac{d^2 k}{(2 \pi)^2} \frac{1}{\bigl( \vk^2 + \mathcal{A} \bigr)^2} 
\eeq
where $\mathcal{A} \equiv x (1-x) \vp^{\,2} + \k^2$.
Neglecting constant terms not relevant in MS scheme, a direct evaluation gives 
\beq
J_2 (\vec{p}) = \frac{1}{4 \pi \mathcal{A}} \simeq 0 
\eeq

\paragraph{Integral $\mathbf{J_3}$.} 
We consider
\beq
\begin{aligned}
\label{eq:integral_J_3}
J_3 (\O,\vp,\vp_1,\vp_2) & \equiv
\int \frac{d^2 k}{(2 \pi)^2} \,
\frac{1}{
\bigl[- \frac12 (\vp_1-\vk)^2 - \frac12 (\vp_2-\vk)^2 + m \O + i \e \bigr]
\bigl[ - \vk^2 + i \e \bigr]
\bigl[ - (\vp - \vk)^2 + i \e \bigr]}  \\
& \hspace{-0.3cm} \longrightarrow - \int \frac{d^2 k}{(2 \pi)^2} \,
\frac{1}{
\bigl[\frac12 (\vp_1-\vk)^2 + \frac12 (\vp_2-\vk)^2 - m \O + \k^2 \bigr]
\bigl[ \vk^2 + \k^2 \bigr]
\bigl[ (\vp - \vk)^2 + \k^2 \bigr]
} 
\end{aligned}
\eeq
where we have introduced a unique IR regulator $\kappa$ in the second line.

Using again the Feynman parametrization for the cubic denominator, we bring the integral to the following form 
\beq
J_3 (\O,\vp,\vp_1,\vp_2) = - 2 \int_0^1 d x \int_0^{1-x} d y \int \frac{d^2 k}{(2 \pi)^2} \, \frac{1}{\bigl( \vk^2 + \mathcal{B} \bigr)^3} 
\eeq
with
\beq
\mathcal{B} \equiv x (1-x) \vp^{\,2} + \frac{y}{2} \bigl( \vp_1^{\,2} + \vp_2^{\,2} - 2 m \O \bigr) - \frac{y^2}{4} \bigl( \vp_1+\vp_2 \bigr)^2 - x y \, \vp \cdot \bigl( \vp_1+\vp_2 \bigr) + \k^2 
\eeq
We eventually find
\beq
J_3 (\O,\vp,\vp_1,\vp_2) = - \frac{1}{4 \pi} \int_0^1 d x \int_0^{1-x} d y \, \frac{1}{\mathcal{B}^2} \simeq 0
\eeq
that is, the result is UV finite.

\section{Additional one-loop computations for the SGED action}
\label{app:additional_loop_corrections}

In this appendix, we collect additional details about the one-loop radiative corrections to the SGED model \eqref{eq:SGED_action}.
In particular, we show that vertices containing more than two (anti)chiral superfields are not generated at quantum level.

\begin{figure}[ht]
    \centering
    \subfigure[]{\label{subfig:4ptchiral1}  \includegraphics[scale=0.62]{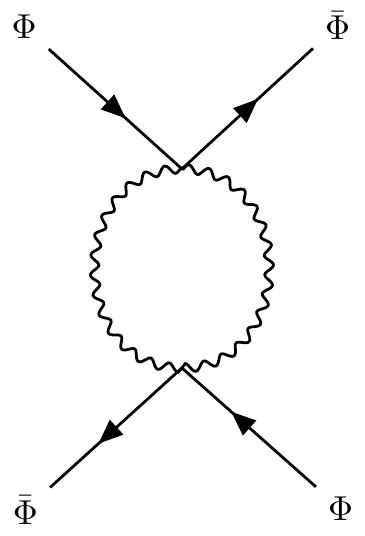}} \quad
       \subfigure[]{\label{subfig:4ptchiral3} 
      \includegraphics[scale=0.5]{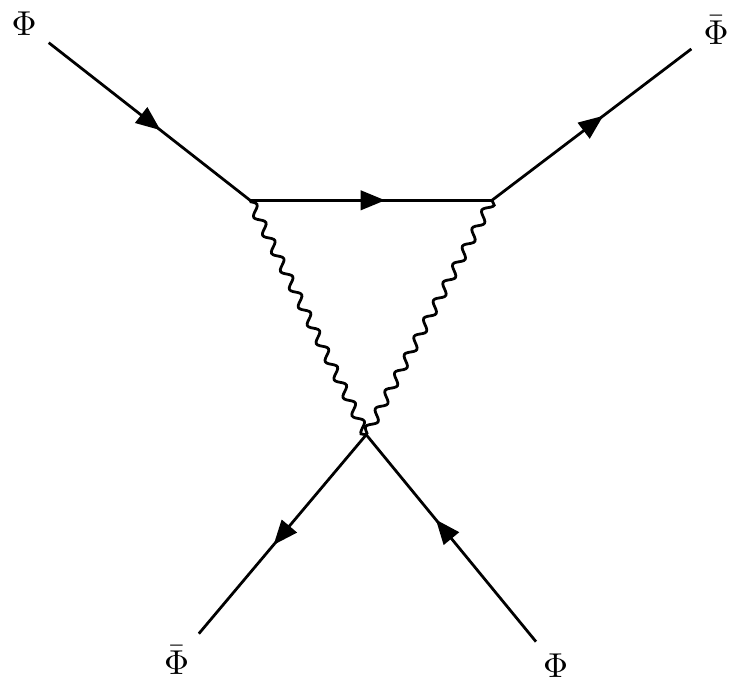}} \quad
        \subfigure[]{\label{subfig:4ptchiral4}  \includegraphics[scale=0.6]{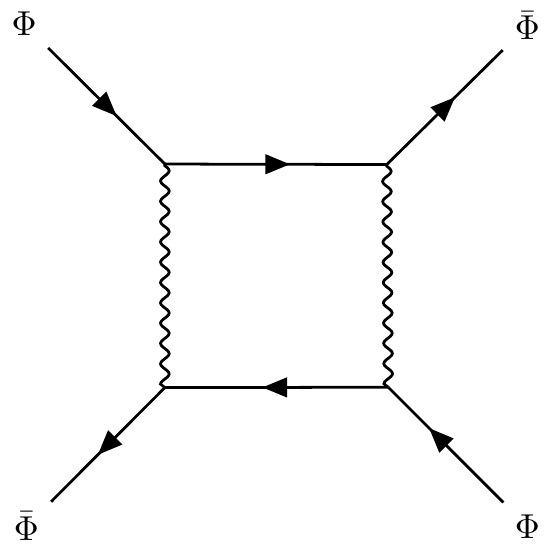}} \quad
          \subfigure[]{\label{subfig:4ptchiral5}  \includegraphics[scale=0.53]{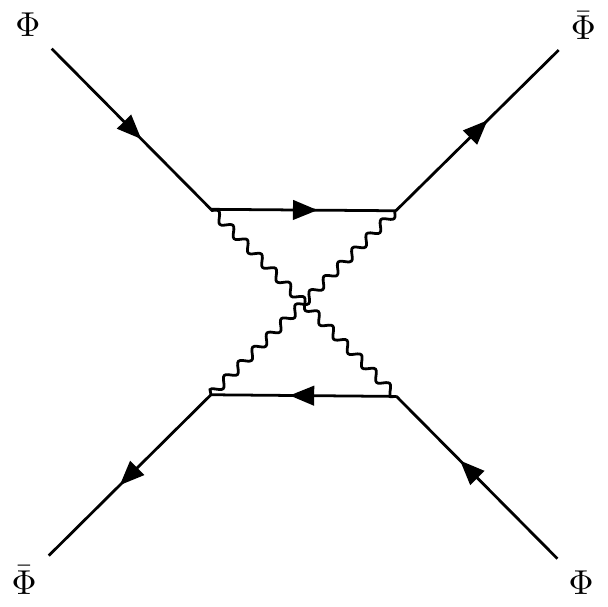}} 
    \caption{Feynman supergraphs contributing to the computation of the four-point vertex with only (anti)chiral external lines.}
    \label{fig:4pt_chiral}
\end{figure}

The first non-trivial case corresponds to the four-point vertex $\langle \bar{\Phi}^2 \Phi^2 \rangle ,$ whose one-loop Feynman supergraphs are collected in fig.~\ref{fig:4pt_chiral}.

\noindent
After performing the D-algebra, several contributions vanish by application of selection rule \ref{sel_rule1}.
The integration over the energy $\omega$ of the remaining ones can be shown to be proportional to the quantity $I$ defined in eq.~\eqref{eq:integral_I}, while the integration along the spatial momenta are proportional to either $J_2$, defined in eq.~\eqref{eq:integral_J_2}, or $J_3,$ defined in eq.~\eqref{eq:integral_J_3}.
The crucial observation is that all these results are finite, therefore no UV divergences arise, and the corresponding vertices are not generated at quantum level.

One can show that this pattern is preserved for an arbitrary number of external (anti)chiral lines. In fact:
\begin{itemize}
    \item Any additional chiral superpropagator \eqref{eq:chiral_superpropagator} brings a factor of covariant derivatives $\bar{D}^2 D^2.$
    In the worst scenario for the UV behaviour of the integral, the contribution to the superficial degree of divergence from this insertion is $\Delta_{\omega}=0$ for the frequency $\omega,$ and $\Delta_{\vec{p}}=0$ for the spatial momenta $\vec{p}.$
    In both cases, it is sufficient to repeat the computation performed for the four-point function to obtain a UV convergent integral.
    \item Any additional vector superpropagator brings superficial degree of divergence $\Delta_{\omega}=0$ and $\Delta_{\vec{p}}=-2,$ since it does not carry any factor of covariant derivatives. Therefore, it can only improve the UV behaviour of the integral, and by repeating the computation of the four-point vertex, we obtain a finite result.
\end{itemize}
The previous arguments also apply in the presence of higher loops, since the discussion on the insertion of internal superpropagators (without changing the number of external lines) follow the same reasoning.
This shows that no vertices with only (anti)chiral superfields are generated at quantum level, at any loop order.

Finally, using the same arguments listed above, it is easy to show that the addition of any number of external vector superfields does not affect the convergence properties of these diagrams.
Therefore, we conclude that no quantum corrections involving more than one chiral and one antichiral superfields are generated by perturbative computations, no matter the number of external prepotentials is.

\section{Details on the covariant approach}
\label{app:details_covariant}

As discussed in section \ref{sec:renormalizable_SGED}, background field method together with gauge covariant $\nabla$- algebra on background field supergraphs provides a very efficient way to perform perturbation theory in supersymmetric gauge theories \cite{Grisaru:1984ja}. In this appendix we collect several  details on the covariant approach, suitably adapted to the non-relativistic superspace.

\subsection{Gauge-covariant derivatives}

In the covariant approach, we introduce derivatives defined by the requirement that under a supergauge transformation \eqref{eq:gaugetransf} they transform as 
\beq
\nabla_A \rightarrow  e^{i \Lambda} \nabla_A e^{-i \Lambda}   
\eeq
The corresponding covariant derivatives in vector representation read 
\beq
\nabla_A \equiv D_A - i \Gamma_A = (\nabla_{\alpha}, \bar{\nabla}_{\beta}, \nabla_{\alpha \beta}) 
=
(e^{-\frac{V}{2}} D_{\alpha} e^{\frac{V}{2}} , e^{\frac{V}{2}}  \bar{D}_{\beta} e^{-\frac{V}{2}} , i \lbrace \nabla_{\alpha}, \bar{\nabla}_{\beta} \rbrace) 
\label{eq:vector_def_covdiv}
\eeq 
where $\Gamma_A$ are the connection superfields. 

When we perform background-quantum splitting, definition \eqref{eq:vector_def_covdiv} refers to background covariant derivatives with $V$ being the background prepotential. In this case, covariant derivatives, superconnections and superfield strengths that we introduce throughout refer to background fields. 

In the Abelian case, the superconnections are linear in the prepotential. Precisely, from \eqref{eq:vector_def_covdiv} we read
\beq
\Gamma_{\alpha} = \frac{i}{2}  D_{\alpha} V \, , \qquad
\bar{\Gamma}_{\alpha} = - \frac{i}{2}  \bar{D}_{\alpha} V
\eeq
\beq
 \Gamma_{\alpha \beta} = i D_{\alpha} \bar{\Gamma}_{\beta} + i \bar{D}_{\beta} \Gamma_{\alpha} =
- \bar{D}_{\beta} D_{\alpha} V - \frac{i}{2}  \p_{\alpha \beta} V 
\label{eq:components_connection}
\eeq
We contract covariant derivatives according to the following conventions
\beq
\nabla^2 \equiv \frac{1}{2} \nabla^{\alpha} \nabla_{\alpha}  \, , \qquad
\bar{\nabla}^2 \equiv \frac{1}{2} \bar{\nabla}_{\alpha} \bar{\nabla}^{\alpha} \, , \qquad
\square \equiv -\frac{1}{2} \nabla^{\alpha\beta} \nabla_{\alpha\beta}
\label{eq:combinations_covdiv}
\eeq
These are the covariant generalization of the differential operators defined in eq.~\eqref{eq:combinations_susy_der} for the SUSY derivatives. In particular, the covariant Schr\"{o}dinger operator can be expanded in terms of the ordinary flat Schr\"{o}dinger operator $\square_0$, as
\beq
\square = \square_0 + i \Gamma^{\alpha\beta} \partial_{\alpha\beta} + \frac{i}{2} (\partial^{\alpha\beta} \Gamma_{\alpha\beta}) + \frac{1}{2} \Gamma^{\alpha\beta} \Gamma_{\alpha\beta} 
\label{eq:cov_flat_diff_operators}
\eeq

Since the action for gauge theories is gauge-invariant, 
prescription \eqref{Berezin integration null reduction} for performing Berezin integrations turns conveniently to
\beq
 \int d^3x d^4\theta \, \tilde{\Psi}  = \int d^3x D^2 \bar{D}^2 \tilde{\Psi} \Big|_{\theta = \bar{\theta}=0} = \int d^3x \nabla^2 \bar{\nabla}^2 \tilde{\Psi} \Big|_{\theta = \bar{\theta}=0} 
 \label{eq:covariant_Berezin_integration}
\eeq
This has the advantage that covariant combinations \eqref{eq:combinations_covdiv} naturally arise.

\subsection{Covariant superpropagators}
\label{app:manipulations_cov_div}

In the covariant approach, an important role in the evaluation of quantum corrections is played by the expansion of covariant superpropagators in terms of the standard scalar one, $1/\square_0$.  

Covariant propagators for (anti)chiral superfields are expressed in terms of the differential operators $\square_{\pm}$ defined as
\beq
\{ \nabla^2 , \bar{\nabla}^2 \}  = \square_- + \nabla^\a \bar{\nabla}^2 \nabla_\a = \square_+  + \bar{\nabla}_\a \nabla^2 \bar{\nabla}^\a
\eeq
They can be explicitly written as
\beq
\square_+ = \square - i W^{\alpha} \nabla_{\alpha} - \frac{i}{2} (\nabla^{\alpha} W_{\alpha})  \, ,  \qquad
\square_- = \square - i \bar{W}_{\alpha} \bar{\nabla}^{\alpha} - \frac{i}{2} (\bar{\nabla}_{\alpha} \bar{W}^{\alpha} ) 
\label{eq:prop_pm}
\eeq
where $W_{\alpha} = i \bar{D}^2 D_{\alpha} V $ is the (background) field strength.

In the Fermi-Feynman gauge $\zeta=1$, the vector propagator is the inverse of the quadratic differential operator 
\beq
\hat{\square} = \square - i W^{\alpha} \nabla_{\alpha} - i \bar{W}_{\alpha} \bar{\nabla}^{\alpha} 
\label{eq:prop_hat}
\eeq
The differential operators introduced in eqs.~\eqref{eq:prop_pm} and \eqref{eq:prop_hat} can be further expanded in terms of the flat Schr\"{o}dinger operator by means of eq.~\eqref{eq:cov_flat_diff_operators}.

A set of useful identities that allow to efficiently perform $\nabla$-algebra is 
\bea\label{eq:boxdelidentities}
\nonumber
& \lbrace \nabla_{\alpha}, \bar{\nabla}_{\beta}  \rbrace = - i \nabla_{\alpha \beta}  \, , \qquad
 \bar{\nabla}^2 \nabla^2 \bar{\nabla}^2 = \bar{\nabla}^2 \square_+ \, , \qquad
\nabla^2 \bar{\nabla}^2 \nabla^2 = \nabla^2 \square_- \, ,
 & \\
& [ \nabla^2, \square_-] = 0 \, , \qquad
[\bar{\nabla}^2, \square_+] = 0 \, , &
 \label{eq:Dalgebra_covdiv}\\
&  \square_- \nabla^2 \bar{\nabla}^2 = \nabla^2 \bar{\nabla}^2 \square_+ \, , \qquad \square_+ \bar{\nabla}^2 \nabla^2 = \bar{\nabla}^2 \nabla^2 \square_- 
\nonumber
\eea
\bea\label{eq:inverseboxdelidentities}
&\qquad
\frac{1}{\square_-} \nabla^2 \bar{\nabla}^2 = \nabla^2 \bar{\nabla}^2 \frac{1}{\square_+} \, , \qquad
\frac{1}{\square_+} \bar{\nabla}^2 \nabla^2 = \bar{\nabla}^2 \nabla^2 \frac{1}{\square_-} \nonumber \\
& ~~ \nonumber \\
& 
\left[ \nabla_{\alpha}, \frac{1}{\square} \right] = - \frac{1}{\square} [\nabla_{\alpha}, \square] \frac{1}{\square} = 
- \frac{1}{\square} \le \bar{W}^{\beta} \nabla_{\alpha \beta} + \frac{1}{2} \nabla_{\alpha \beta} \bar{W}^{\beta}  \ri \frac{1}{\square} 
\eea

In the background field method, $\nabla$-algebra on Feynman supergraphs requires first to expand the covariant propagators in powers of the background connections and superfield strengths appearing in eqs.~\eqref{eq:prop_pm} and \eqref{eq:prop_hat}.

The expansion of the vector propagator $1/\hat{\square}$ reads
\beq
\begin{aligned}
\frac{1}{\hat{\square}} & = \frac{1}{\square} + \frac{1}{\square} \le i W^{\alpha} \nabla_{\alpha} + i \bar{W}_{\alpha} \bar{\nabla}^{\alpha}  \ri \frac{1}{\hat{\square}} = \\
& = \frac{1}{\square} + \frac{1}{\square}  i W^{\alpha} \frac{1}{\hat{\square}} \nabla_{\alpha} + \frac{1}{\square}  i \bar{W}_{\alpha} \frac{1}{\hat{\square}} \bar{\nabla}^{\alpha} + \frac{1}{\square}  i W^{\alpha} \left[ \nabla_{\alpha}, \frac{1}{\hat{\square}}  \right] + \frac{1}{\square}  i \bar{W}_{\alpha} \left[ \bar{\nabla}^{\alpha}, \frac{1}{\hat{\square}}  \right] = \\
& = \frac{1}{\square} + \frac{1}{\square}  i W^{\alpha} \frac{1}{\square} \nabla_{\alpha} + \frac{1}{\square}  i \bar{W}_{\alpha} \frac{1}{\square} \bar{\nabla}^{\alpha} + \mathcal{O} \le \frac{1}{\square} \ri^3 
\end{aligned}
\label{eq:expansion_squarehat}
\eeq
We note that the commutators in the second line include higher-order insertions,  by virtue of identity \eqref{eq:inverseboxdelidentities}.

Pictorially, the expansion in eq.~\eqref{eq:expansion_squarehat} can be depicted as follows
\beq
\includegraphics[scale=1]{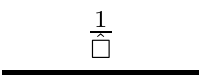} = 
\includegraphics[scale=1]{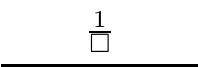} + 
\includegraphics[scale=1]{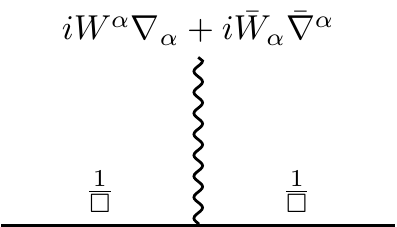} + \mathcal{O} \le \frac{1}{\square} \ri^3 
\label{eq:pictorial_expansion_hatprop}
\eeq

Using this expansion, a single covariant supergraph with an internal vector propagator gives rise to an infinite sum of supergraphs on which one still needs to perform $\nabla$-algebra. However, 
it is clear that starting from the second term, new insertions of background fields get generated, each of them leading to the addition of one extra covariant propagator $1/\square$. In general, as discussed in the main text and in appendix \eqref{app:convergence_vec_insertions}, this increasingly improves the degree of convergence of the corresponding momentum integral. Therefore, as long as we are interested only in divergent diagrams, this expansion typically stops at a given order. 

Similar manipulations of the (anti)chiral propagators give
\beq
\frac{1}{\square_+} = \frac{1}{\square} + \frac{1}{\square}  i \bar{W}_{\alpha} \frac{1}{\square} \bar{\nabla}^{\alpha} + \frac{1}{\square}  \frac{i}{2} (\bar{\nabla}_{\alpha} \bar{W}^{\alpha}) \frac{1}{\square} + \mathcal{O} \le \frac{1}{\square} \ri^3 
\label{eq:expansion_squarep}
\eeq
\beq
\frac{1}{\square_-} = \frac{1}{\square} + \frac{1}{\square}  i W^{\alpha} \frac{1}{\square} \nabla_{\alpha} + \frac{1}{\square}  \frac{i}{2} (\nabla^{\alpha} {W}_{\alpha} ) \frac{1}{\square} + \mathcal{O} \le \frac{1}{\square} \ri^3 
\label{eq:expansion_squarem}
\eeq
At graphic level, these expansions can be represented in the same way as in eq.~\eqref{eq:pictorial_expansion_hatprop}, the only difference being the precise superfield configurations appearing on the external legs. According to eqs.~\eqref{eq:expansion_squarep} and \eqref{eq:expansion_squarem}, now the insertions bring factors
$
i \bar{W}_{\alpha} \bar{\nabla}^{\alpha} + \frac{i}{2} (\bar{\nabla}_{\alpha} \bar{W}^{\alpha})$ and 
$i W^{\alpha} \nabla_{\alpha} + \frac{i}{2} (\nabla^{\alpha} W_{\alpha})$, respectively.

After completion of $\nabla$-algebra we are left with a set of supergraphs with internal lines corresponding to $1/\square$. As a last step, we need  to further expand the covariant propagator in terms of the flat Schr\"{o}dinger one, using
\beq
\frac{1}{\square} = \frac{1}{\square_0} - \frac{1}{\square_0} \le i \Gamma^{\alpha\beta} \p_{\alpha\beta} + \frac{i}{2} (\p_{\alpha\beta} \Gamma^{\alpha\beta}) + \frac{1}{2} \Gamma^{\alpha\beta} \Gamma_{\alpha\beta}  \ri \frac{1}{\square_0} + \mathcal{O} \le \frac{1}{\square_0} \ri^3 
\label{eq:expansion_cov_dAlembert}
\eeq
Pictorially, this can be represented by
\beq\label{eq:pictorial_chiral_expansion}
\begin{aligned}
\includegraphics[scale=1]{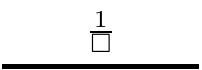} = 
\includegraphics[scale=1]{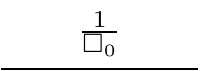} +  \includegraphics[scale=1]{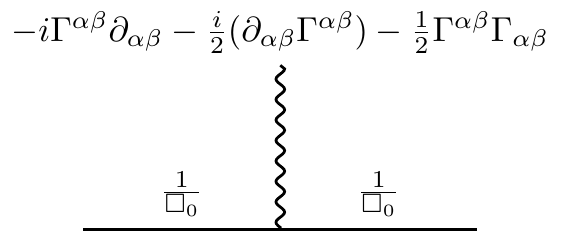} + \mathcal{O} \le \frac{1}{\square_0} \ri^3 
\end{aligned}
\eeq
As a result, we obtain a set of Feynman diagrams with ordinary propagators that give rise to ordinary Feynman integrals, either in configuration or momentum space.

\subsection{Convergence of vector insertions for the self-energy computation}
\label{app:convergence_vec_insertions}

We apply the techniques developed in appendix \ref{app:manipulations_cov_div} to show that the expansion of the covariant vector propagator $1/\hat{\square}$ in eq.~\eqref{eq:expansion_squarehat} produces only convergent one-loop corrections to the self-energy of the (anti)chiral superfields studied in section \ref{sec:original_cov_radiative_corrections}.

Let's start with the power counting of the momenta for each propagator insertion from eq.~\eqref{eq:expansion_cov_dAlembert}.
Since the vector has vanishing mass, $M=0,$ its propagator is purely spatial. It follows that any additional propagator insertion adds two factors of spatial momenta at denominator, and none along the temporal direction $\omega.$
On the other hand, any additional factor of the connection brings factors of momenta at numerator, which depend on the specific component, according to eq.~\eqref{eq:components_connection}.
The worst scenario for the convergence corresponds to the case where we select the component $\Gamma^{11} \partial_{11},$ which in momentum space carries one factor of $\omega$ at numerator, according to eq.~\eqref{eq:3d_derivatives}.
Putting all the external momenta to zero (they do not affect the UV behaviour), these contributions read 
\beq
\begin{aligned}
 -g^2 \, \int d^4 \t \int \frac{d \o \, d^2 k}{(2 \pi)^3} \, 
\frac{\tilde{\Phi}(\Omega,\vec{p}, \theta) \bar{\tilde{\Phi}} (\Omega, \vec{p}, \bar{\theta})}{(- \vec{k}^2+ i \varepsilon) \left[ 2 m (\Omega- \omega) - (\vec{p} - \vec{k})^2 + i \varepsilon \right]}  
\sum_{n=0}^{\infty}  \le  \frac{\sqrt{2} g \omega \bar{D}_2 D_2 V_0}{- \vec{k}^2 + i \varepsilon} \ri^n 
\end{aligned}
\eeq
In this case it is not possible to apply immediately  the residue theorem or Sokhotski-Plemelj formula, as done for the other Feynman diagrams considered in this work.
Instead, we perform the integration over spatial momenta first, applying Feynman parametrization
\beq
\frac{1}{AB^{n+1}} = \int_0^1 dx \, \frac{(n+1)(1-x)^{n}}{[xA + (1-x)B]^{n+2}} 
\eeq
and we define
\beq
A \equiv 2m (\Omega-\omega) - (\vec{p} - \vec{k})^2 + i \varepsilon \, , \qquad
B \equiv - \vec{k}^2 + i \varepsilon \, ,
\eeq
\beq
xA + (1-x) B = - (\vec{k} -x \vec{p})^2 + \tilde{\Delta} \, , \qquad
\tilde{\Delta} \equiv 2m x (\Omega-\omega) + x (1-x) \vec{p}^2 + i \varepsilon 
\eeq
Using polar coordinates along the two-dimensional spatial momenta, we compute
\beq
\int_0^1 dx \, (n+1) (1-x)^n \int_0^{\infty} \frac{2 \pi k dk}{(2 \pi)^2} \frac{(-1)^{n+2}}{(\vec{k}^2 - \tilde{\Delta})^{n+2}}  = \int_0^1 dx \, \frac{(1-x)^n}{2 \pi}  \frac{1}{\tilde{\Delta}^{n+1}}
\eeq
There is one $\omega$ integration left, which is of the form
\beq
  \int_0^1 dx \, (1-x)^n  \int_{-\infty}^{\infty} \frac{d\omega}{2\pi}  \frac{(\sqrt{2} g \omega)^n}{ [ 2 m (\Omega- \omega) - (\vec{p} - \vec{k})^2 + i \varepsilon ] [ 2m x (\Omega-\omega) + x (1-x) \vec{p}^2 + i \varepsilon]^{n+1}} 
\eeq
This contribution is vanishing, thanks to Jordan's lemma and residua theorem.
This shows that the inclusion of insertions on the vector propagator always improves the convergence properties of the diagram, and the only divergent terms arise from the expansion of the chiral superpropagators, computed in section \ref{sec:cov_radiative_corrections}.

\bibliographystyle{JHEP}
\bibliography{bibliography}

\end{document}